\documentclass[12pt]{aastex61}
\usepackage{amsmath}

\begin{document}

\title{Astrophysical background and dark matter implication based on latest AMS-02 data} 
\correspondingauthor{Hong-Bo Jin}\email{hbjin@bao.ac.cn, hbjin@itp.ac.cn}

\author{Hong-Bo Jin}
\affiliation{CAS Key Laboratory  of Theoretical Physics, Chinese Academy of Sciences}
\affiliation{National Astronomical Observatories, Chinese Academy of Sciences, Beijing, 100012, China}
\author{Yue-Liang Wu}
\affiliation{CAS Key Laboratory  of Theoretical Physics, Chinese Academy of Sciences}
\affiliation{University of Chinese Academy of Sciences, Beijing 100190, China}
\affiliation{Institute of Theoretical Physics, Chinese Academy of Sciences, Beijing 100190, China}
\author{Yu-Feng Zhou}
\affiliation{CAS Key Laboratory  of Theoretical Physics, Chinese Academy of Sciences}
\affiliation{Institute of Theoretical Physics, Chinese Academy of Sciences, Beijing 100190, China}

\begin{abstract}
The cosmic ray(CR)  positrons and antiprotons are often regarded as the collision products of CR nucleons with the interstellar medium. However this conclusion is challenged by recent experimental data. In this work, we choose the latest AMS-02 data to analyze the astrophysical background of CR positrons and antiprotons with the GALPROP code for CR propagation and the QGSJET-II-04 model for the hadronic interactions. The results show that in low energies the predicted fluxes of CR antiprotons and positrons are consistent with AMS-02 data in the diffusion model combining the re-acceleration and convection terms (DCR model). Thus, the excess of CR antiprotons in the low energy are not justified.  
In that model, the predicted flux of CR protons is consistent with AMS-02 data whose hardening feature above 330 GeV is also matched. Based on that model, the total fluxes of CR electrons and positrons from AMS-02 are also used to analyzed the interpretation of dark matter annihilation about the positron excess.
The predicted best-fit masses of dark matter are from 400 GeV to 4 TeV. The DCR model does not give the unusual interpretation of the positron excess.
\end{abstract}
\section{Introduction}
In the origin of Galactic CRs, the primary particles, such as nucleons and electrons, are commonly regarded as the injection of Supernova relics(SNRs), in which, CRs are accelerated by the diffusive shock \citet{Fermi:1949ee,1977DoSSR.234.1306K,Blandford:1978ky,Drury:1983zz,Blandford:1987pw}. 
In that mechanism, the injection spectra of CRs are a power law below Knee, which are verified by the experimental data  \citet{Amenomori:2008aa}. 
The other particles, called as the secondary CRs, are mainly produced in the collision between the primary particles and the interstellar medium in the Galaxy. The spectra of the secondary CRs is looked as the astrophysical background from that origin.
The fluxes of the secondary CRs are positively relevant to the collision cross-sections, which are often calculated with the hadronic interaction models.  
Since the spectra of the primary CRs are a single power law, the spectral features of the secondary CRs are mainly relevant to the hadronic interaction models. 
Besides the hadronic interaction models, the exclusive cross-section of the produced particles are also calculated with the empirical parameterizations of the Accelerator data. 
In practice, these calculations are developed into the parameterization model package, such as FLUKA  \citet{Ferrari2005,bohlen2014the}, QGSJET-II-04  \citet{Ostapchenko:2010vb}, EPOS-LHC  \citet{Pierog:2013ria}, etc. In this work, QGSJET-II-04 is chosen to calculate the products of the secondary antiproton.
    
When the charged particles of CRs propagate in the Galaxy, they may be accelerated via the interaction with the turbulent interstellar magnetic field. The processes are also called as re-acceleration in contrast to DSA. CRs also commit the energy loss when propagating in the Galactic winds, which are blowing outwards from the Galactic disc. That is called as convection in CR propagation models.  
Besides the Galactic diffusion of CRs, these interactions alter the original structure of the spectra of CRs. As a result, the measured spectra of CRs are different from the injection of the sources. 
Thus, the secondary particle spectra are also relevant to the propagation models of CRs. In the conventional model, CR production and propagation are governed by the same mechanism below $10^{17}$ eV, and CR propagation is often described by the diffusion equation \citet{Ginzburg:1990sk}. As a whole, the uncertainties of the secondary particle spectra are analyzed  with the experimental data of the primary particles of CRs in the propagation models and the hadronic interaction models. The spectral uncertainties of the secondary particles are relevant to the propagation parameters of CRs, which are constrained from the fitting to the experimental data of CRs.    

Recently, AMS-02 has reported their observed results of CRs. Below TeV, the spectra of CR protons can be described by the high precision data \citet{Aguilar:2015ooa}. CR positrons \citet{Aguilar:2014mma}, antiprotons \citet{Aguilar:2016kjl} and B/C \citet{Aguilar:2016vqr} have also the precise measurements.
In the previous analysis \citet{Jin:2014ica}, we have performed a global analysis of the propagation parameters with the AMS-02 data. The propagation parameters are well determined by the only CR proton and B/C data from AMS-02. That strategy of fitting parameters are different from the two ratios of CRs, such as Be$^{10}$/Be$^{9}$ and B/C. 
As a result, at the 95\% confidence level(C.L.), the constrained parameters are in the more narrow range than the Be$^{10}$/Be$^{9}$ and B/C case.
This situation originates from the high precision data from AMS-02. The spectral structure of CR protons has represented the alterations of the injection spectra, which is a simple power law, by the propagation processes, such as the diffusion and re-acceleration. 
In detail, the spectral feature of CR protons is also sensitive to the propagation parameters: V${}_{A}$ (Alfv$\grave{\mbox{e}}$n velocity), D${}_{0}$ (diffusion coefficient), Z${}_h$ (the height of the diffusion halo) etc. The combination of proton flux plus B/C ratio breaks the degeneracies between those parameters, which are often done by the ratio data, such as Be$^{10}$/Be$^{9}$ and B/C. The details are found in Figure 1 of the paper \citet{Jin:2014ica}.

With the propagation parameter models constrained by the measured data of CRs, the astrophysical background of the secondary particles are naturally calculated. 
Since the spectra of the secondary particles, such as CR positrons \citet{Aguilar:2014mma} and antiprotons \citet{Aguilar:2016kjl}, already have the measured data with high precision, the calculated spectra may be used to verify the propagation and the hadronic interaction model and explore the origin of the experimental data. 
The positron excess has been discovered by PAMELA \citet{Adriani:2010ib}, Fermi-LAT \citet{FermiLAT:2011ab} and AMS-02 \citet{Accardo:2014lma} in the past years, which indicated that above 10 GeV the measured fluxes of CR positrons are greater than the astrophysical fluxes predicted by the CR propagation models and the hadronic interaction models. 
The phenomenological implications can be found in refs. \citet{Jin:2013nta,Liu:2013vha,Chen:2015uha,Jin:2014ica,Zhou:2016eul}.
Below 10 GeV the calculated fluxes of CR positrons and antiprotons are also both consistent incompletely with the latest data of AMS-02, i.e. the flux of CR positrons is greater than the measured data and the flux of antiprotons is less.
For an example, in the paper \citet{Trotta:2010mx}, the fluxes of the CR positrons and antiprotons predicted from a global bayesian analysis by using \software{GALPROP v54} package \citet{Strong:1998fr} are inconsistent with the experimental data below 10 GeV. That paper indicated that the under-predicted antiprotons may result from a general feature of the re-acceleration models. The hadronic interaction model of CR antiproton product is from the parameterizations of Tan \& Ng and Duperray et al. \citet{Trotta:2010mx}, which is called as the Conventional model. That is default in GALPROP package.
Thus, if the above inconsistence is anlysized, the alternative propagation and the hadronic interaction models need be considered obviously.  

In ref. \citet{Kachelriess:2015wpa}, authors have compared the antiproton yields from the Conventional model, QGSJET-II-04 \citet{Ostapchenko:2010vb} and EPOS-LHC \citet{Pierog:2013ria}.  The result shows in the Conventional model, the antiproton yield is less than the others below 10 GeV. And in the ranges from 10 GeV to 100 GeV, EPOS-LHC model has more contribution to the antiproton yields than the Conventional and QGSJET-II-04 model. Thus, in this work QGSJET-II-04 is moderate model to describe the hadronic interaction from the low to high energy, which is chosen as the hadronic interaction model to enhance the flux of CR antiprotons in the low energy. In the GALPROP package, the codes relevant to producing CR antiprotons are replaced with QGSJET-II-04 model by us. The modified version of GALRPOP is used in our analysis.

In this paper, the analysis strategy of CR positrons and antiprotons is the following. 
The calculations of the cross-section relevant to producing antiprotons between the interstellar medium and the primary CRs, are performed by Monte Carlo (MC) generator QGSJET-II-04 \citet{Ostapchenko:2010vb}. That has been parameterized as the Z-factors in ref. \citet{Kachelriess:2015wpa}.
In order to compare the difference between the Conventional and QGSJET-II-04 models, the Z-factors are also recalculated for CR antiprotons and positrons using \software{CRMC v1.6.0} package \citet{crmc,Baus:2015sez}.
In the source terms of propagation equation, the injection spectra of the primary CRs are characterized by a continuous function in the referred rigidity of CRs, which analytically express a simple power law with the same indices below/above the referred rigidity. That is also used in the analysis of the features of the measured spectra of CRs by Voyager \citet{Stone2013,Corti:2015bqi}. 
In the propagation models, the re-acceleration and convection of the Galactic CRs are both taken into account. 
The propagation equation of CRs is solved numerically using GALPROP package \citet{Strong:1998fr}. The effect of the solar modulation of CRs is considered using the model of force-field approximation \citet{Gleeson:1968zza}.

Based on the above strategy, the fluxes of the CR positrons and antiprotons are predicted. In the re-acceleration diffusion (DR) model the under-predicted flux of CR antiprotons has been intensified to be consistent with the AMS-02 data \citet{Aguilar:2016kjl}. 
The more product of antiprotons from QGSJET-II-04 model enhances the flux of CR antiprotons. 
While the predicted flux of CR positrons below 10 GeV keeps inconsistence with the AMS-02 data \citet{Aguilar:2014mma}. When the product of CR positrons is also based on the QGSJET-II-04 model, the calculated flux of CR positrons is not enhanced to be consistent with AMS-02 data. In the low energies the product of positrons are less slightly in the QGSJET-II-04 model than the Conventional model and not enough to depress the over-predicted flux. The Z-factor comparison is found in the right of Figure 1.  
In the strategy of the exclusive potential of the solar modulation for CR positrons, the over-predicted flux of CR positrons below 10 GeV cannot be reduced. If The propagation parameters constrained by the experimental data of CR protons and B/C ratio are chosen in the large C.L. , the consistent flux of CR positrons does not be also found. When the strategy is used by combining the convection and re-acceleration diffusion (DCR) models , the over-predicted flux of CR positrons can be depressed. In the meantime the flux of CR antiprotons is also consistent with the AMS-02 data.
In the propagation of CRs, the flux of CR positrons are decreased by Galaxy wind, which is described as the convection term in the propagation equation.  
Thus, DCR model is more available than DR model in the prediction of the secondary CRs. In this work the QGSJET-II-04 model and the DCR model are used to explore the dark matter contribution to positron excess antiproton excess above 100 GeV.  

In the compatible analysis with ACE data (Be$^{10}$/Be$^9$) \citet{yanasak2001}, the propagation parameters constrained by the CR protons, antiprotons, positron and B/C from AMS-02 experiment do not well predict the values of Be$^{10}$/Be$^9$ compatible with ACE data($\chi^2$/N=38/4). 
However, if the measured data of CR positrons and antiprotons are not included, the constrained propagation parameters can predict the consistent values completely with the experimental data ($\chi^2$/N$<2$ for the three kind of experimental data: CR protons, Be$^{10}$/Be$^9$ and B/C). Thus, it implied that the predicted fluxes of secondary antiprotons and positrons have some tensions with the predicted values of Be$^{10}$/Be$^9$.
   
This paper is organized as follows. In section 2, we outline the framework for the calculation of 
the propagation of the cosmic-ray particles and the cross-section of interaction between nucleons. In section 3, we describe the data selection and the strategy of the data fitting in a number of propagation models. The numerical results are presented in section 4. Our conclusions are given in section 5.

\section{Cosmic ray propagation and hadronic interaction model}\label{sec:CRpropagation}
\subsection{Cosmic ray propagation equation and parameters}
In the conventional model, CR production and propagation are governed by the same mechanism at energies below $10^{17}$ eV. CR propagation is often described by the diffusion equation \citet{Ginzburg:1990sk}:
\begin{align}\label{eq:propagation}
  \frac{\partial \psi}{\partial t} =&
  \nabla (D_{xx}\nabla \psi -\mathbf{V}_{c} \psi)
  +\frac{\partial}{\partial p}p^{2} D_{pp}\frac{\partial}{\partial p} \frac{1}{p^{2}}\psi
  -\frac{\partial}{\partial p} \left[ \dot{p} \psi -\frac{p}{3}(\nabla\cdot \mathbf{V}_{c})\psi \right]
  \nonumber \\
  & -\frac{1}{\tau_{f}}\psi
  -\frac{1}{\tau_{r}}\psi
  +q(\mathbf{r},p)  ,
\end{align}
where $\psi(\mathbf{r},p,t)$ is  the number density per unit of total particle momentum, which is related to the phase space density $f(\mathbf{r},p, t)$ as $\psi(\mathbf{r},p,t)=4\pi p^{2}f(\mathbf{r},p,t) $.
$D_{xx}$ is the spatial diffusion coefficient parametrized as
\begin{align}
\label{eq:Dxx}
D_{xx}=\beta D_{0} \left( \frac{\rho}{\rho_{0}} \right)^{\delta_{1,2}}  ,
\end{align}
where $\rho=p/(Ze)$ is the rigidity of the CR particles, and $\delta_{1(2)}$ is the index below (above) a reference rigidity $\rho_{0}$.  
The parameter $D_{0}$ is a normalization constant and $\beta=v/c$ is the ratio of the velocity $v$ of the CR particles to the speed $c$ of light.
$\mathbf{V}_{c}$ is the convection velocity related to the drift of CR particles from the Galactic disc due to the Galactic wind. 
The diffusion in the momentum space is described by  the re-acceleration parameter $D_{pp}$ related to the  Alfv$\grave{\mbox{e}}$n speed $V_{a}$, i.e. the velocity of turbulences in the hydrodynamical plasma, whose level is characterized as $\omega$ \citet{Ginzburg:1990sk,Seo:1994aug}:
\begin{align}
D_{pp}=
\frac{4V_{a}^{2} p^{2}}
{3D_{xx}\delta_{i}
\left(4-\delta_{i}^{2}\right)
\left(4-\delta_{i}\right)\omega},
\end{align}
where $\delta_{i}=\delta_{1}$ or $\delta_{2}$ is the index of the spatial diffusion coefficient. 
$\dot{p}$, $\tau_{f}$ and $\tau_{r}$ are the momentum loss rate, the time scales for fragmentation and the time scales for radioactive decay, respectively.
The momentum loss rate of CR electrons is not the same as CR nucleons, and the relevant expressions are found in the APPENDIX C of paper  \citet{Strong:1998pw}.

The convection term in the equation\eqref{eq:propagation} is used to describe the Galactic wind blowing outwards from the Galactic disc, and in  GALPROP \citet{Strong:1998pw}, the wind velocity $\mathbf{V}_{c}$ is expressed as  \citet{Strong:1998pw}:
\begin{align}
\label{eq:Vc}
V_{c}=V_0+z\frac{dV_c}{dz},
\end{align} 
z is the height perpendicular to the Galactic disc and also appears in the next equation \eqref{eq:source}.
 
The source $q(\mathbf{r},p) $ of the primary particles is often described as a broken power law spectrum multiplied by the assumed spatial distribution described in the cylindrical coordinate (R,z) \citet{Strong:1998pw}:
\begin{align}
\label{eq:source}
q_{A}(R,z)=q_{0}c_{A}
\left( 
\frac{\rho}{\rho_{br}}
\right)^{\gamma_{s}}
 \left( \frac{R}{R_{\odot}} \right)^{\eta}
\exp
\left[
-\xi \frac{R-R_{\odot}}{R_{\odot}}
-\frac{|z|}{0.2~\text{kpc}}
 \right]~,
\end{align}
where $\eta=0.5$, $\xi=1.0$ and the parameter $q_{0}$ is normalized with the propagated flux of CR protons. $c_{A}$ is the relative abundance of the Ath nucleon.
The reference rigidity $\rho_{br}$ is described as the breaks of injection spectrum. $\gamma_{s}$ is the power indices below(above) a reference rigidity. 

In this paper, the injection spectra of the primary CRs are described by the expression of the continuous functions in the referred rigidity of CRs, which may express a simple power law with the same indices below/above the referred rigidity. The spectral index difference between CR species is not considered in the paper. The expression $(\frac{\rho}{\rho_{br}})^{\gamma_{s}}$ is replaced by the following,
\begin{align}
\label{eq:priInjection}
\left(\frac{\rho}{\rho_{br}}\right)^{-\gamma_{1}}\left(1+\left(\frac{\rho}{\rho_{br}}\right)^{\frac{-\gamma_{1}-(-\gamma_{2})}{\alpha}}\right)^{-\alpha},
\end{align}
$\gamma_{1}$ and $\gamma_{2}$ are the spectral indices below/above the referred rigidity, $\alpha$ determines the smoothness of the spectral change in the left and right sides of the referred rigidity, when $\alpha$ is $0$, as a broken power law, the expression is same as the one in the equation \eqref{eq:source}.
In this work, $\alpha$ is taken as a free parameter to analyze the smoothness of the injection spectra of CR nucleons.   
  
The flux of secondary particles is derived from the primary particle's spectra, spatial distribution and interaction with the interstellar medium. The calculation of secondary particle flux is referred to the refs. \citet{Strong:1998pw,Kelner:2006tc}. 

\subsection{the secondary particle production in the interstellar medium and the analytical expression of Z-factors}
CR antiprotons and positrons are produced in the collision with the interstellar medium, and in the propagation equation\eqref{eq:propagation} their source terms are described as follows \citet{Moskalenko:1997gh},
\begin{align}\label{secoundayCRs}
q_{\bar{p},e^{\pm}}(p)=\frac{c}{4\pi}\frac{dn(p)}{dt}=\frac{c}{4\pi}\sum_{i=H,He}n_{i}\sum_{j}\int dp^{'}\beta
n_{j}(p^{'})\frac{d\sigma_{ij}(E_{tot},p^{'})}{dE_{tot}},
\end{align}
$n_{i}$ is the interstellar H and He density, $n_{j}(p^{'})$ is density of No. $j$ CR nucleon, and $\sigma_{ij}$ is cross-section between the No. $j$ CR nucleon and interstellar H and He.

With the cross-sections between the nucleon's collision, which are taken from the theoretical models or the collision experimental data, the equation\eqref{secoundayCRs} is often used to calculate the flux of secondary CRs. Recently, these calculations are improved by using Z-factors \citet{Kachelriess:2014mga,Kachelriess:2015wpa}, where they took a numerical calculation instead of the equation\eqref{secoundayCRs}. The concerned details may be found in the papers \citet{Kachelriess:2014mga,Kachelriess:2015wpa}. As the calculated Z-factors are relative to a simple power law spectrum, in this paper, a broken power law spectrum is chosen to recalculate the Z-factors using CRMC v1.6.0 package \citet{crmc}. The details of CRMC are found in the paper \citet{Baus:2015sez}. In the recalculation of Z-factors, the collision between the particles is more than 8,780,000 times.  

For the Z-factor expression of CR antiproton case, the equation\eqref{secoundayCRs} is modified as:
\begin{align}\label{zFactorsEq}
 q^{ij}_{\bar p}(E_{\bar p}) = n_i \, I_j(E_{\bar p}) \, Z^{ij}_{\bar p}(E_{\bar p},\alpha_j)\,.
\end{align} 
Here, $\alpha_j$ are the power indice of the interstellar spectra $I_j(E_{\bar p})\propto E_{\bar p}^{-\alpha_j}$ of No. $j$ CR nucleon. the $Z$-factor $Z^{ij}_{\bar p}$ is expressed via the inclusive
spectra of antiprotons $d\sigma^{ij\rightarrow \bar p}(E,z_{\bar p})/d z_{\bar p}$ with
$z_{\bar p}=E_{\bar p}/E$, as follows

\begin{align}\label{zFactorsExpression}
 Z^{ij}_{\bar p}(E_{\bar p},\alpha_j) = \int_0^1 d z \, z^{\alpha-1}\,
 \frac{d\sigma^{ij\rightarrow \bar p}(E_{\bar p}/z,z)}{d z}\,.
\end{align}

In Table (1) of paper \citet{Kachelriess:2015wpa}, Z-factors are calculated with the modified QGSJET-II-4 model, whose values are listed discretely in the limited ranges of the energy and the spectral indices. In the calculation of secondary particle's spectra, these numbers are not used conveniently.
In this paper, these numbers are interpolated and fitted to an analytic expression, which is written in a good approximation as continuous functions     
\begin{equation}\label{zFactorsFitExpressionCore}
\begin{array}{rl}
\sigma_j(\epsilon_{\bar p},\gamma_j)&=\gamma_j^{-6}ln^{0.8}[\, 0.16\epsilon_{\bar p}(10-ln\,A_j)+1\, ](1+\frac{1}{\epsilon_{\bar p}}+\frac{a}{\epsilon_{\bar p}^{3}})^{-2},\\
\gamma_j&=\frac{\gamma_{2,j}}{1+\frac{1}{12}(\frac{\rho_{br}}{E_{\bar p}})^{\gamma_{2,j}-\gamma_{1,j}}}.\\
\end{array}
\end{equation}
Where, the dimensionless quantity $\epsilon_{\bar p}=\frac{E_{\bar p}}{\scriptsize\mbox{GeV}/\scriptsize\mbox{n}}$ denotes kinetic energy $E_{\bar p}$ over GeV per nucleon.
The above spectral indice $\alpha_j$ is replaced by $\gamma_j$.
$A^i_{gas}$ and $A_j$ are the nucleon number of the $i$th interstellar gas and the $j$th CR nucleon respectively.
a=0.5 when $A_j\leq 4$ and a=1.0 when $A_j>4$. In the expression of $\gamma_{j}$, the three parameters $\gamma_{1,j}$, $\rho_{br}$ and $\gamma_{2,j}$ are served for the broken power law spectra of the primary CRs.

With the equation \eqref{zFactorsFitExpressionCore}, $Z^{ij}_{\bar p}(\epsilon_{\bar p},\gamma_j)$  is simply rewritten as,
\begin{equation}\label{zFactorsFitExpression}
Z^{ij}_{\bar p}(\epsilon_{\bar p},\gamma_j)=C^{ij}A^i_{gas}A_j\sigma_j(\epsilon_{\bar p}).\\
\end{equation}
Where, $C^{ij}$ is a discrete function expressed as 
\begin{equation}\label{zFactorsFitExpressionCfactor}
C^{ij}=\left\{\begin{array}{lll}
9.44,& & A^i_{gas}=1\\
8.97,& & A^i_{gas}=4, A_j=1\\
7.08,& & A^i_{gas}=4, A_j>1\\
\end{array}\right.
\end{equation}
  
Based on the analytic expression \eqref{zFactorsFitExpression}, the source term of the secondary antiproton is modified as,
\begin{align}\label{antiProtonFlux}
q_{\bar p}(E_{\bar p}) = \sum_{i=H,He}n_i\sum^{A_{max}}_{j=1} \, I_j(E_{\bar p}) \, Z^{ij}_{\bar p}(E_{\bar p},\gamma_j),
\end{align} 
$A_{max}$ is the maximum nucleon number of the chosen particle of CRs, which mainly contribute to the production of CR antiprotons. 
 \begin{table}[htb]
\begin{center}
\begin{tabular}{l|rrrr|rrrr}
  \hline\hline
$E_{\bar p}$&\multicolumn{4}{c|}{($A^i_{gas},A_j$) in this paper Eq.\eqref{zFactorsFitExpression}} &\multicolumn{4}{c}{($A^i_{gas},A_j$) in Ref. \citet{Kachelriess:2015wpa}}\\
&\small(P,P)&\small(P,He)&\small(He,P)&\small(He,CNO)&\small(P,P)&\small(P,He)&\small(He,P)&\small(He,CNO)\\
\hline
1 &0.008&0.028&0.025&0.189&0.00772 &0.0248 &0.0277 &0.196\\
10 &0.094&0.360&0.310&3.617&0.1 &0.35 &0.339 &3.24\\
100 &0.178&0.694&0.587&7.106&0.187 &0.715 &0.612 &7.15\\
1000 &0.244&0.960&0.805&9.902&0.248 &0.978 &0.787 &9.81\\
10000 &0.304&1.199&1.002&12.428&0.307 &1.2 &0.959 &12\\
\hline\hline
\end{tabular}
\end{center}
\caption{The Z-factors from the expression \eqref{zFactorsFitExpression} and Ref. \citet{Kachelriess:2015wpa}. As an example, $\gamma_j$ is chosen as 2.4 without the break. for the injection nucleons $A_j$, the nucleon numbers are from the values in Table 1 of Ref. \citet{Kachelriess:2015wpa}. As an example, the $A_j$ is 14 for CNO. The target nucleons $A^i_{gas}$ are P and He.}
\label{tab:fittingCheck}
\end{table}
In order to check the values of the expression \eqref{zFactorsFitExpression}, the calculated values in an example $\gamma_j$=2.4 are listed in Table \ref{tab:fittingCheck}. If the standard deviation $\sigma$ is 0.02, the total $\chi^2$ over data points calculated by the differences between the values of the expression \eqref{zFactorsFitExpression} and the reference \citet{Kachelriess:2015wpa} is near 1.0 for all cases ($A^i_{gas},A_j$). Thus, the errors may be accepted. 

In this paper, the CR propagation equation \eqref{eq:propagation} is solved by GALPROP v54 package, which is based on a Crank-Nicholson implicit second-order scheme \citet{Strong:1998pw}.
In order to solve the equation, a cylindrically symmetric geometry is assumed. And the spatial boundary conditions assume that the density of CR particles vanishes at the boundaries of radius $R_h$ and half-height $Z_h$. 
The calculation option of the tertiary antiprotons is turn on in GALPROP package. As the flux of the tertiary antiprotons is less than the secondary ones, the production of the tertiary antiprotons does not enhance the total fluxes of CR antiprotons to match the AMS-02 data.

At the top of the atmosphere of the Earth, CR particles are affected by the solar winds and the heliospheric magnetic field. The force-field approximation is used to describe that effect and the solar modulation potential $\phi$ denotes the force field intensity \citet{Gleeson:1968zza}. In this paper, $\phi$ is a free parameter and in some fitting, the difference of that between the experimental data is also taken for granted.
\section{Data selection and fitting schemes}\label{sec:fitSchemes}
\subsection{The propagation parameters and the experimental data selections}
In our previous paper \citet{Jin:2014ica}, it was found that the propagation parameters: half-height $Z_h$,  diffusion parameters $D_{0}$ and $\delta(\delta=\delta_{1,2})$, Alfv$\grave{\mbox{e}}$n speed $V_{A}$, and power indices:  $\gamma_{1,2}^{p}$ below(above) a reference rigidity $\rho_{br}$ of CR protons, can be determined by the AMS-02 data: proton flux (P) and the ratio of Boron to Carbon flux (B/C). In this paper, in order to constrain the fluxes of CR antiprotons and positrons, besides AMS-02 data of P and B/C, we also include the latest released AMS-02 data of CR antiprotons and positrons. It is already known that in the re-acceleration diffusion model the flux of CR positrons below 10 GeV is over-predicted, which perhaps implies that the other effects of CRs propagation and interaction etc. would be considered. These effects concern the solar modulation $\phi$ and the convection, i.e. Galactic winds blowing outwards from the Galactic disc, which is denoted by $V_c$ and $dV_c/dZ_h$. 

\begin{figure}
\includegraphics[width=0.49\textwidth]{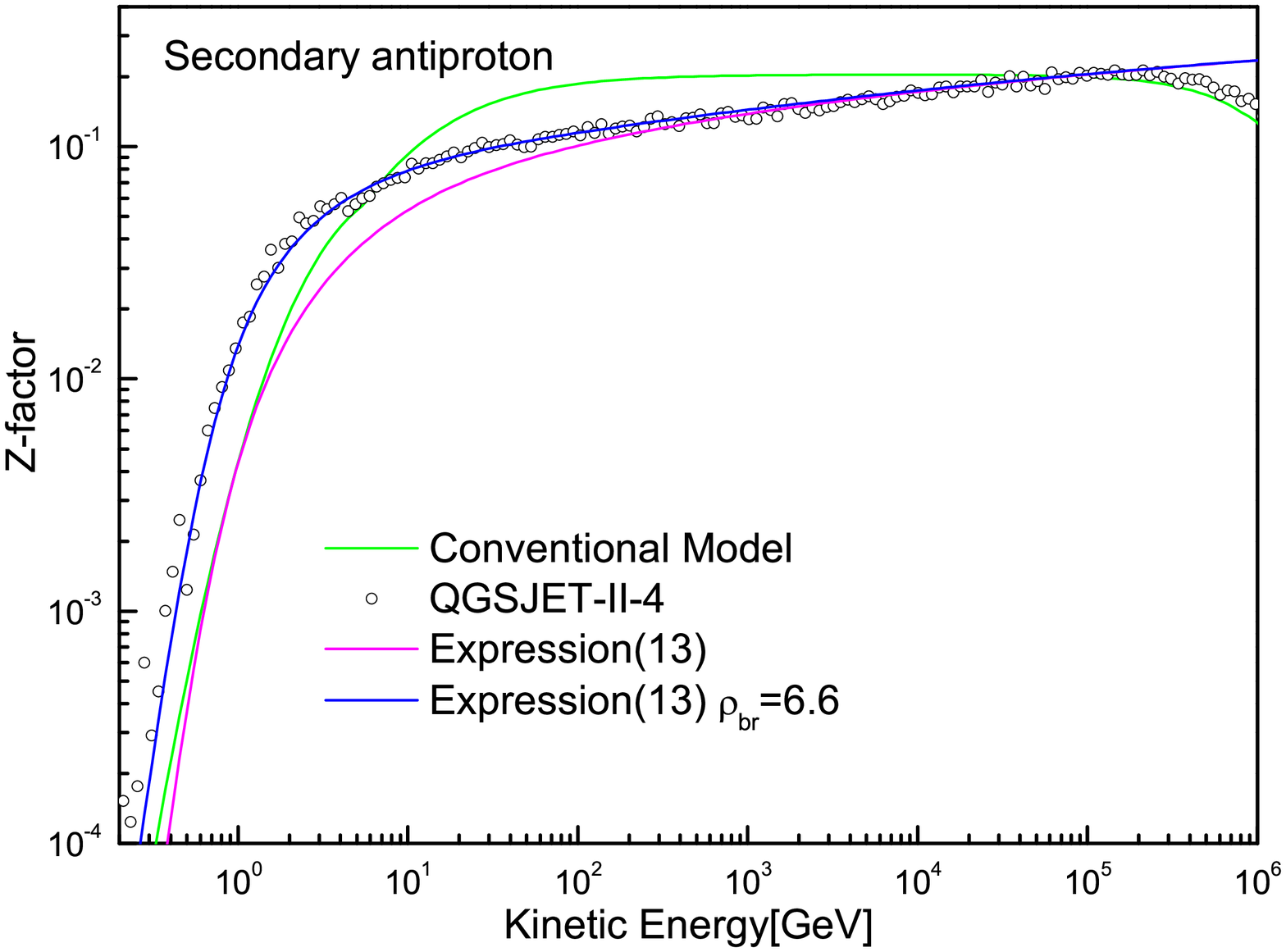}\includegraphics[width=0.49\textwidth]{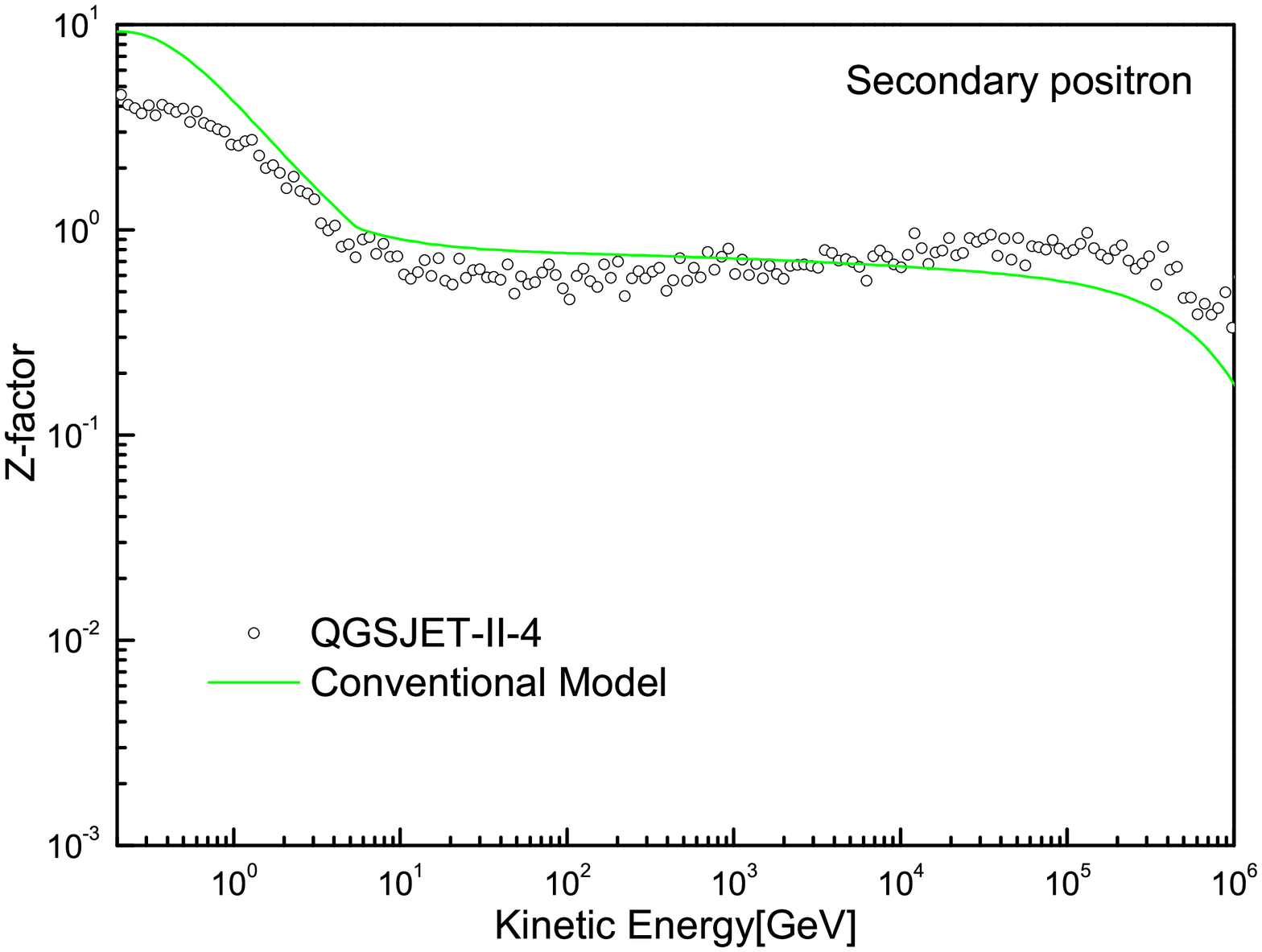}
\caption{In the Conventional and QGSJET-II-4 models, the Z-factors are calculated for the secondary antiproton(left) and positron(right) based on a broken power law spectrum of the primary particles. And the expression\eqref{antiProtonFlux} fitting to QGSJET-II-4 model is denoted with Expression (13)$\rho_{br}=6.6$. In the left figure, as a comparison, the line relevant to a simple power law is drawn and the legend is relative to Expression (13).}
\label{fig:zfactorAll}
\end{figure}

In order to explore the difference between the Conventional and QGSJET-II-4 models, the Z-factors are recalculated for the secondary antiproton and positron based on a broken power law spectrum of the primary particles. The relevant figures are shown in Figure \ref{fig:zfactorAll}. In the left of Figure \ref{fig:zfactorAll}, it is seen that the Z-factors relevant to QGSJET-II-4 model are greater than the Conventional model in the low energies. For CR positrons, the Z-factors relevant to QGSJET-II-4 model are just slightly less than the Conventional model, which is seen in the right of Figure \ref{fig:zfactorAll}.

In the chis-square fitting of combining the secondary antiproton and positron data, in order to avoid the propagation parameters deviating the determined ranges by CR protons and B/C data, the sensitive intervals of the energy of CRs, which are relevant to the interaction model (for antiproton) and source (for positron), are necessary to be chosen. For CR positrons, the energy range of the fitting to CR positron data of AMS-02 is limited in 0.6 GeV - 6.0 GeV. Above  6.0 GeV, the positron excess begins to appear and increases with energy raising in the CR positron data of AMS-02 experiment. For CR antiprotons, the cross point of the Z-factors between the conventional model and QGSJET-II-4 is 6 GeV, which is found in the figure \ref{fig:zfactorAll}. From the comparison between the fitting line and MC points,Below  2 GeV, the uncertainties of the calculated z-factors are large which is found in the figure \ref{fig:zfactorAll}. Since from 6.0 GeV to 100 GeV CR antiproton flux begin to be consistent with AMS-02 data, which is not issue in many paper for past time. Thus, the energy range of CR antiproton flux is restricted in 2.0 GeV - 6.0 GeV. 
In order to explore the energy range sensitive to the propagation parameters, the chis-square fitting for all of CR antiproton data from AMS-02 are done as a test. The result shows the overlarge convention velocity (3050 km/s) is favored to match the minimal chis-square for CR antiprotons ($\chi^2$/N=94/57). Above 100 GeV the predicted flux of CR antiprotons is closest to AMS-02 data and includes the part of the excess flux from dark matter annihilation. That is shown in the right of Figure \ref{fig:BeComparison}. It implies that the other sources need be considered to contribute to CR antiprotons so as to tune the propagation parameters to the reasonable bounds. For the estimation of the astrophysical background, the flux of CR antiprotons should be predicted in the energy range sensitive to the propagation parameters. 

Besides the smoothness parameter $\alpha$ in the Equation \eqref{eq:priInjection}, there are in general 12 fitting parameters: $Z_h, D_{0}$, $\delta$, $V_{A}$, $\gamma_{1}^{p}$,$\gamma_{2}^{p}$, $\rho^p_{br},N_p$(Normalization of CR flux), $V_c$, $dV_c/dZ_h$, $\phi$ and $\alpha$, which are determined from fitting four groups of AMS-02 data: Proton \citet{Aguilar:2015ooa}, B/C \citet{Aguilar:2016vqr}, antiproton \citet{Aguilar:2016kjl} and positron \citet{Aguilar:2014mma}.  Be$^{10}$/Be$^{9}$data from ACE experiment \citet{yanasak2001} are also used to analyze the compatibility with AMS-02 data.

With the 12 parameters and four groups of AMS-02 data, the parameter models are constructed to analyze the constrained flux of CR antiprotons and positrons from AMS-02 data. The results are presented in Tables \ref{tab:parameters} and \ref{tab:chisquares}. 
In Table \ref{tab:parameters} and \ref{tab:chisquares}, DCR denotes for a diffusion model including the convection and re-acceleration effects, DR for a diffusion model involving only the re-acceleration effect, and DC for a diffusion model containing only the convection effect. 
In the DR model, the solar modulation potential $\phi$ is taken with different values for CR positrons and CR nucleons. 
In the DCR model, some hadronic models are improved to replace the equation \eqref{zFactorsFitExpression}. 
In order to compare the recalculated flux of CR antiprotons, a DCR model denoted as DCR${}_{0}$ model is included, which is relevant to the conventional hadronic model \citet{Moskalenko:2001ya}. 
In the DCR model, Alfv$\grave{\mbox{e}}$n speed $V_{A}$ is not different between CR positrons and nucleons. 
The DCR${}_{V}$ model is constructed to verify whether re-acceleration has the different effect between CR positrons and nucleons. 

In order to check the value of $\mbox{Be}^{10}/\mbox{Be}^{9}$ compatible with ACE data, the propagation parameters are re-fitted by including the ACE data in the DCR model. 
In the DCR${}_1$ model, the five groups of the experimental data: CR protons, antiprotons, positrons, B/C and Be$^{10}$/Be$^{9}$ 
are included to fit the best-fit parameters. 
In the DCR${}_2$ model, the fluxes of CR antiprotons and positrons are not predicted and only calculated with the chosen parameters, which are the scanned values relevant to $\chi^2/\mbox{N} <2$ for CR protons, B/C and Be$^{10}$/Be$^{9}$ from the DCR${}_1$ model.

\subsection{The fitting schemes of dark matter implications}
For the dark matter implications of the AMS-02 data, the prediction of mass and annihilation cross-section of dark matter are based on the background of the total CR electrons and antiprotons, whose fluxes are calculated with the best-fit parameters in the DCR model of the context. 
The fluxes of CR electrons are divided into the primary and secondary ones. 
In ref. \citet{Chen:2014nea}, in the DR model, the flux of primary electrons are predicted by the difference between CR electron and positron of the AMS-02 data. 
In this paper, the basic propagation parameters are chosen fixedly from the DCR Model in Table \ref{tab:parameters} and the parameters relevant to the injection strustrue of the primary electrons are fitted freely with the constraints of AMS-02 data. The best-fit $\chi^2/n$ is 21.74/72. The concerned parameters are given in Table \ref{tab:para_electronMinusPositon}. 
As seen in the table, above GeV there are three spectral indices to describe the spectra of the primary electrons, which indicates that the primary electrons have a complex feature connected with the other origin. That has been discussed in the paper \citet{Chen:2014nea}.
\begin{table}[htb]
\begin{center}
\begin{tabular}{lllllllll}
\hline\hline
$N_e$&$\gamma^{e}_{1}$&$\rho^{e}_{br1}$&$\gamma^{e}_{2}$&$\rho^{e}_{br2}$&$\gamma^{e}_{3}$\\
\hline
0.4167&0.5718&1.4623&2.6743&93.32&2.4768\\	
\hline\hline
\end{tabular}
\end{center}
\caption{Best-fit values of the parameters with the constraints of AMS-02 data. The unit of normalization $N_e$ is $10^{-9}\mbox{MeV}^{-1}\mbox{cm}^{-2}\mbox{sr}^{-1}\mbox{s}^{-1}$, and the referred energy of CR electrons is 34.5 GeV. The breaks of injection electron spectra are $\rho^{e}_{br1,2}$, whose unit is GV. $\gamma^{e}_{1,2,3}$ are spectral indices below/above the breaks, respectively.}
\label{tab:para_electronMinusPositon}
\end{table}

In the fitting of the annihilation cross-section of the dark matter, the annihilation channels of the dark matter contributing to CR electrons, positrons and antiprotons involve the following particle states: 
\begin{itemize}
        \item $2\mu, 4\mu, 2\tau, 4\tau, W^{+}W^{-}, b\bar{b}, ZZ, q\bar{q},$ hh and $t\bar{t}$ for CR positrons and electrons
        \item $W^{+}W^{-}, b\bar{b}, ZZ, hh, q\bar{q}$ and $t\bar{t}$ for CR antiprotons
\end{itemize}
In the fitting strategy, the limits on annihilation cross-section are calculated with the best-fit parameters based on the given mass of dark matter, which is called as the mass-fixed best-fit and also the more tensive in the annihilation cross-section limits. A global minimum in any values of dark matter mass is relevant to the minimal chi-square.

The annihilation spectra of Majorana dark matter particles via these channels are calculated using the numerical package \software{PYTHIA v8.175} \citet{Sjostrand:2007gs}.
In the paper \citet{Ciafaloni:2010ti}, it is known that for the dark matter particles with the mass in the TeV range, the electroweak corrections are important in particular for annihilation/decay in leptonic channels. In this work, the electroweak corrections are always turned on for the calculation of the annihilation spectra of the dark matter. The analysis from dark matter decays is found in the paper \citet{Mambrini:2015sia}.

Through the global $\chi^2$-fit using the \software{MINUIT} package, all the best-fit values of the propagation parameters and the CR spectra are derived from the minimized $\chi^2$.  
In Table \ref{tab:parameters}, the best-fit parameters of each model are listed. 
In Table \ref{tab:chisquares}, the corresponding relations between the models and their concerned experimental data are presented, which shows the best-fit $\chi^2$ values relevant to the propagation models and the experimental data concerned. 

\begin{table}[htb]
\begin{center}
\begin{tabular}{crrrrrrrrrrrr}
\hline\hline
Para.&DCR&DR&DC&DCR${}_{V}$&DCR${}_{0}$&DCR${}_{1}$&DCR${}_{2}$\\
\hline
$\alpha$&0.0&0.238 &8.0E-3&0.0&0.0&0.0&0.0\\
$\phi(\phi^{e^+})$&701.9&597.4(817.6)&394.8&603.5&601.2&664.8&463.2\\
\hline
$V_A(V_A^{e^+})$&83.12 &44.46&&156.37(178.43)&68.44&83.2&43.9\\
$V_{c0}$&496.75&&32.21&4673.1&435.8&483.9&114.2\\
$\frac{dV_c}{dz}$&90.99&&2.57&26.2&61.78&75.5&44.7\\
\hline
$Z_h$&2.695&3.7&4.0&4.013&4.1&3.0&3.0 \\
$D_0/Z_h$&2.726&1.636&1.164&21.166&1.891&2.946&1.222\\
$\delta$&0.305&0.321&0.393&0.144&0.366&0.291&0.403\\
$N_p$&4.609&4.533&4.53&4.564&4.649&4.583&4.571 \\
$\rho_{br}$&6.6&10.153&8.41&5.032&4.873&6.6&6.6 \\
$\gamma^{p}_{1}$&1.8398&1.772&1.748&1.654&1.586&1.808&1.538 \\
$\gamma^{p}_{2}$&2.4125&2.447&2.427&2.411&2.398&2.408&2.386 \\
\hline\hline
\end{tabular}
\end{center}
\caption{The best-fit parameters of models DCR, DR and DC. The unit of $N_p$ is $10^{-9}\mbox{MeV}^{-1}\mbox{cm}^{-2}\mbox{sr}^{-1}\mbox{s}^{-1}$. $\phi$, $\rho_{br}$, $Z_h$, $D_0/Z_h$, $v_{A}$ and $V_{c0}$ are in units of MV, GV, kpc, $10^{28}\mbox{cm}^2\mbox{s}^{-1}/\mbox{kpc}$, km $\mbox{s}^{-1}$ and km $\mbox{s}^{-1}$. In the DR model, the bracketed $\phi^{e^+}$ or $V_A^{e^+}$ is only used for CR positrons. The rest propagation parameters are referred to the example 01 of GALPROP WebRun \citet{Vladimirov:2010aq}.}
\label{tab:parameters}
\end{table}

\begin{figure}
\includegraphics[width=0.19\textwidth]{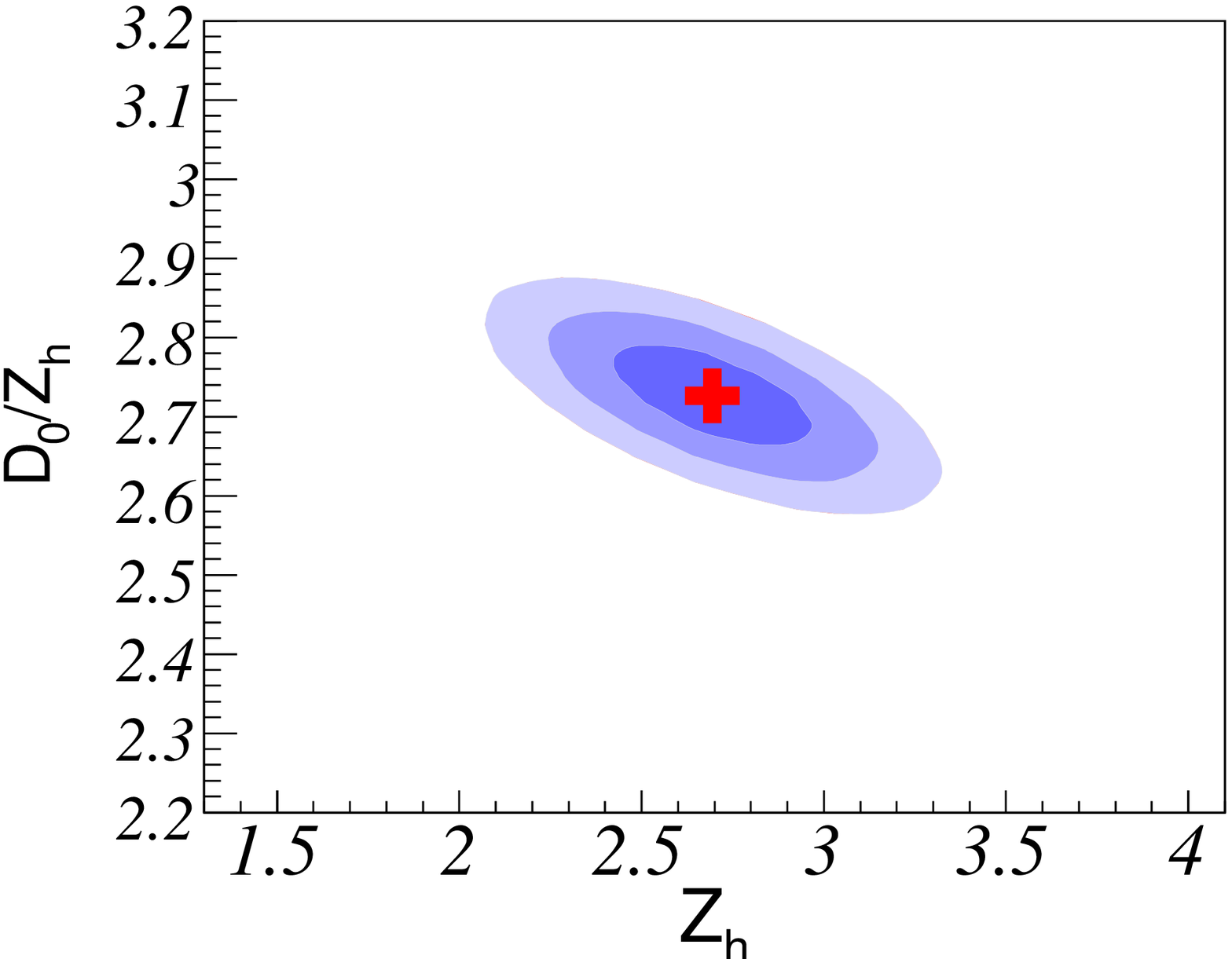}
\includegraphics[width=0.19\textwidth]{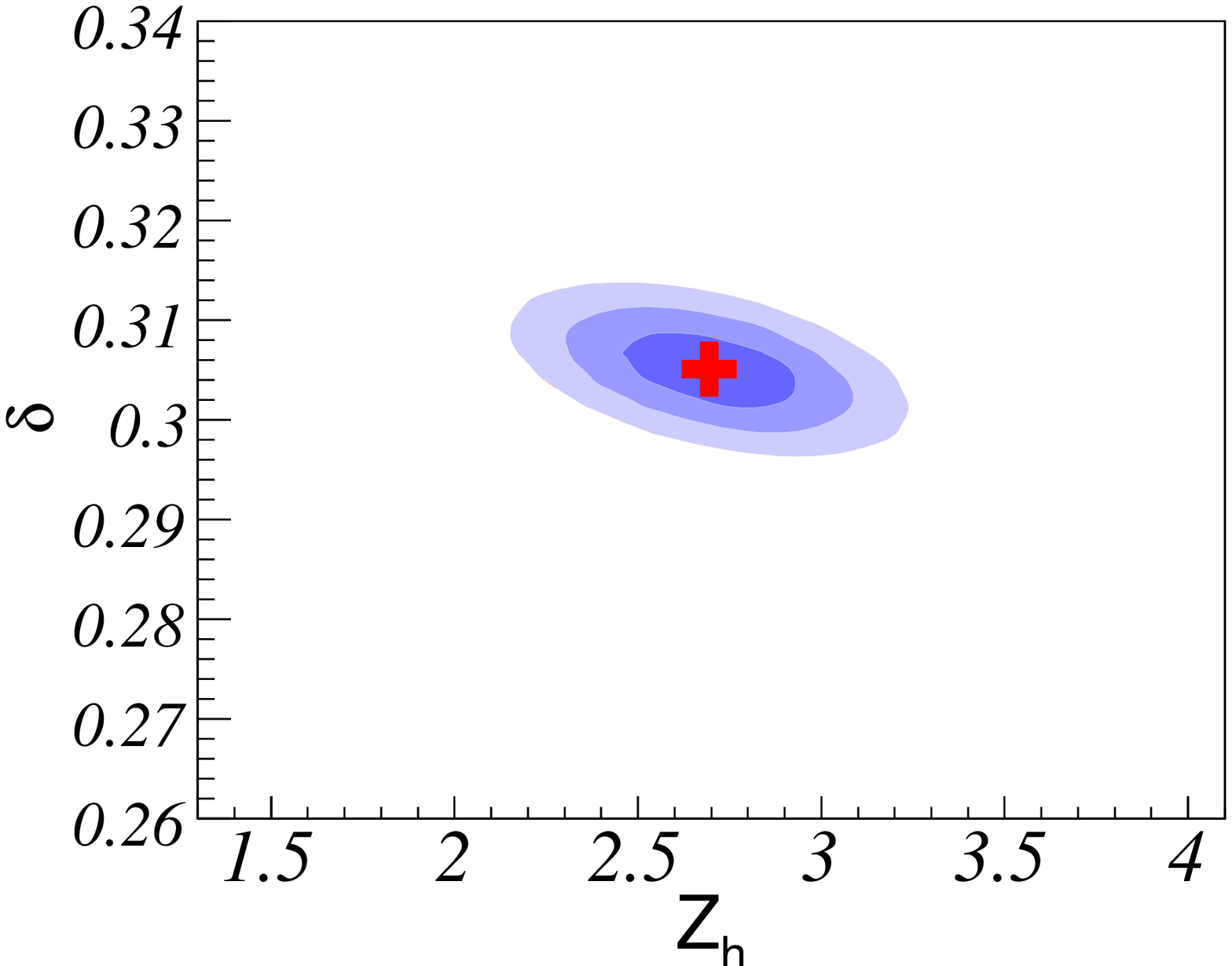}
\includegraphics[width=0.19\textwidth]{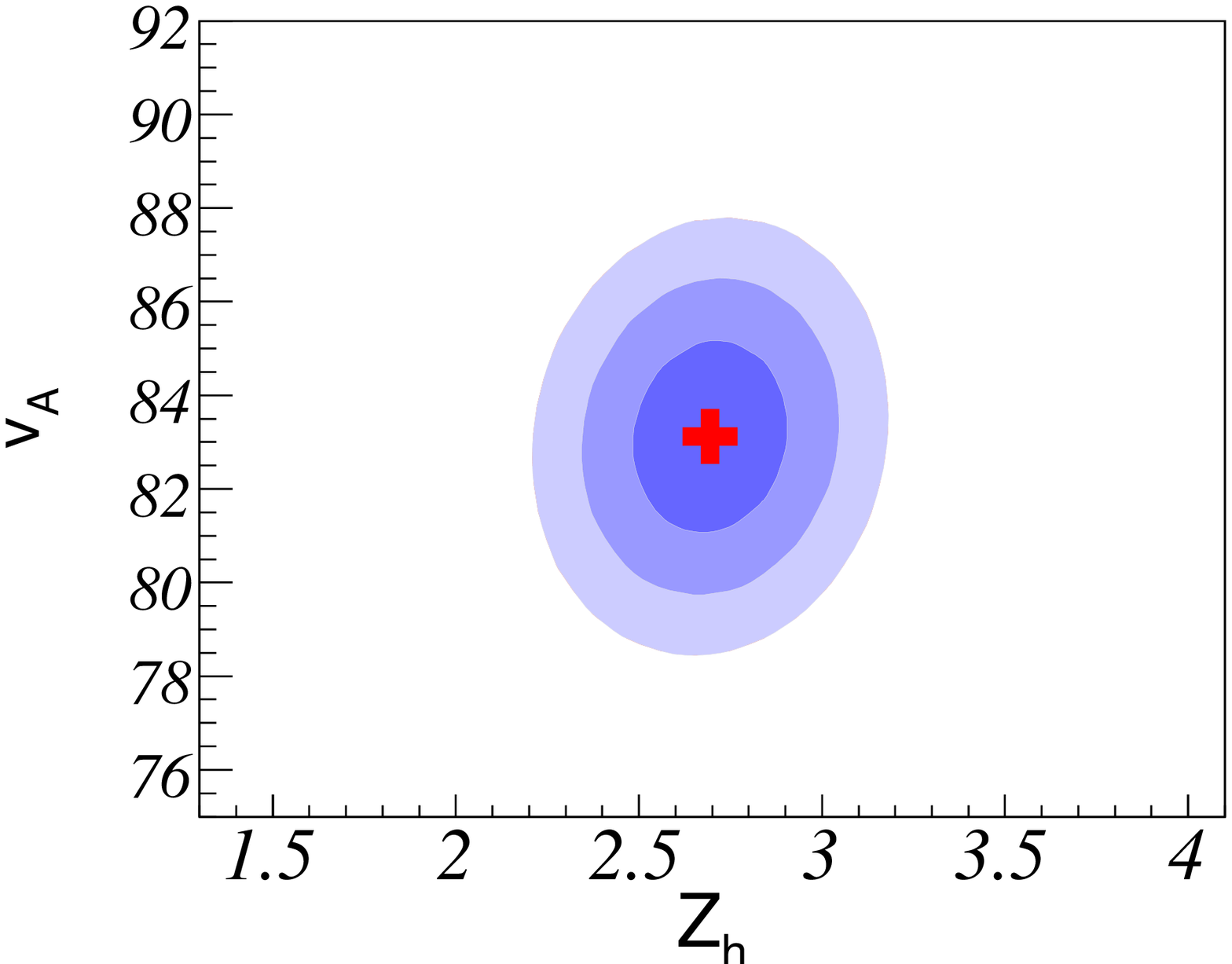}
\includegraphics[width=0.19\textwidth]{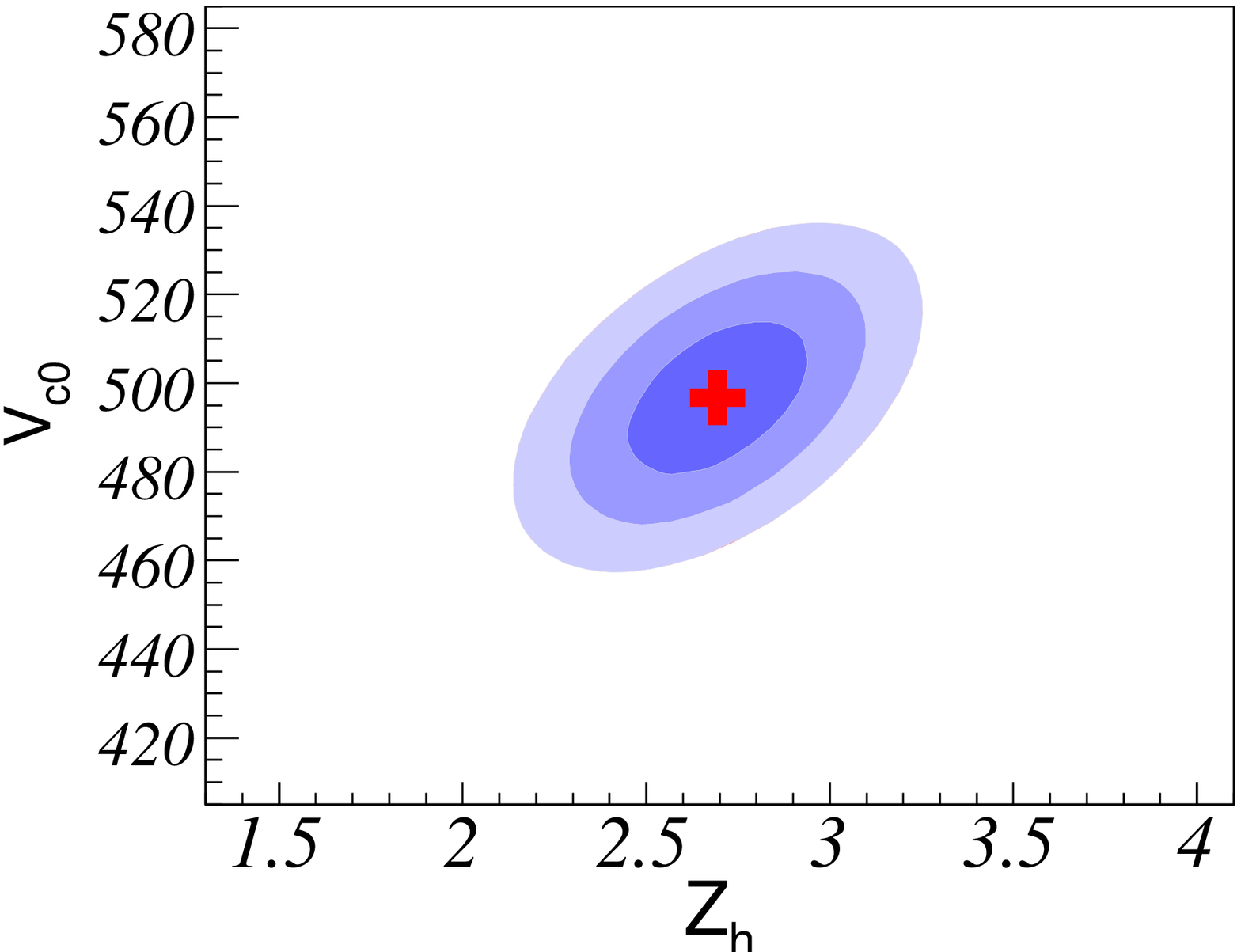}
\includegraphics[width=0.19\textwidth]{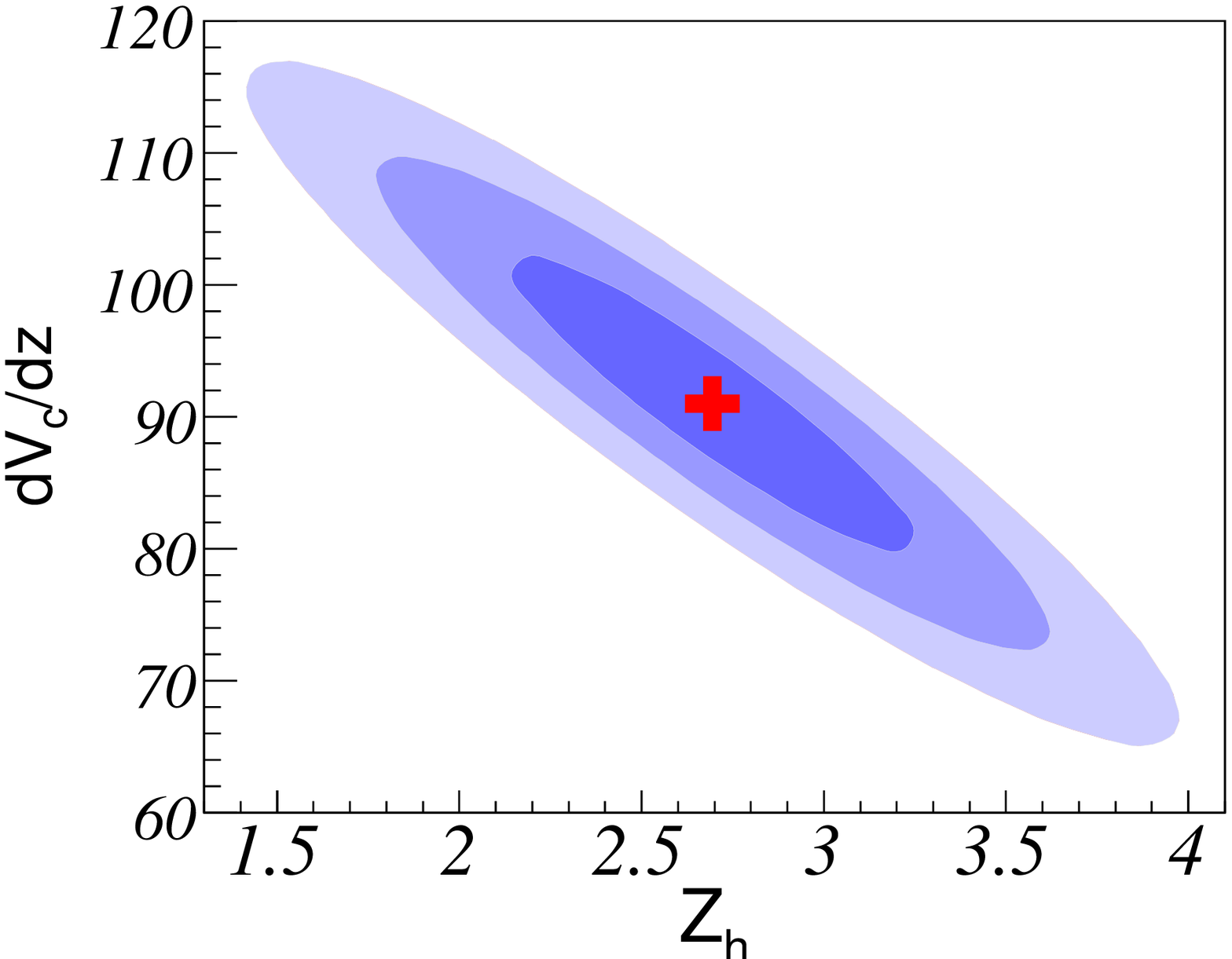}
\\
\includegraphics[width=0.19\textwidth]{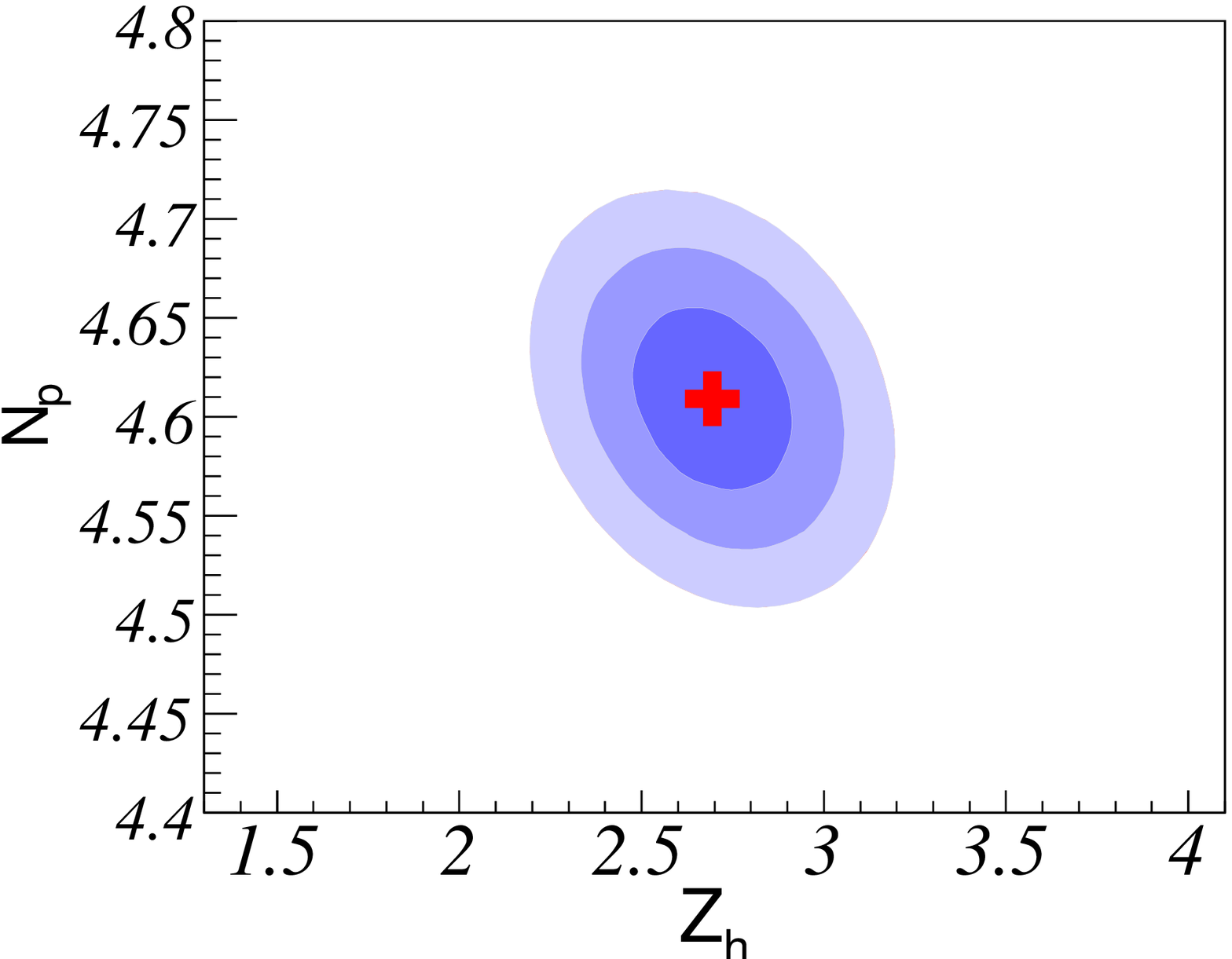}
\includegraphics[width=0.19\textwidth]{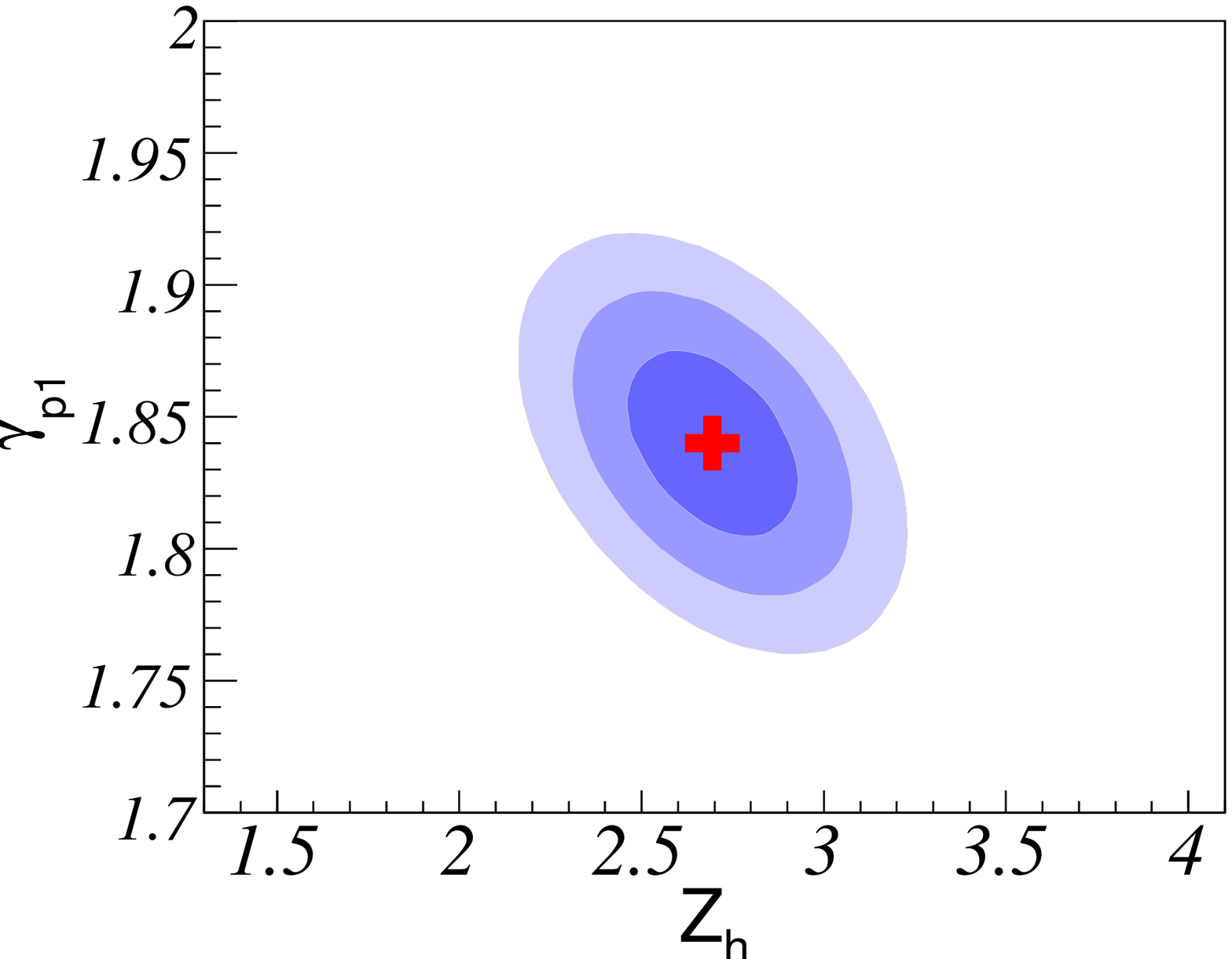}
\includegraphics[width=0.19\textwidth]{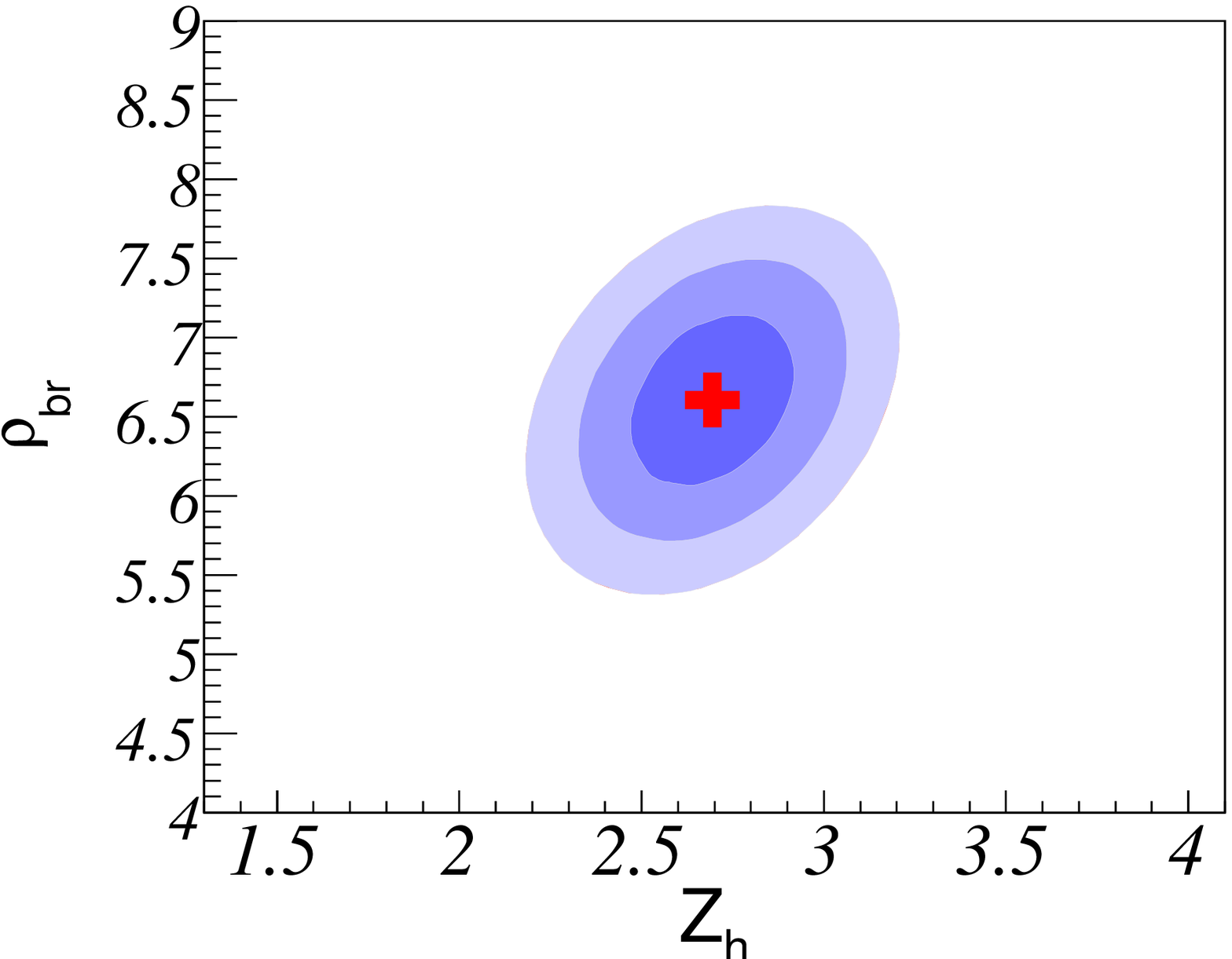}
\includegraphics[width=0.19\textwidth]{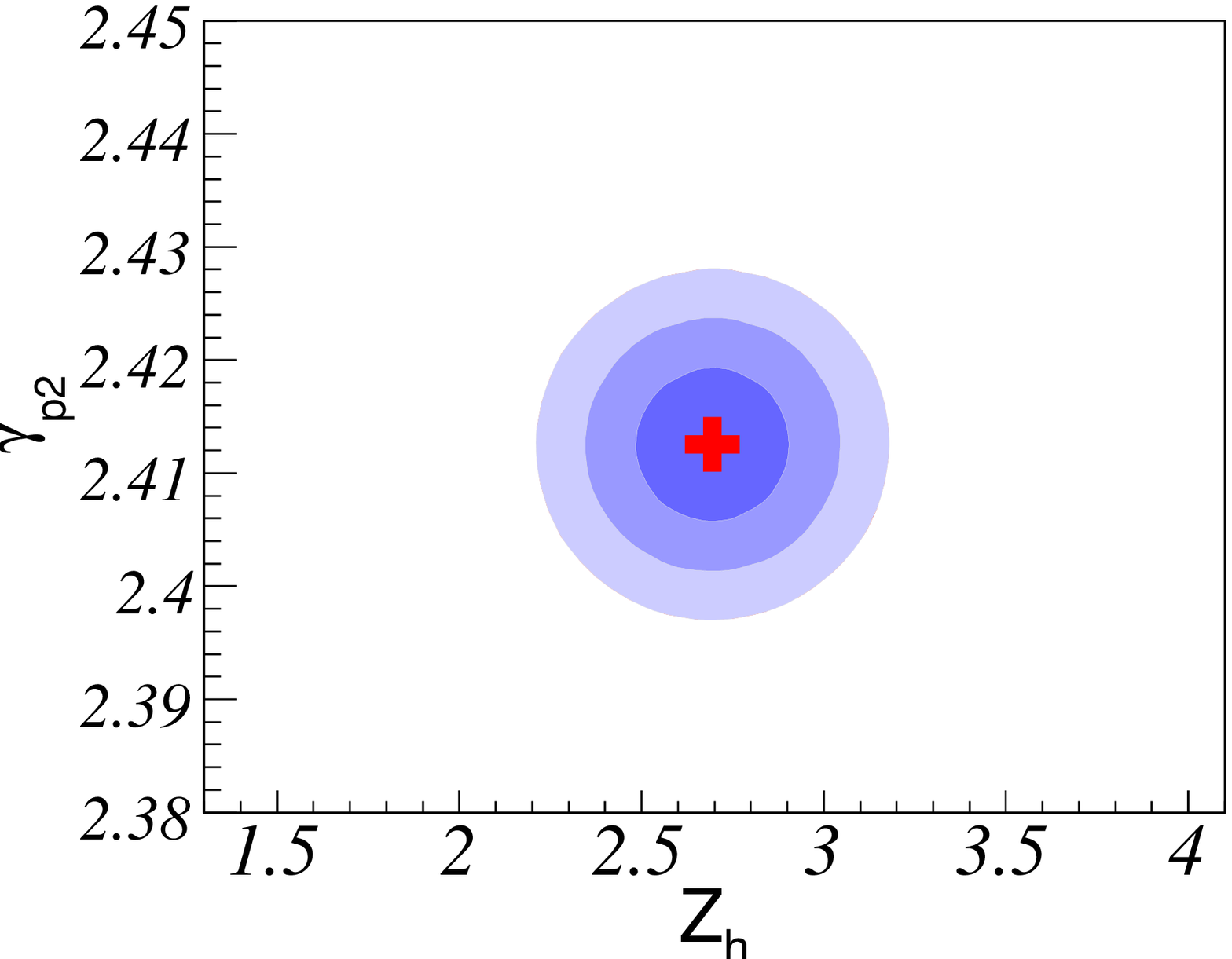}
\includegraphics[width=0.19\textwidth]{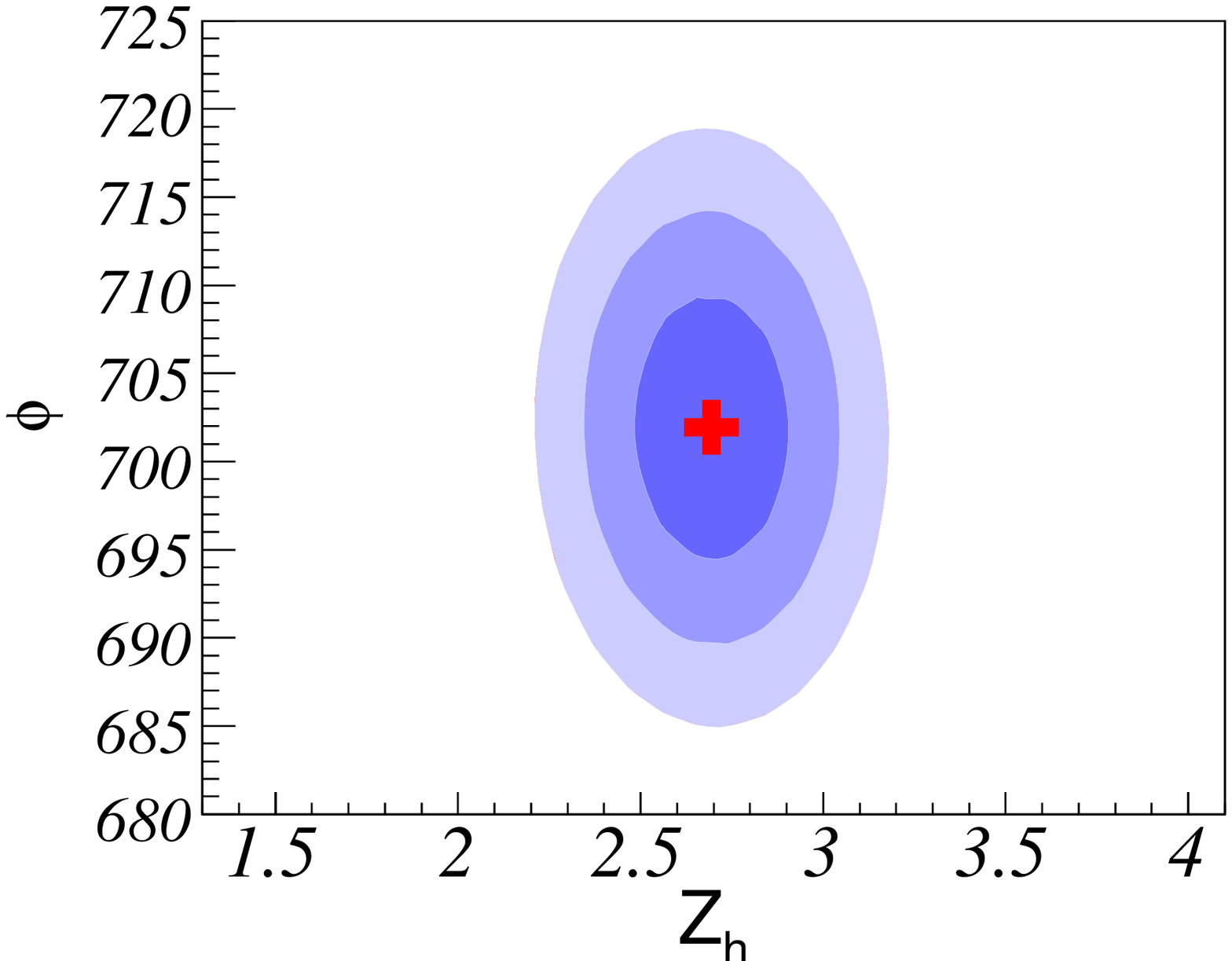}
\\
\includegraphics[width=0.19\textwidth]{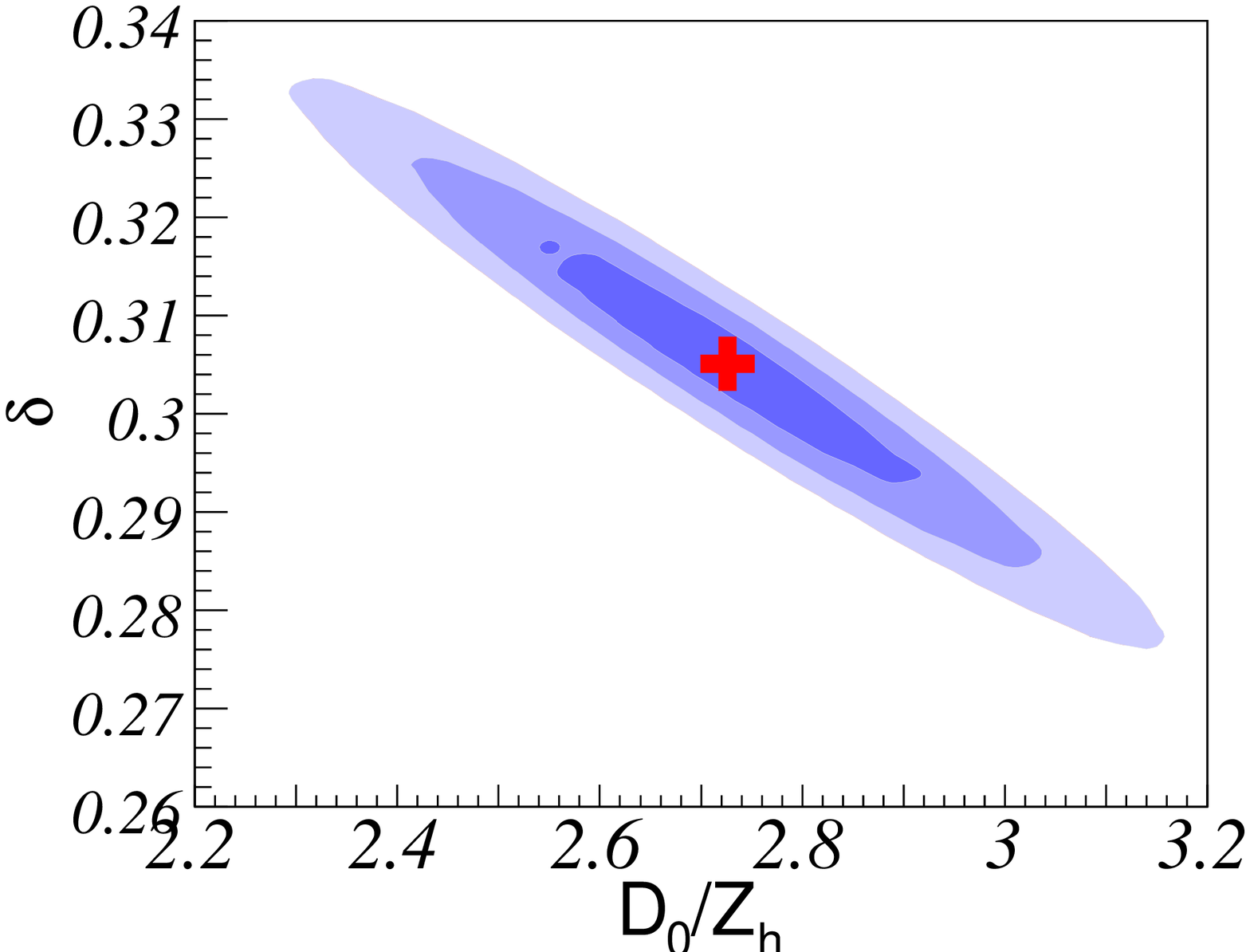}
\includegraphics[width=0.19\textwidth]{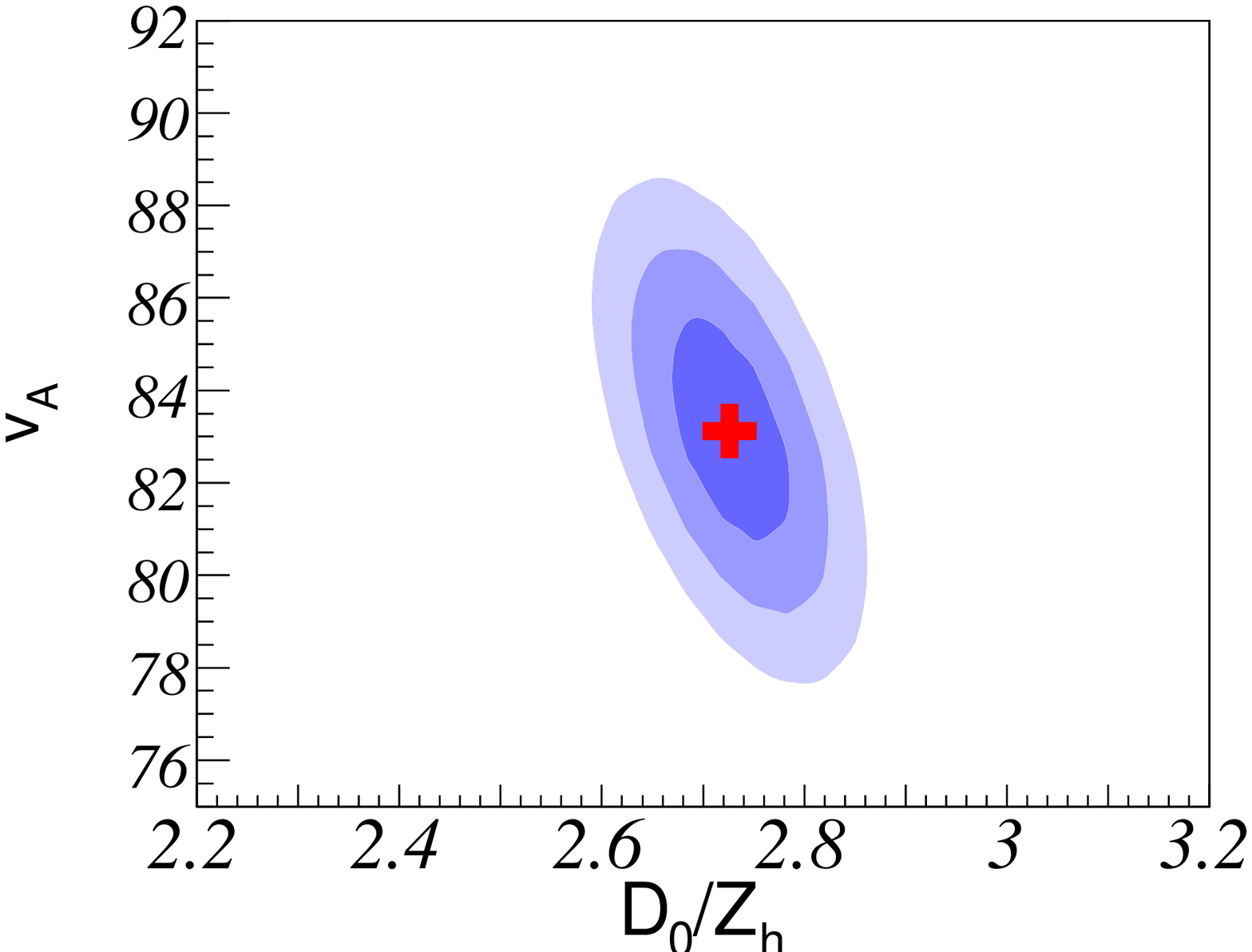}
\includegraphics[width=0.19\textwidth]{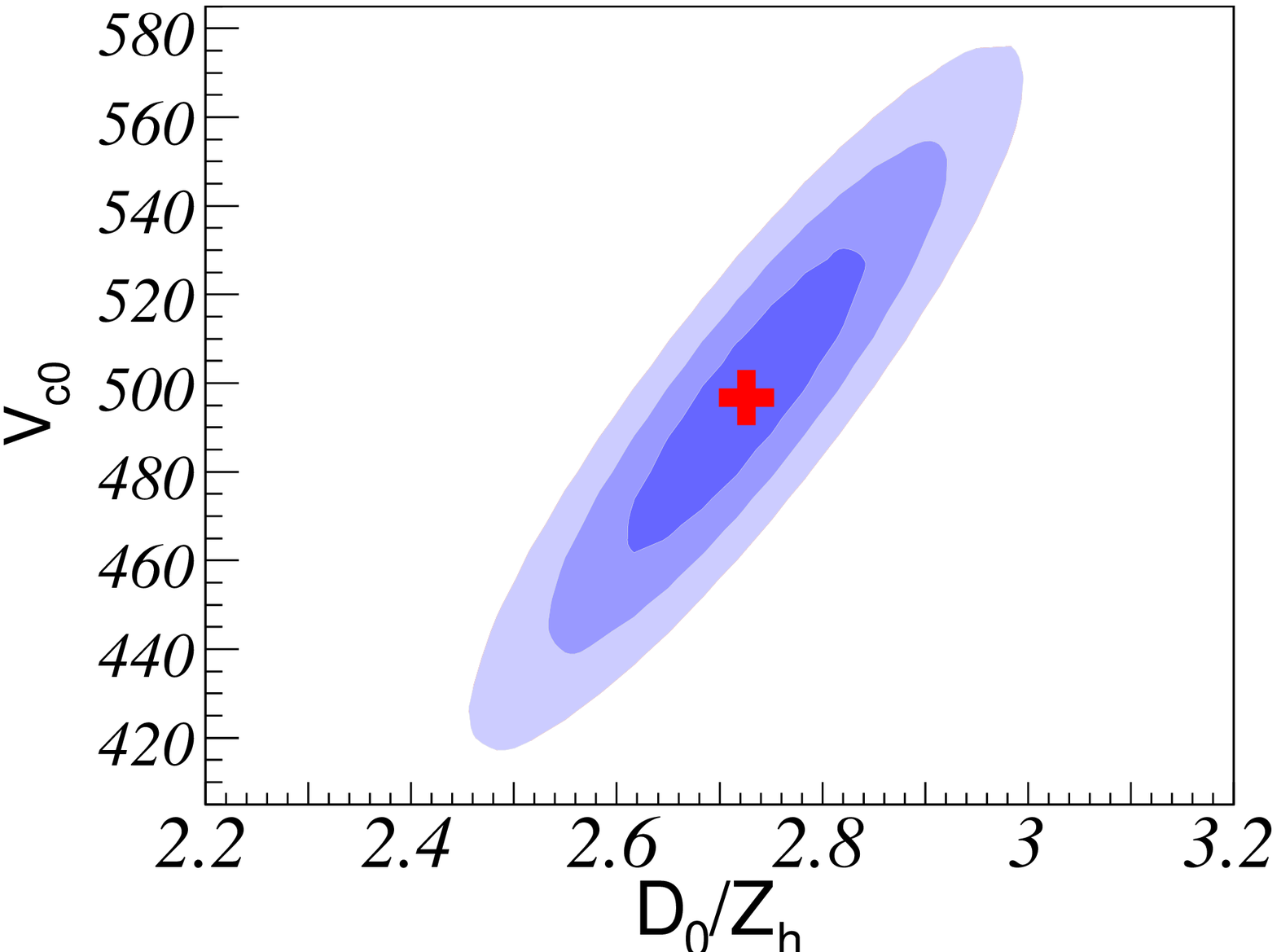}
\includegraphics[width=0.19\textwidth]{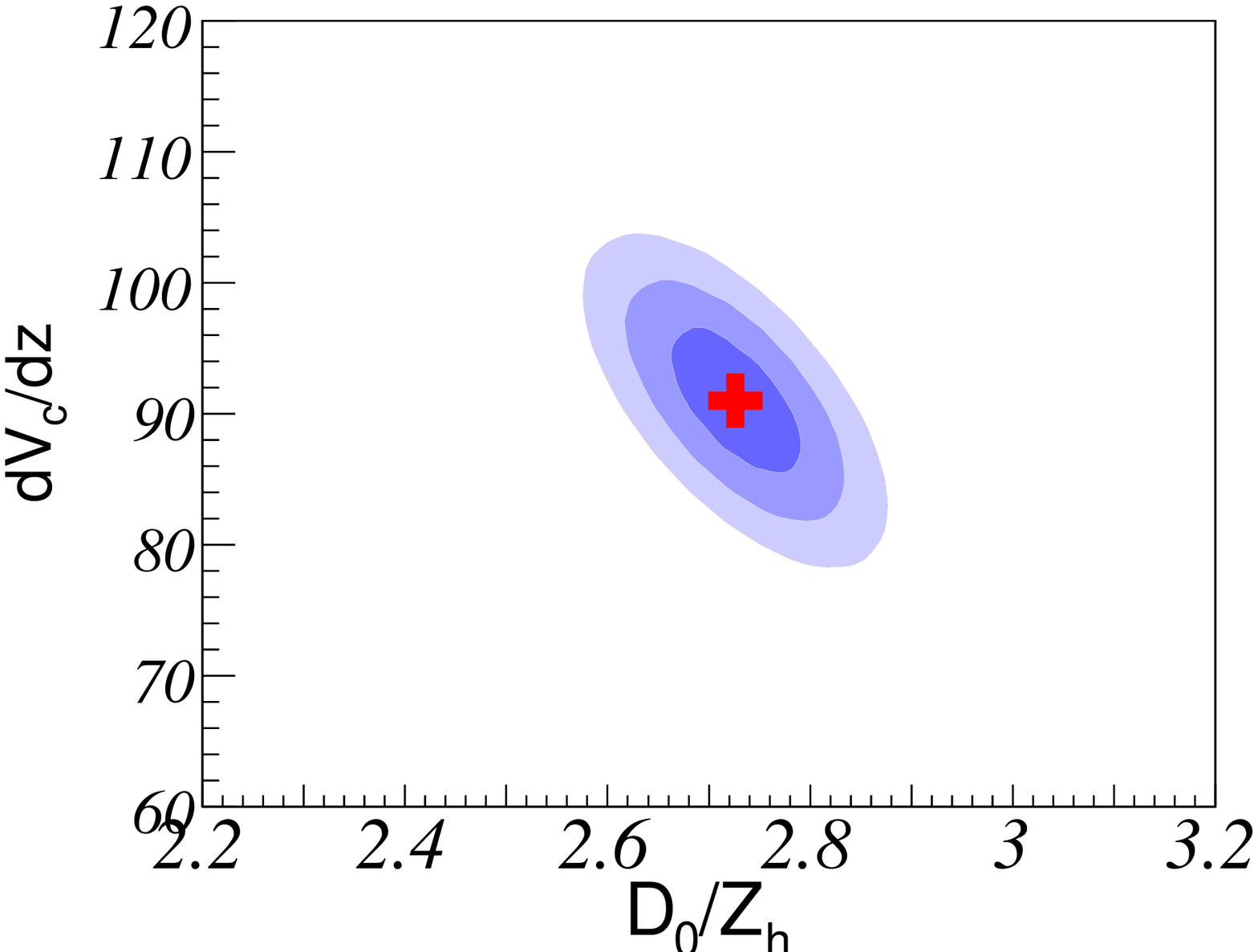}
\includegraphics[width=0.19\textwidth]{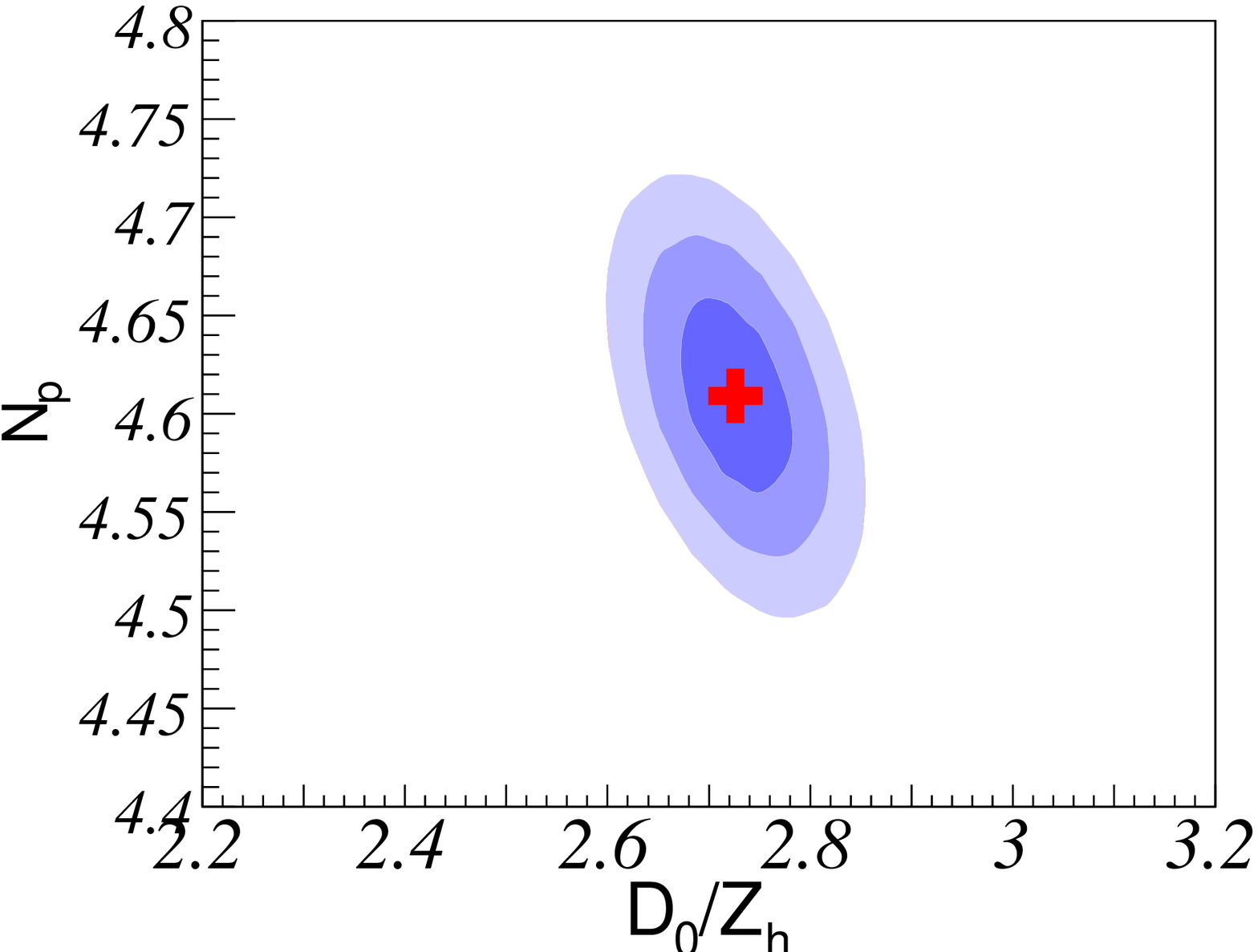}
\\
\includegraphics[width=0.19\textwidth]{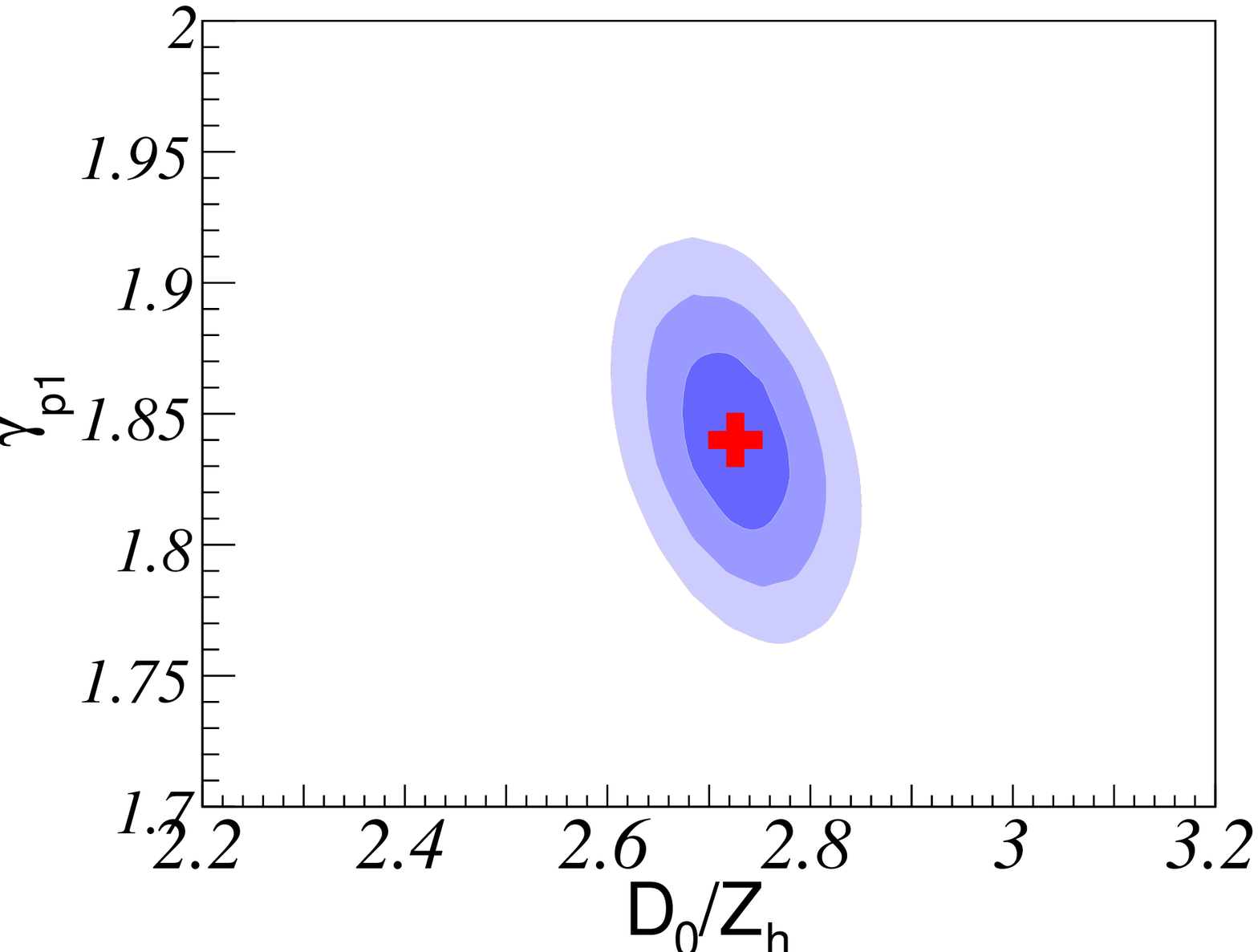}
\includegraphics[width=0.19\textwidth]{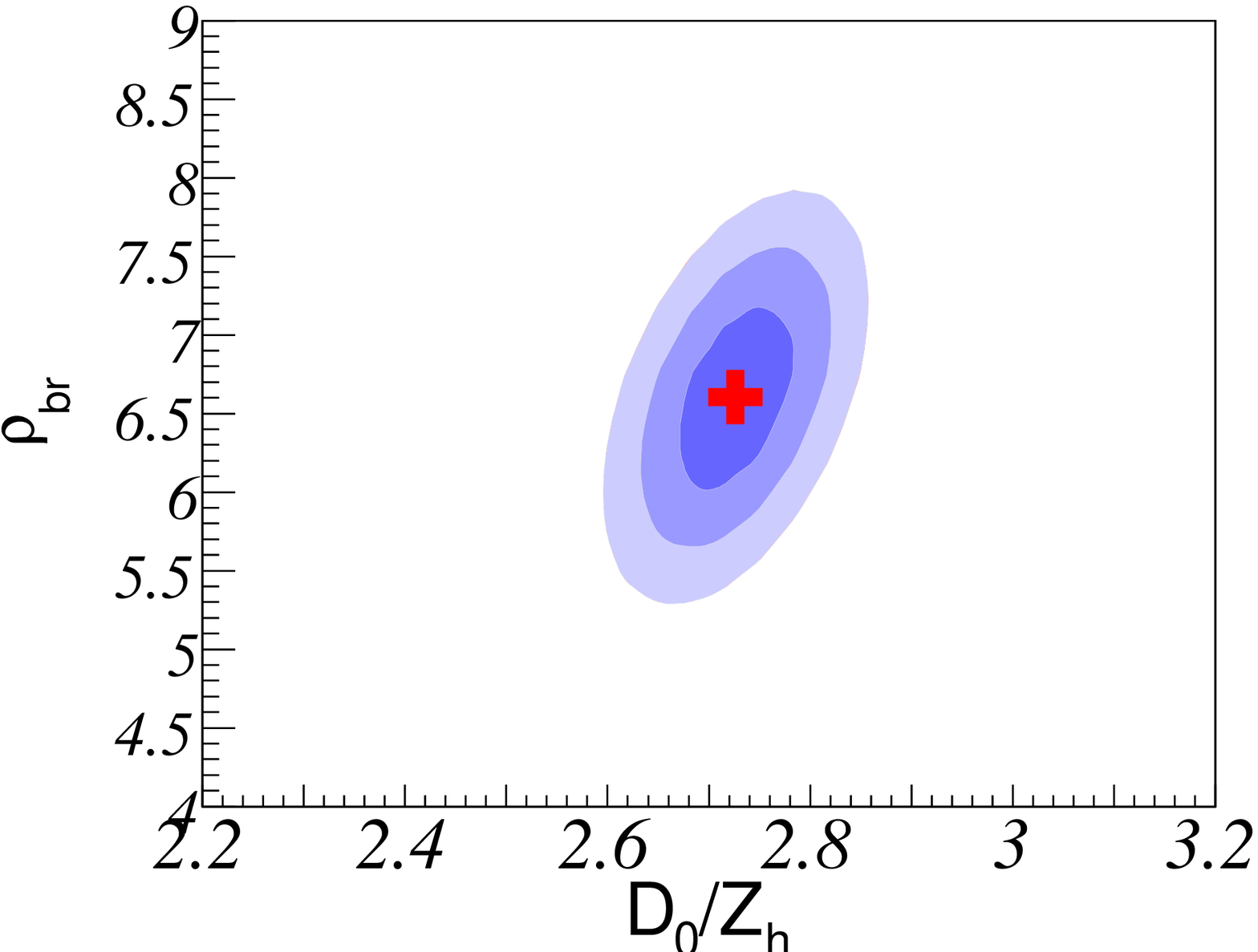}
\includegraphics[width=0.19\textwidth]{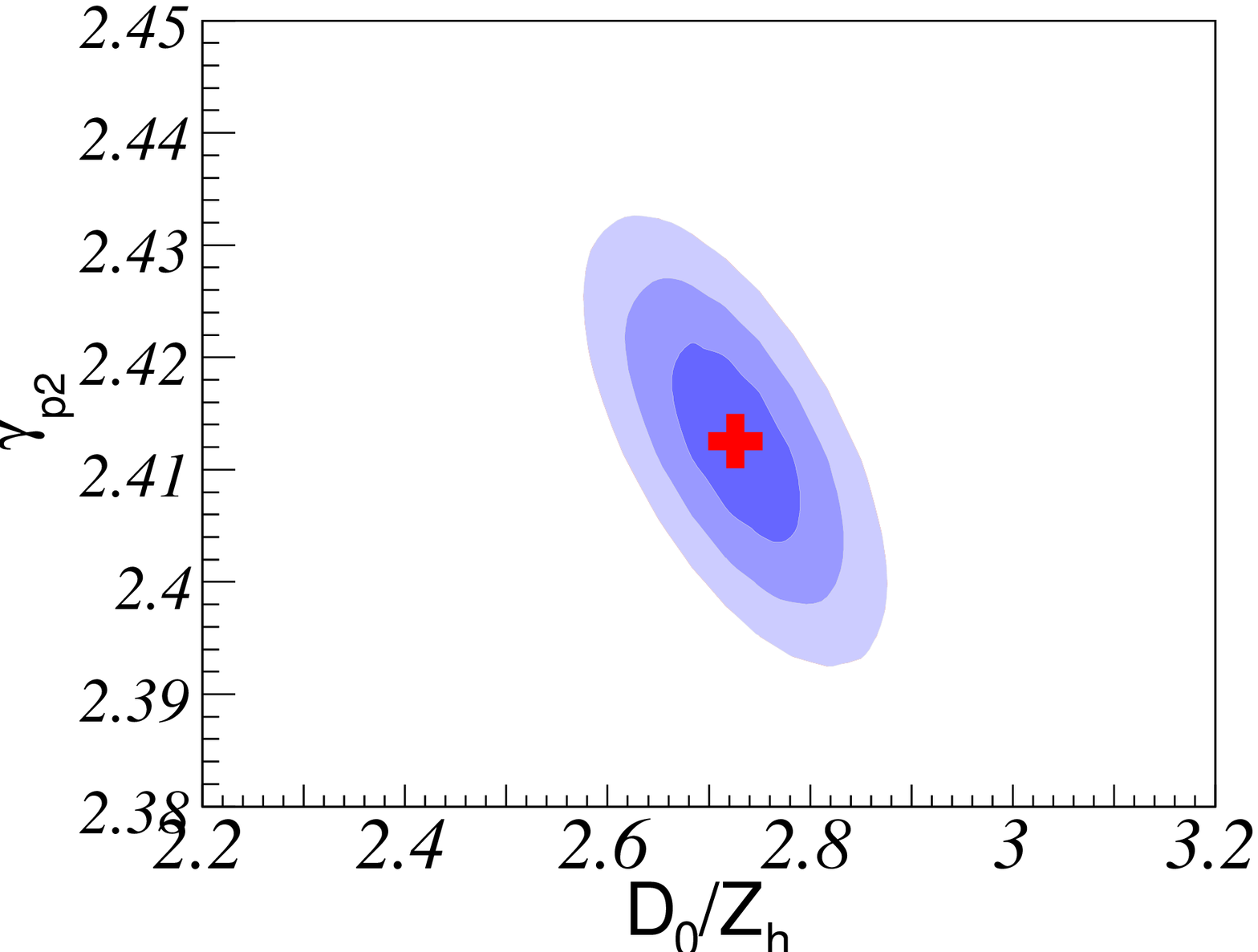}
\includegraphics[width=0.19\textwidth]{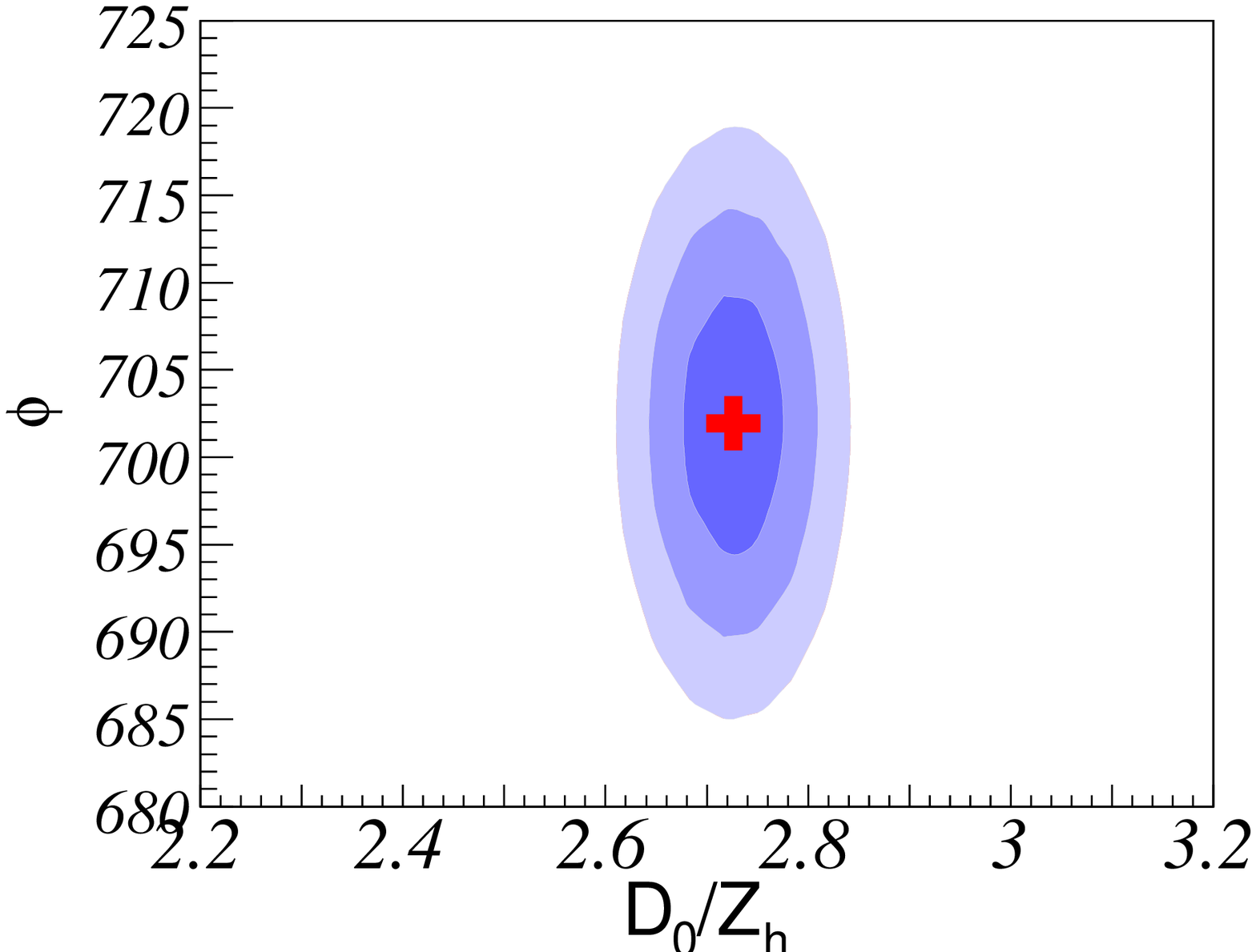}
\includegraphics[width=0.19\textwidth]{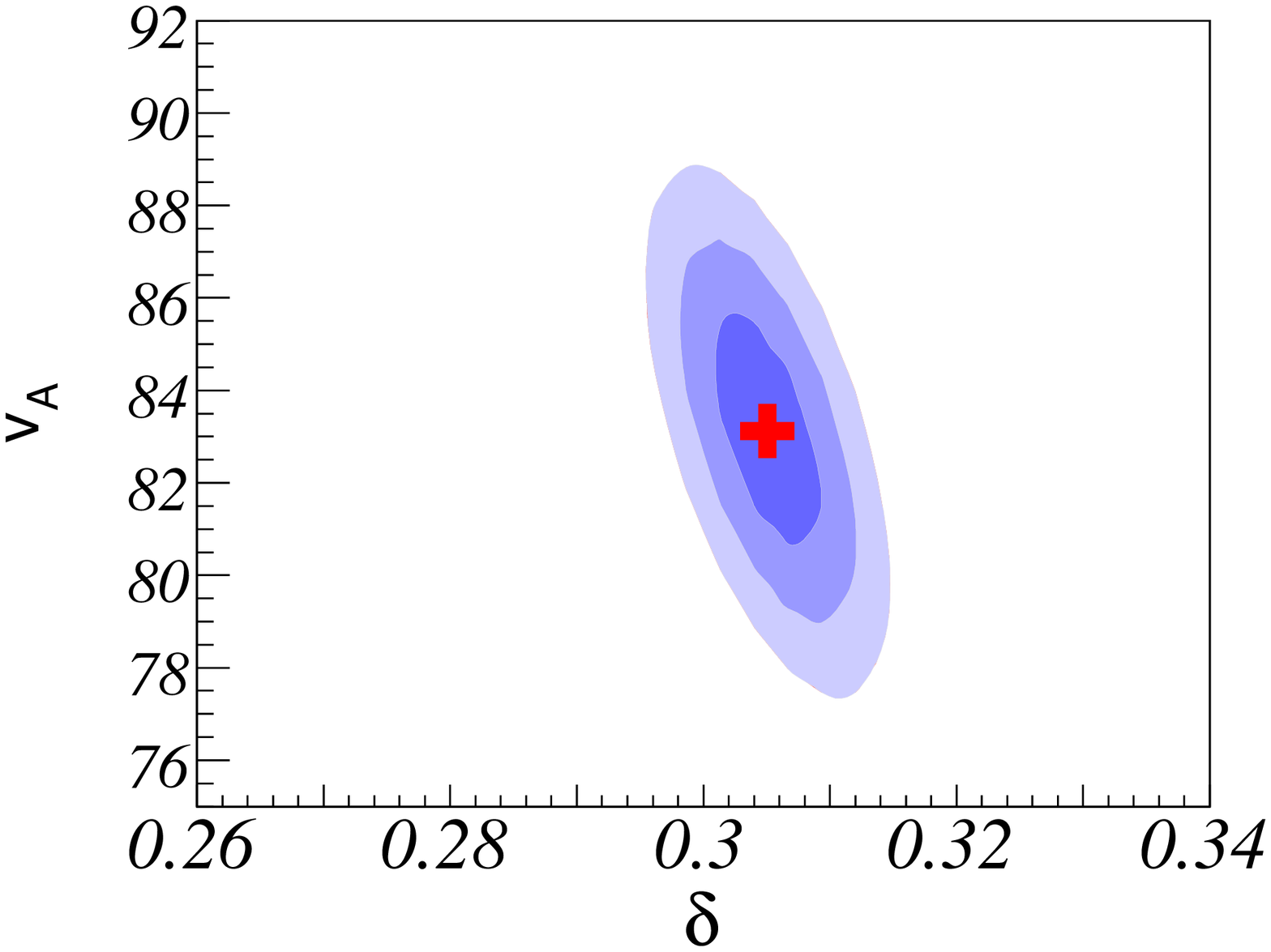}
\\
\includegraphics[width=0.19\textwidth]{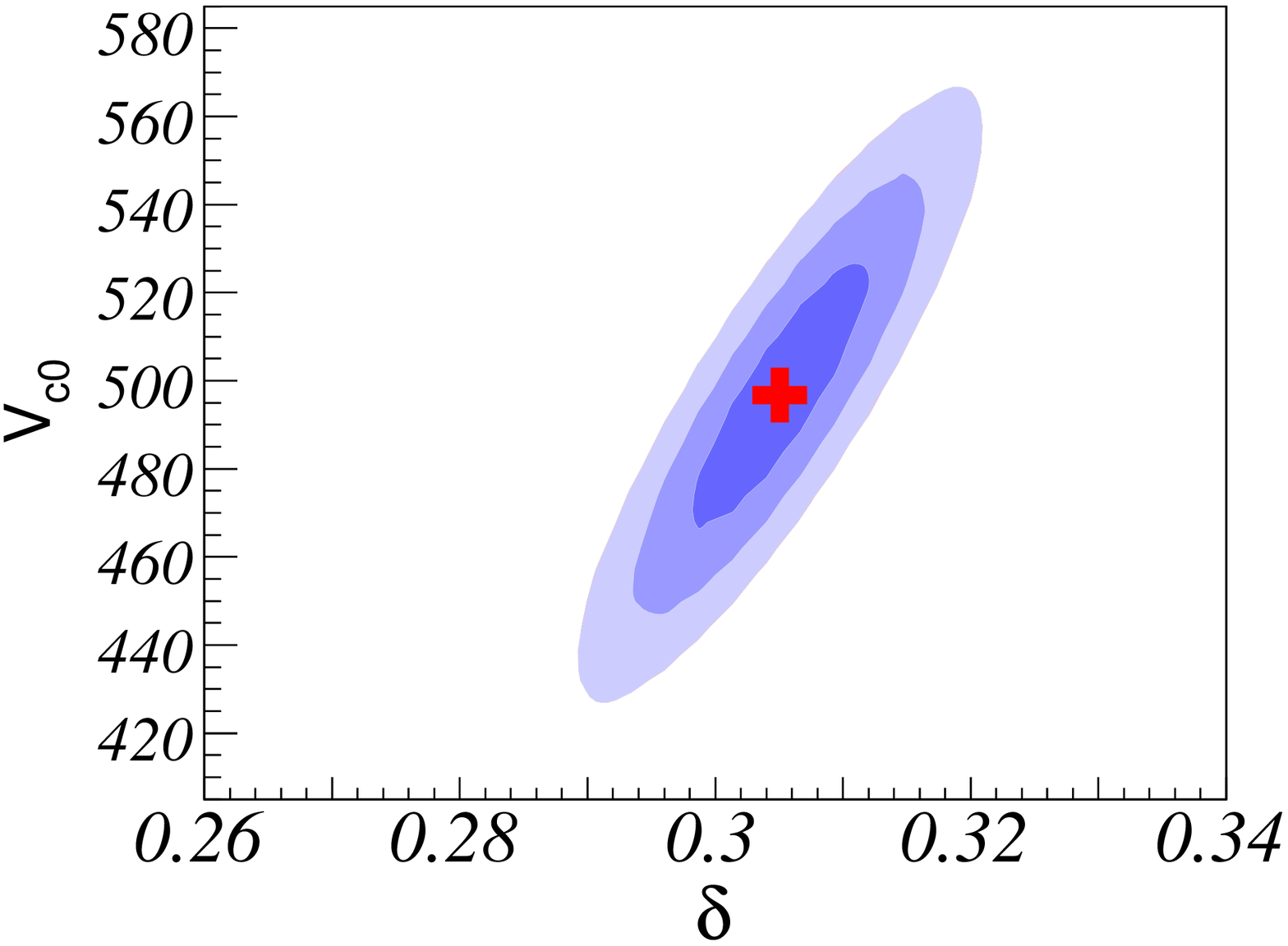}
\includegraphics[width=0.19\textwidth]{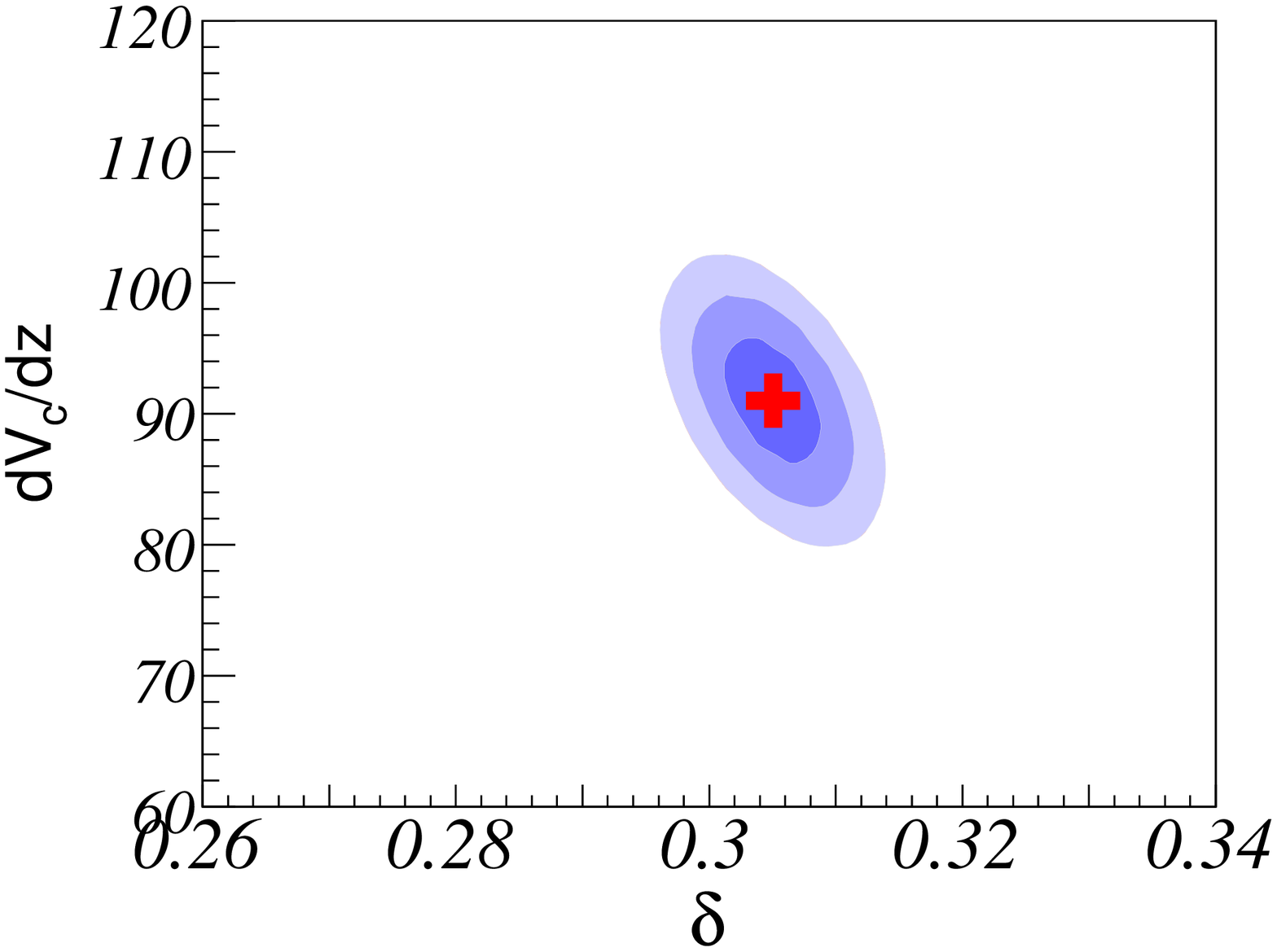}
\includegraphics[width=0.19\textwidth]{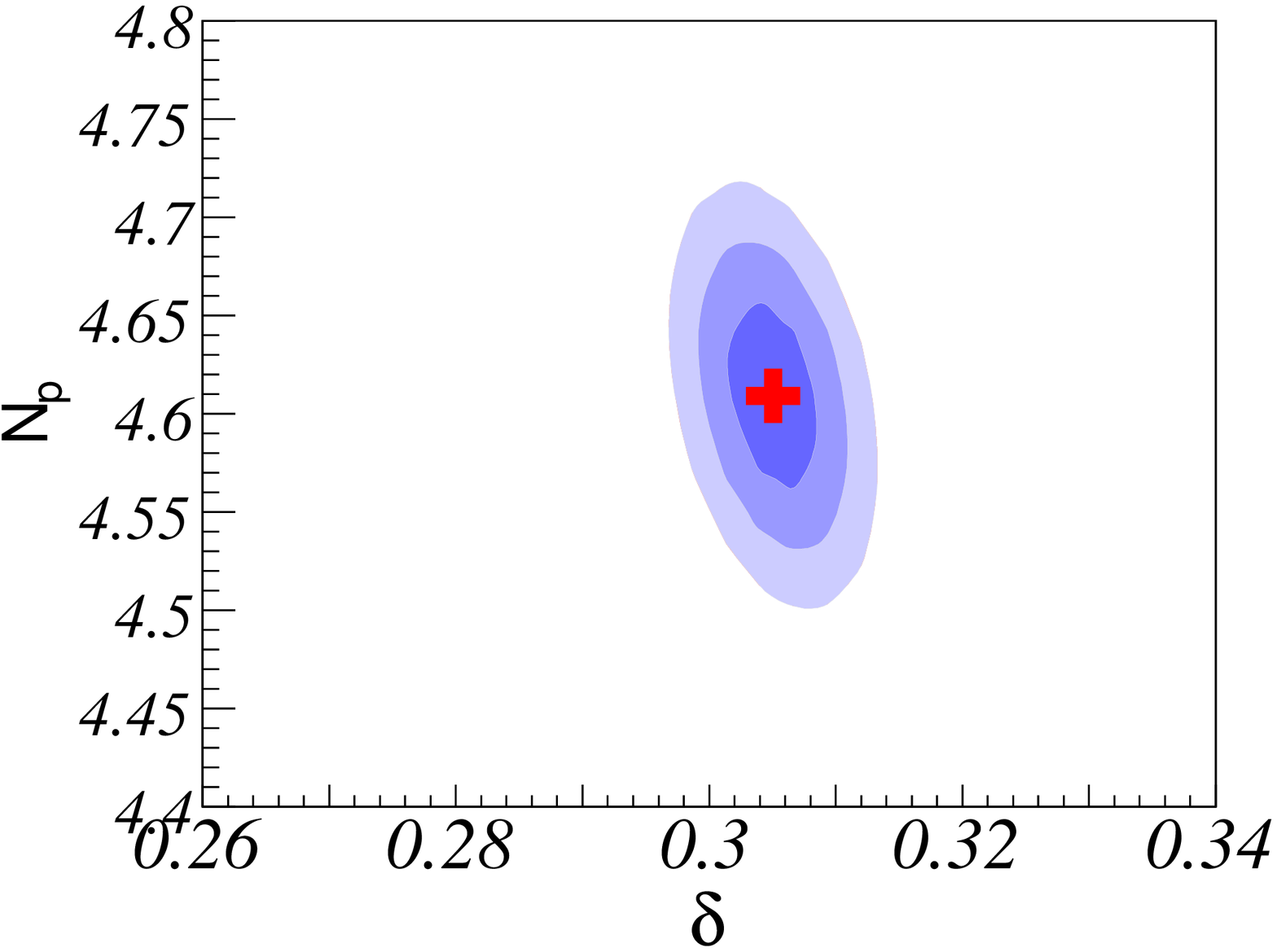}
\includegraphics[width=0.19\textwidth]{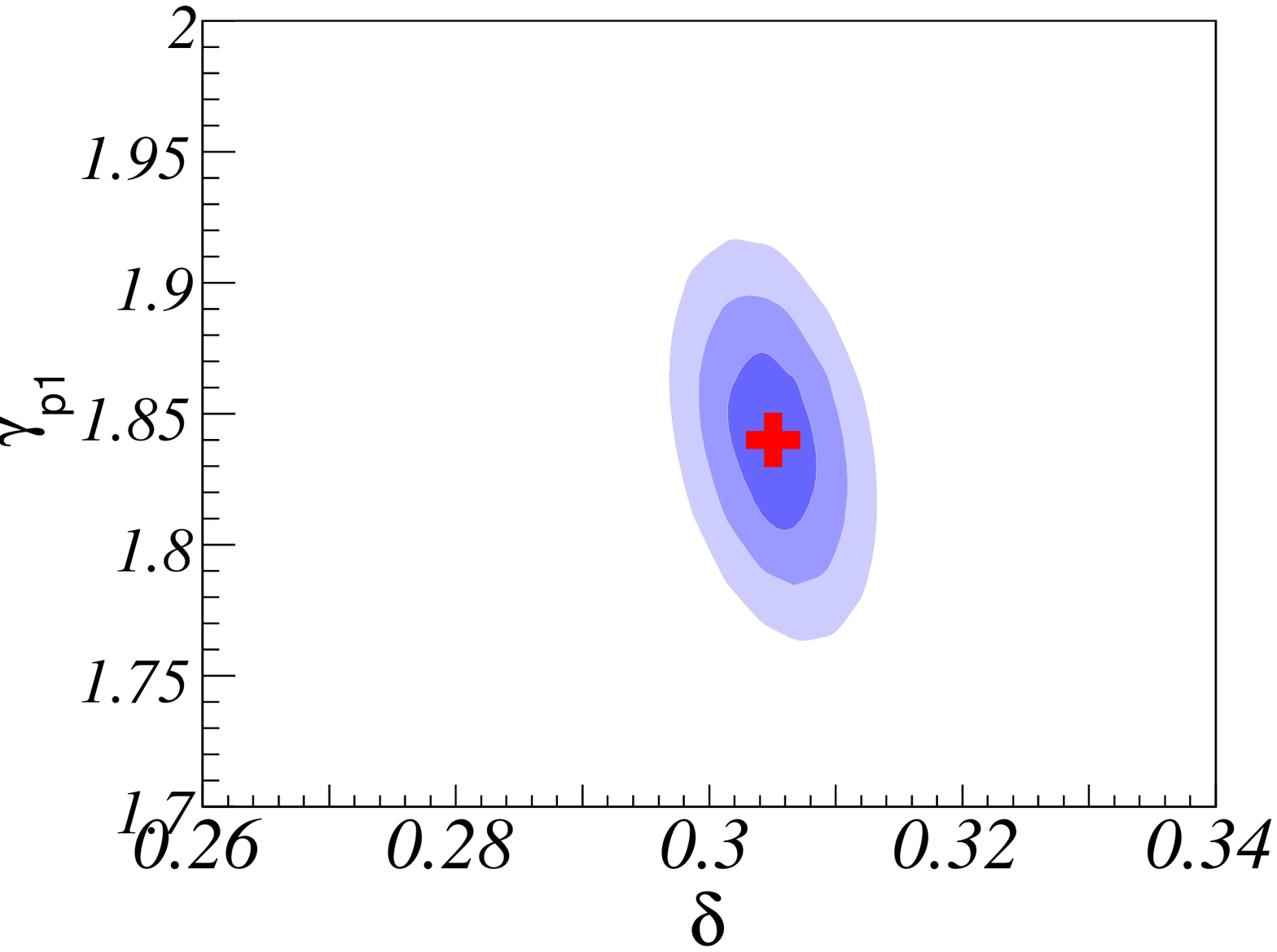}
\includegraphics[width=0.19\textwidth]{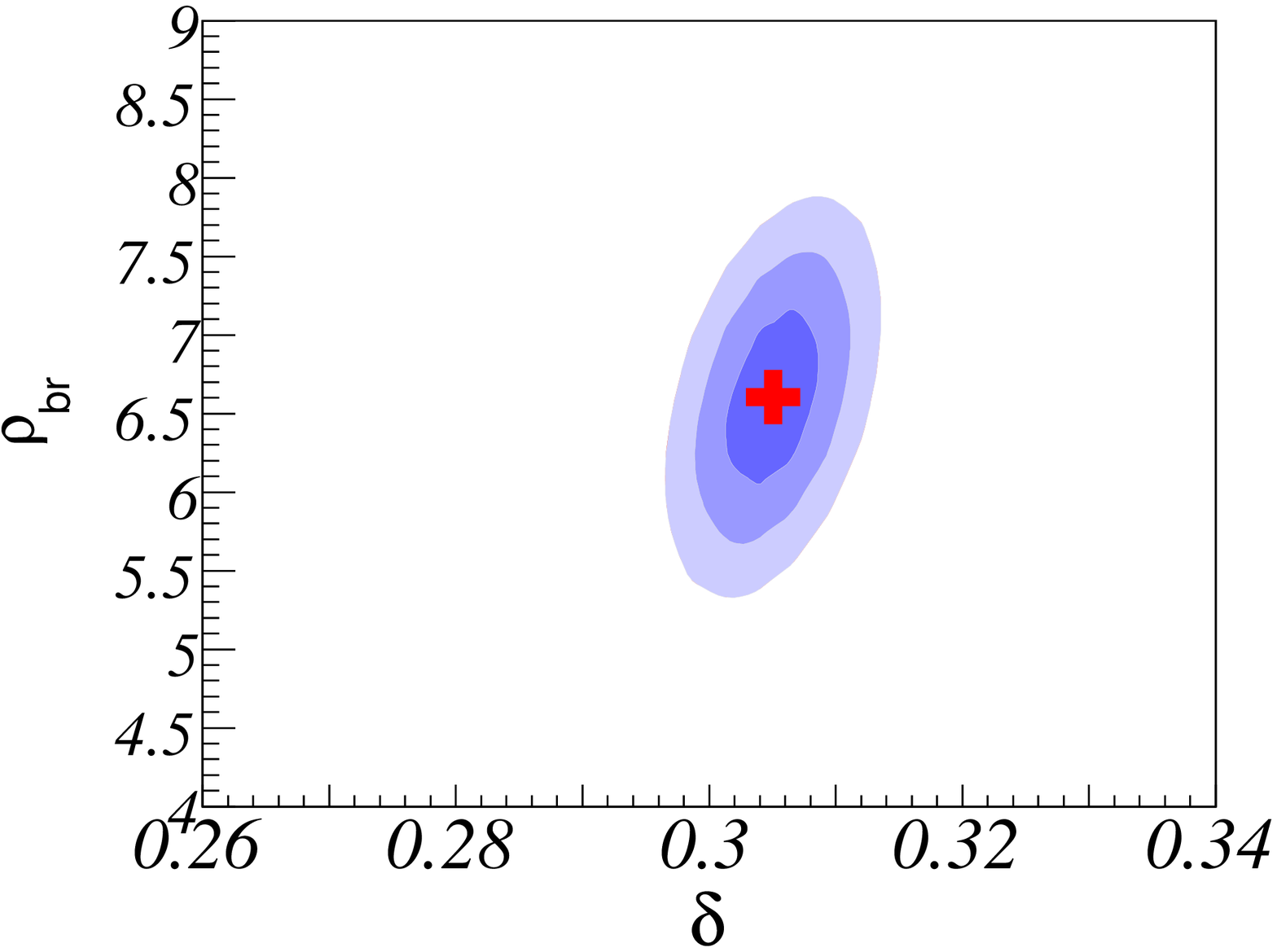}
\\
\includegraphics[width=0.19\textwidth]{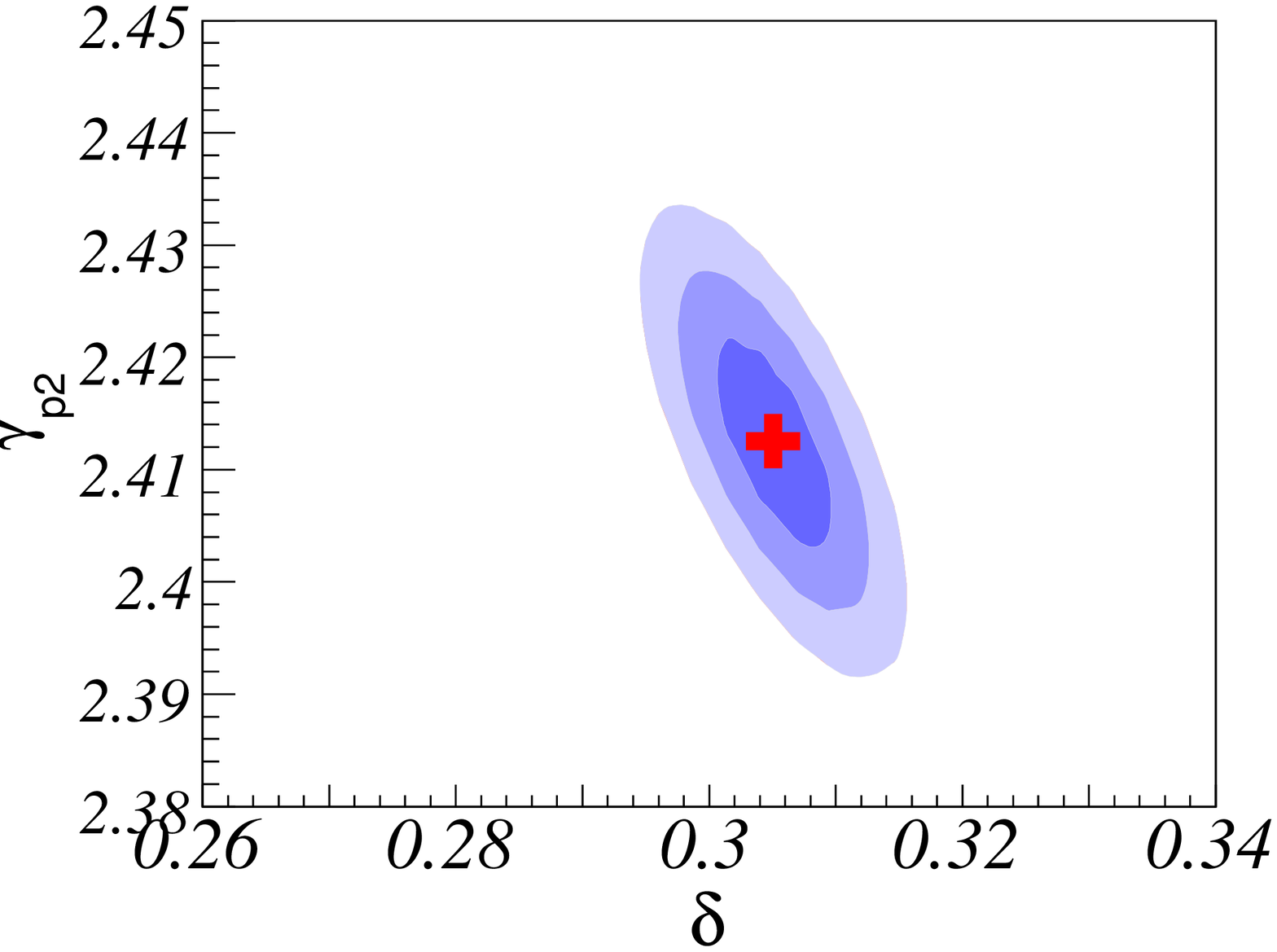}
\includegraphics[width=0.19\textwidth]{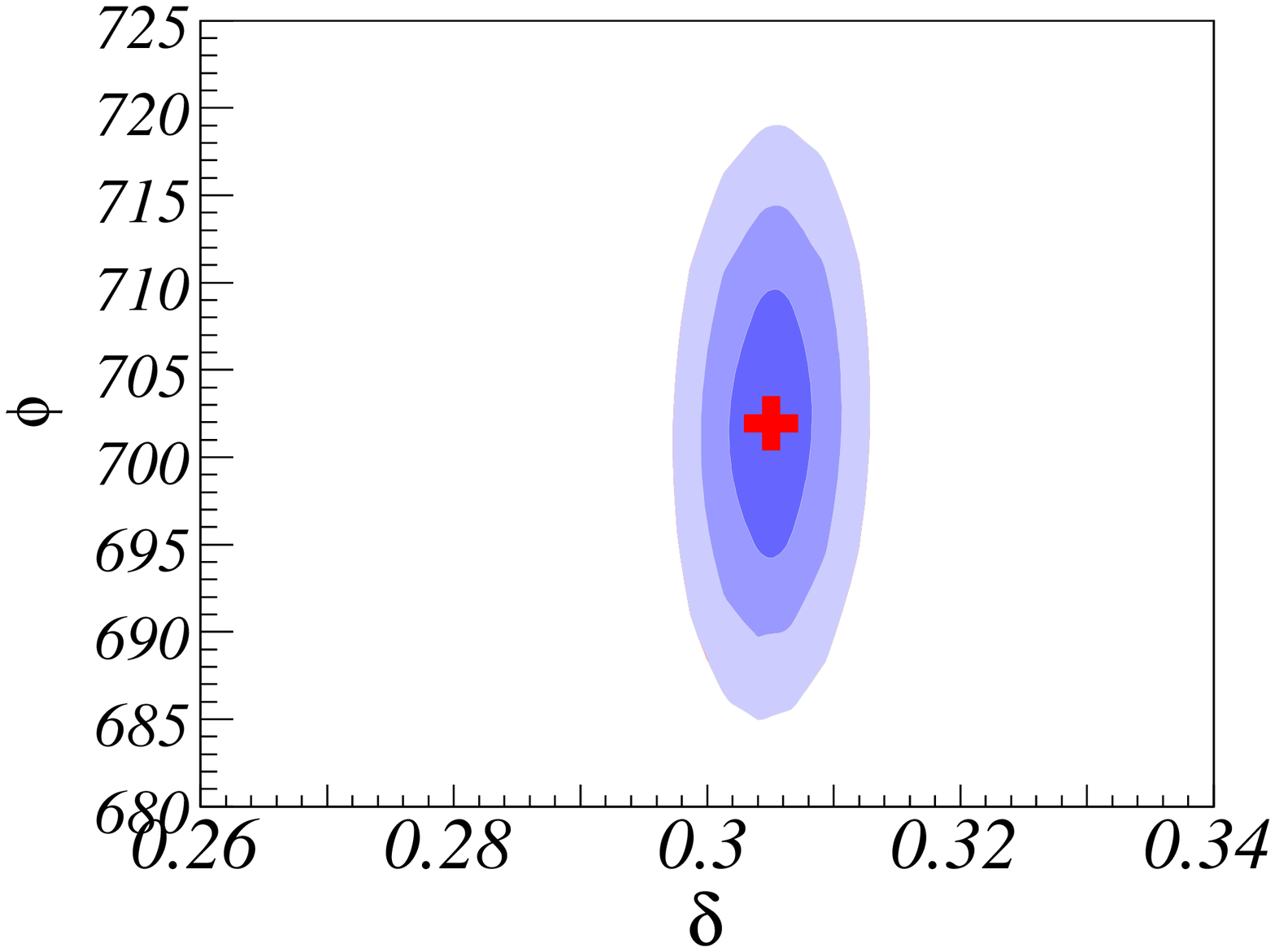}
\includegraphics[width=0.19\textwidth]{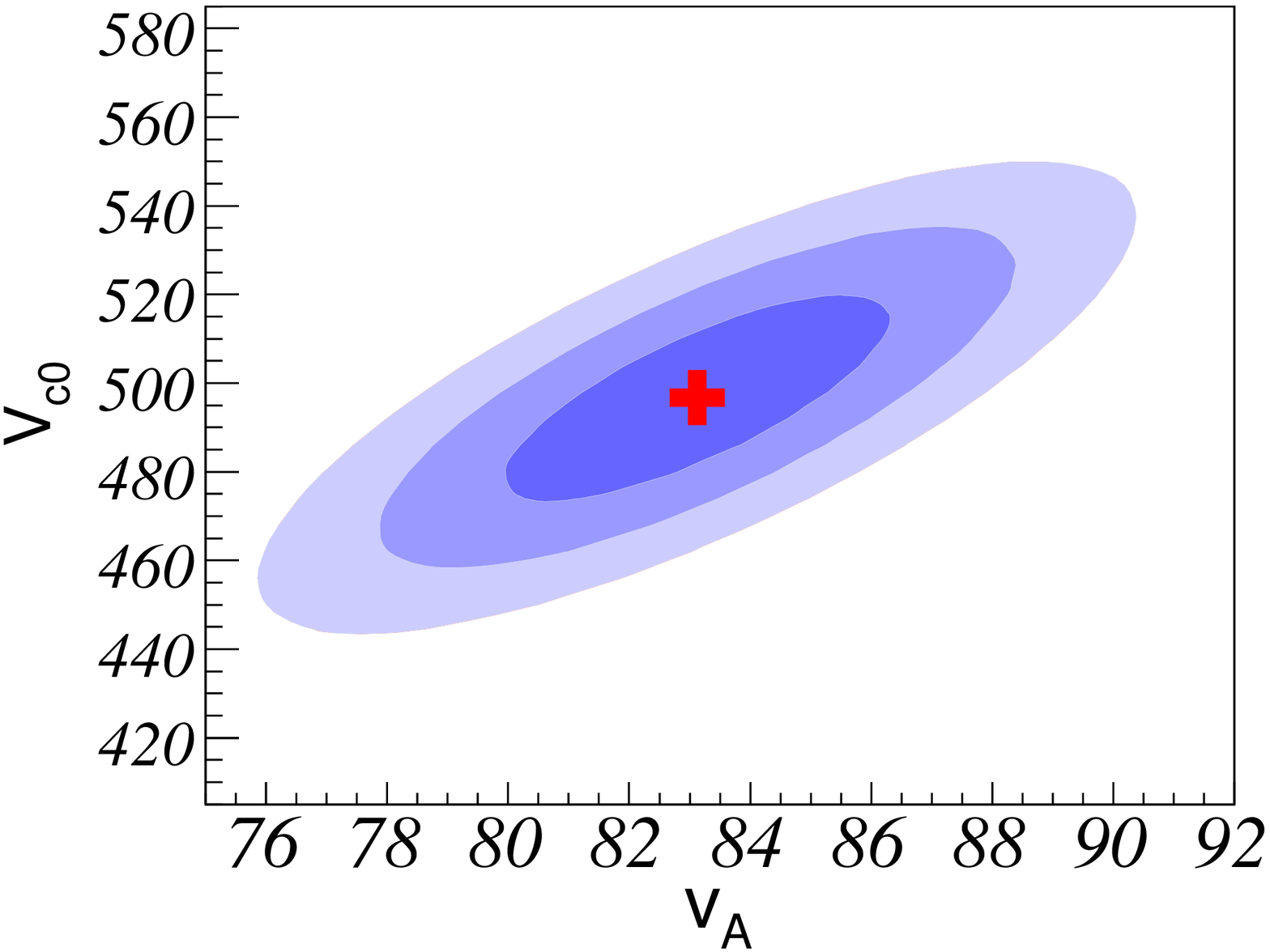}
\includegraphics[width=0.19\textwidth]{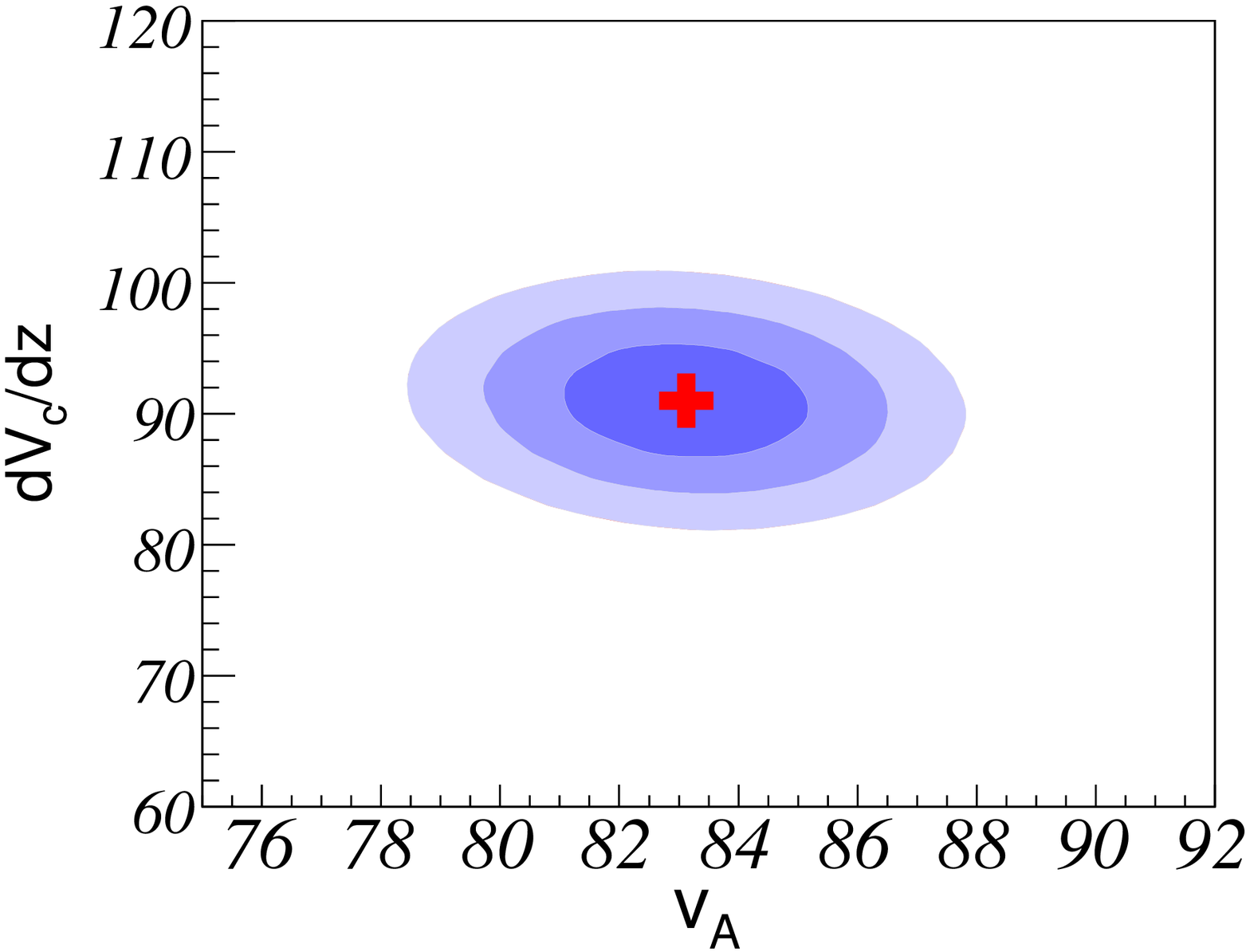}
\includegraphics[width=0.19\textwidth]{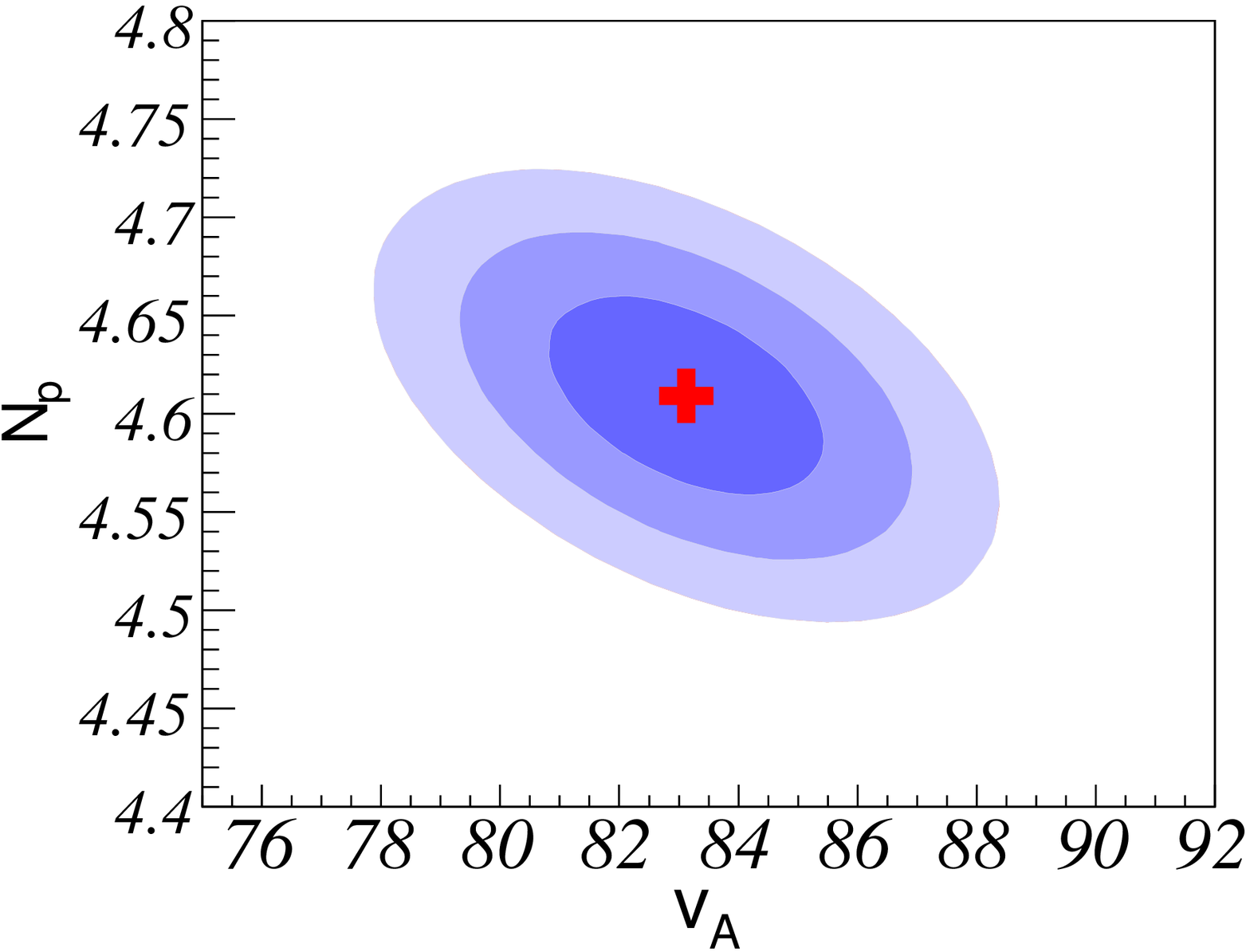}
\\
\includegraphics[width=0.19\textwidth]{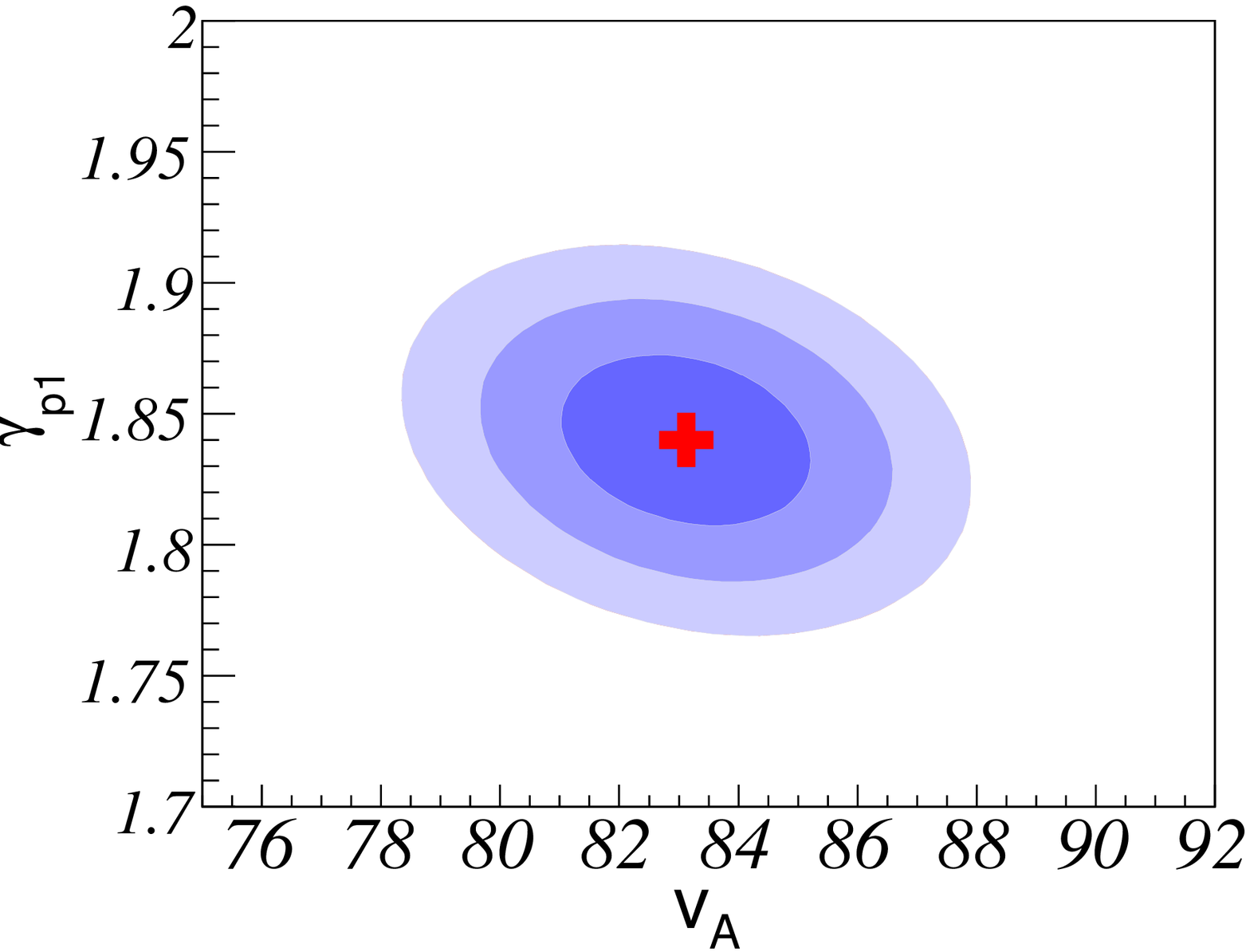}
\includegraphics[width=0.19\textwidth]{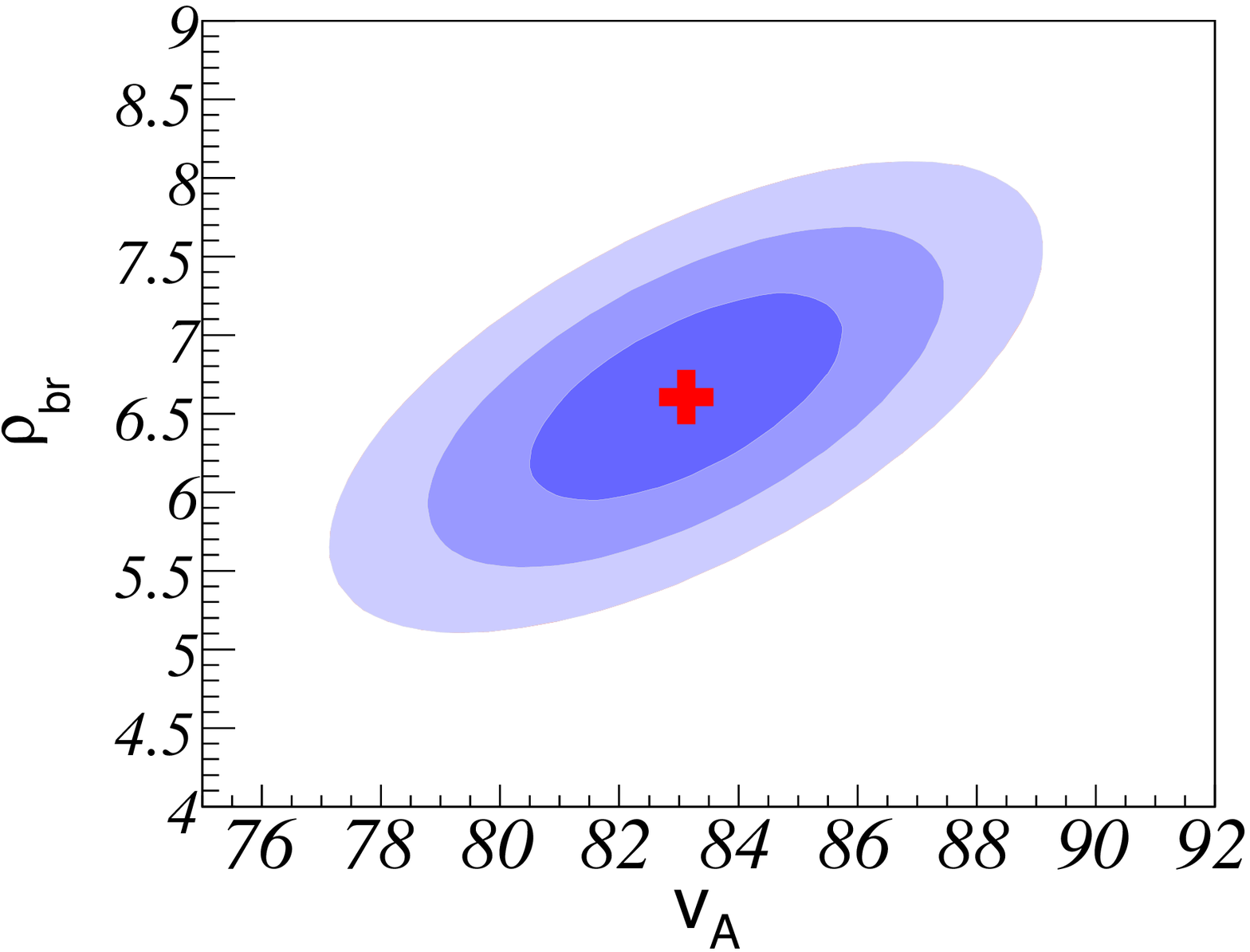}
\includegraphics[width=0.19\textwidth]{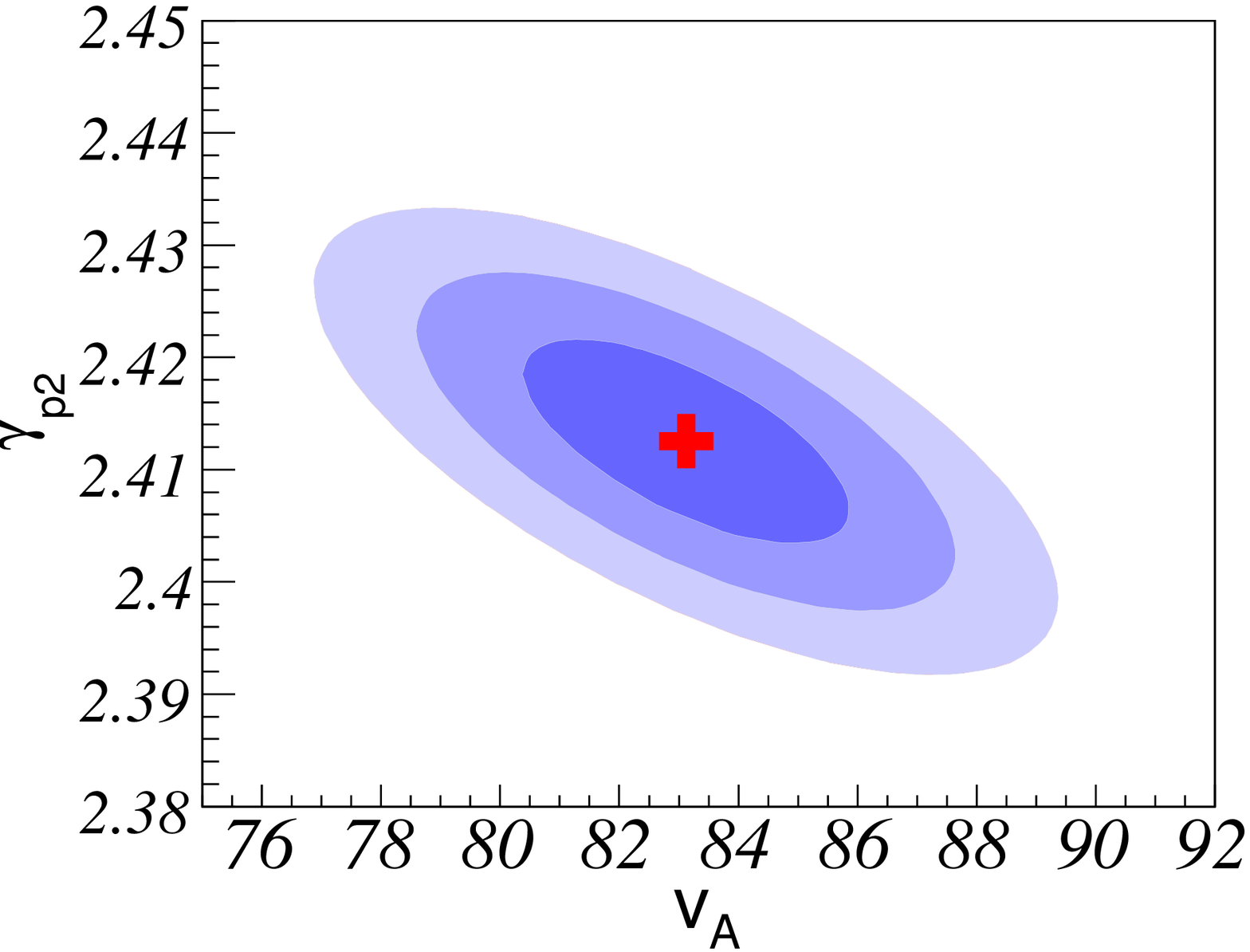}
\includegraphics[width=0.19\textwidth]{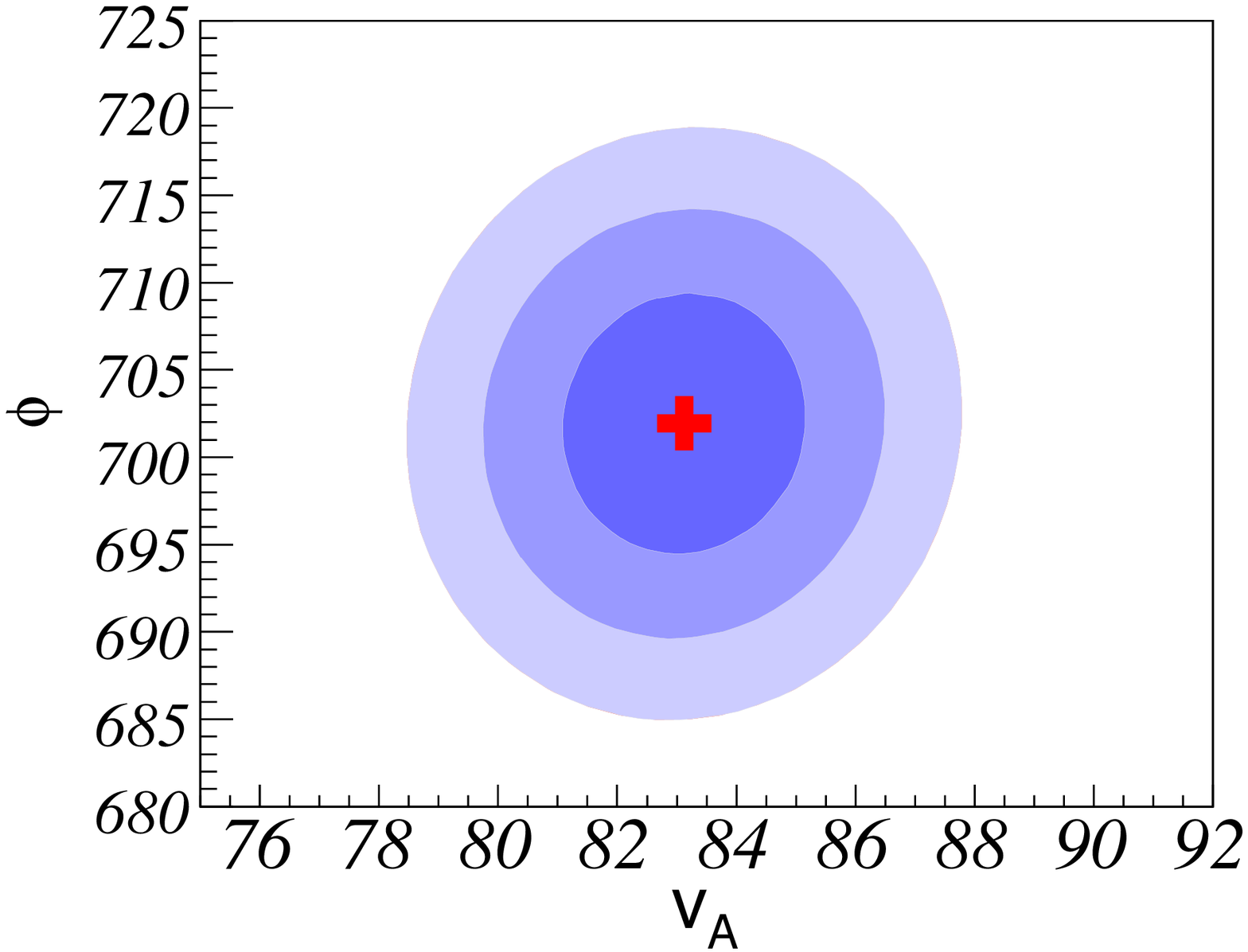}
\includegraphics[width=0.19\textwidth]{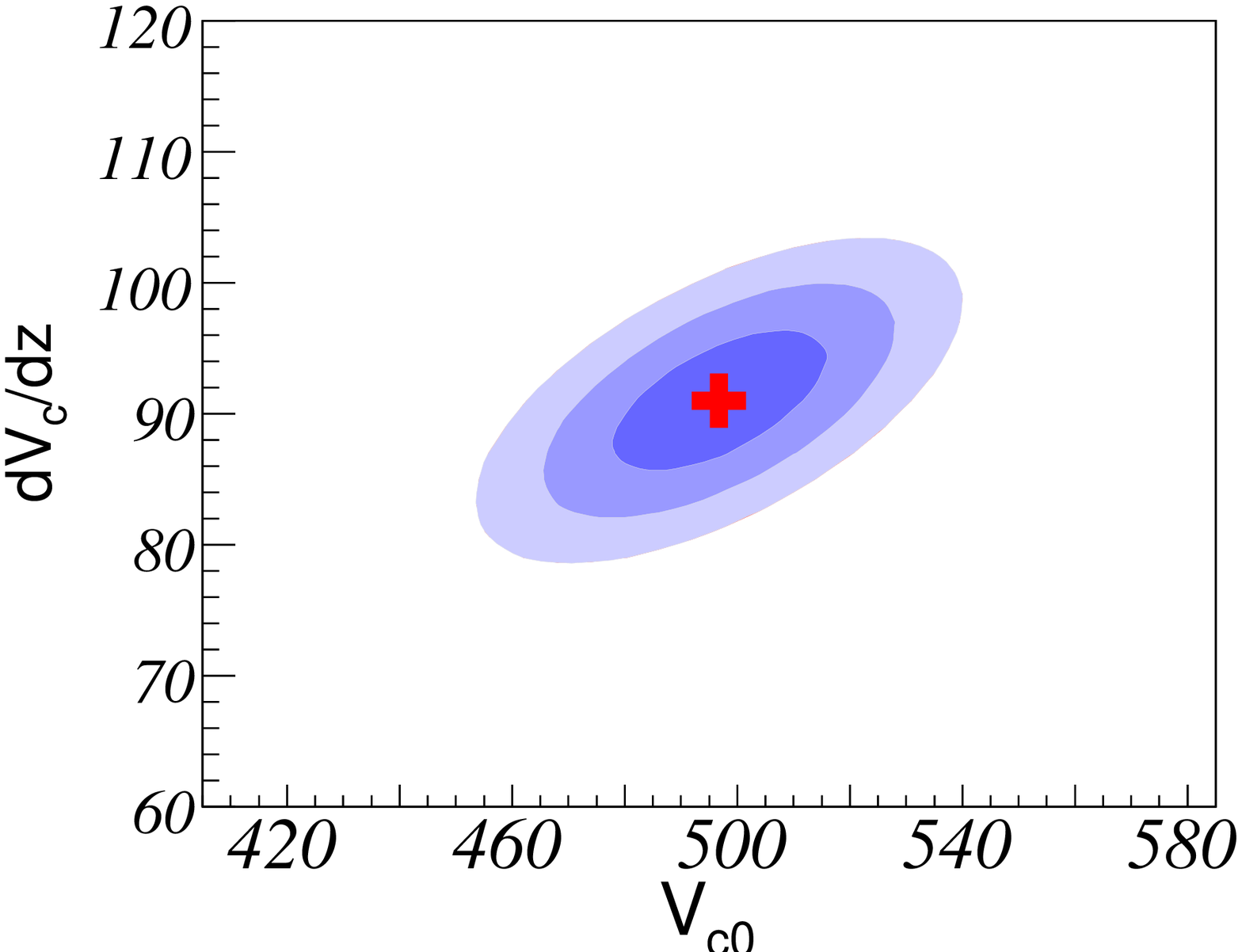}
\\
\includegraphics[width=0.19\textwidth]{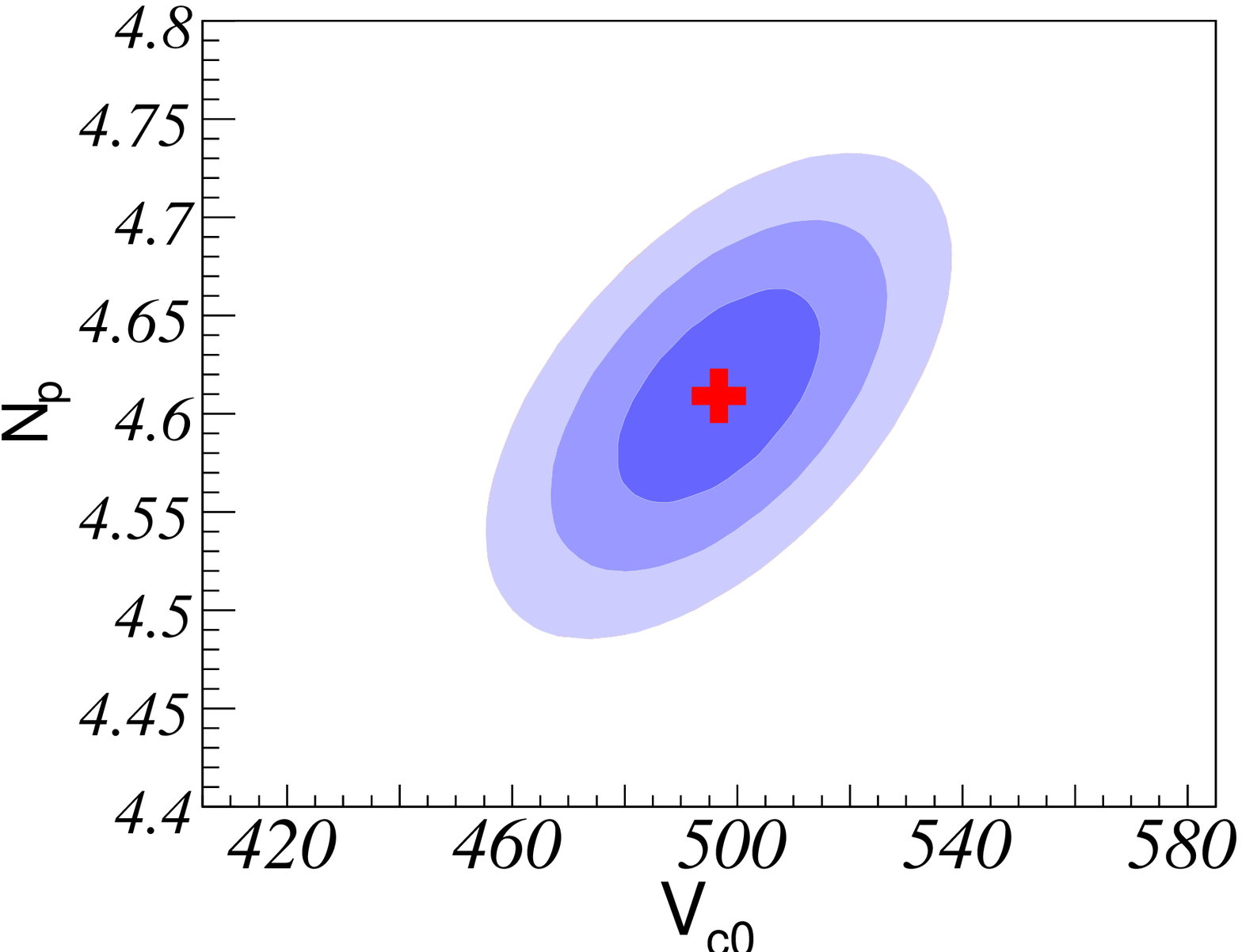}
\includegraphics[width=0.19\textwidth]{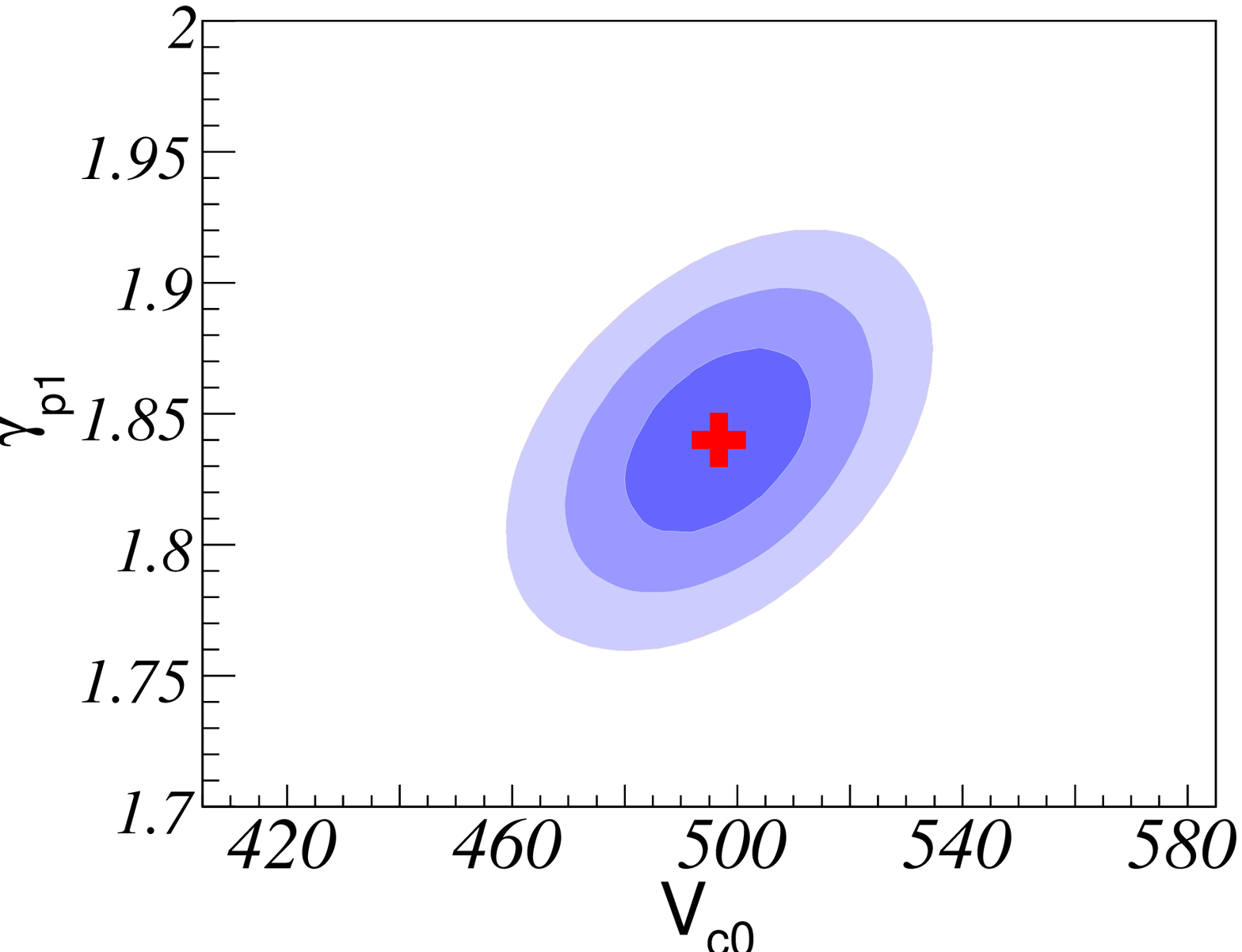}
\includegraphics[width=0.19\textwidth]{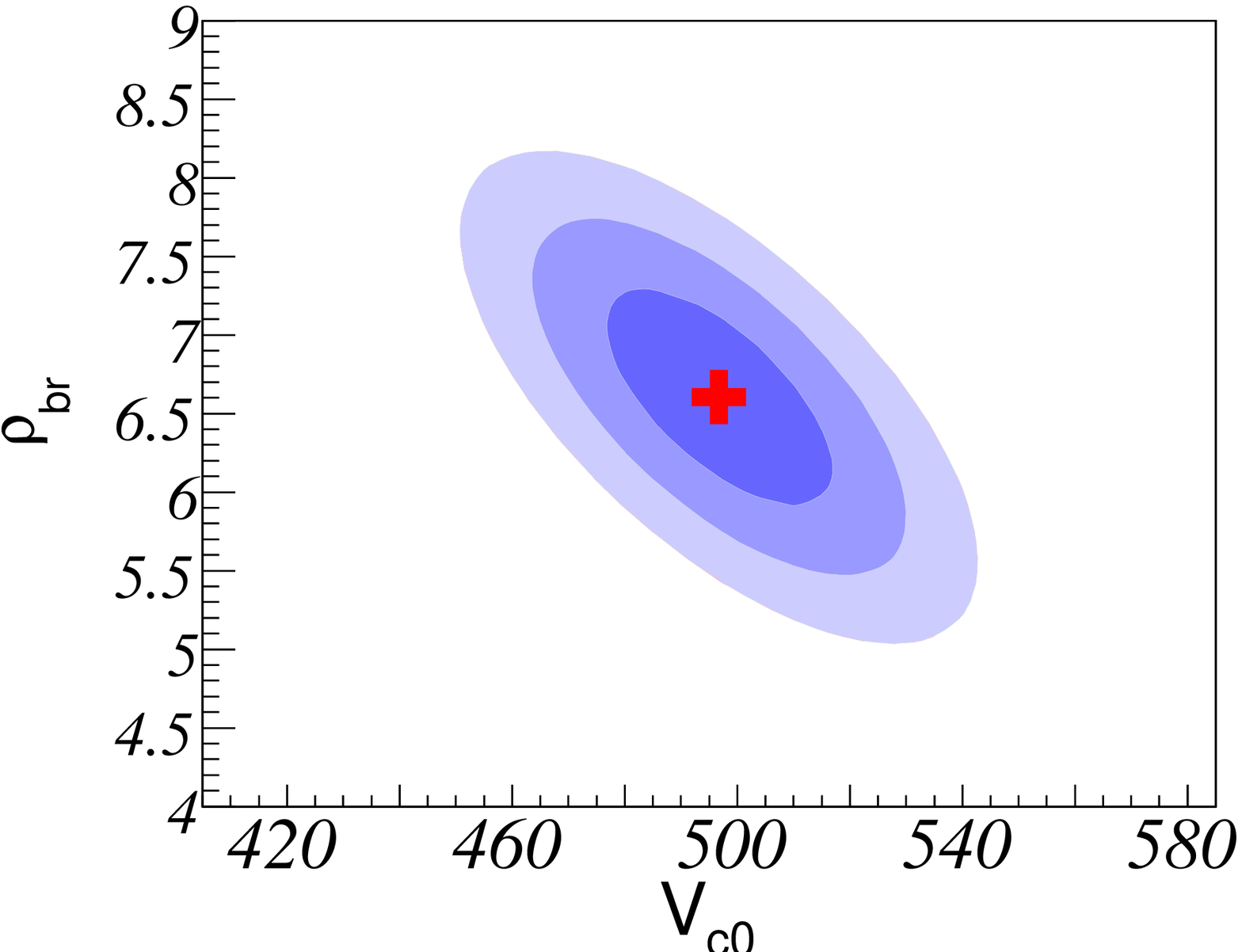}
\includegraphics[width=0.19\textwidth]{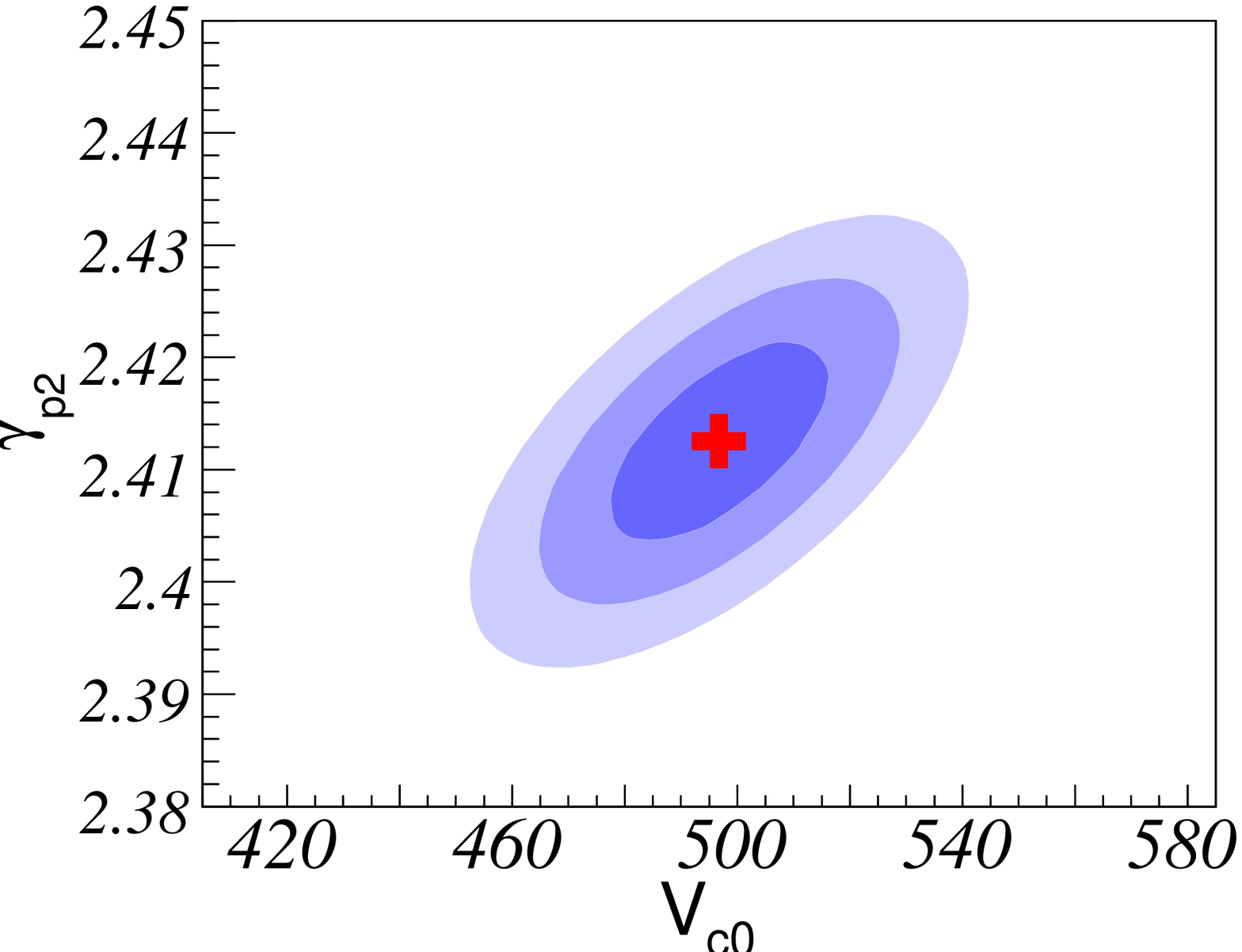}
\includegraphics[width=0.19\textwidth]{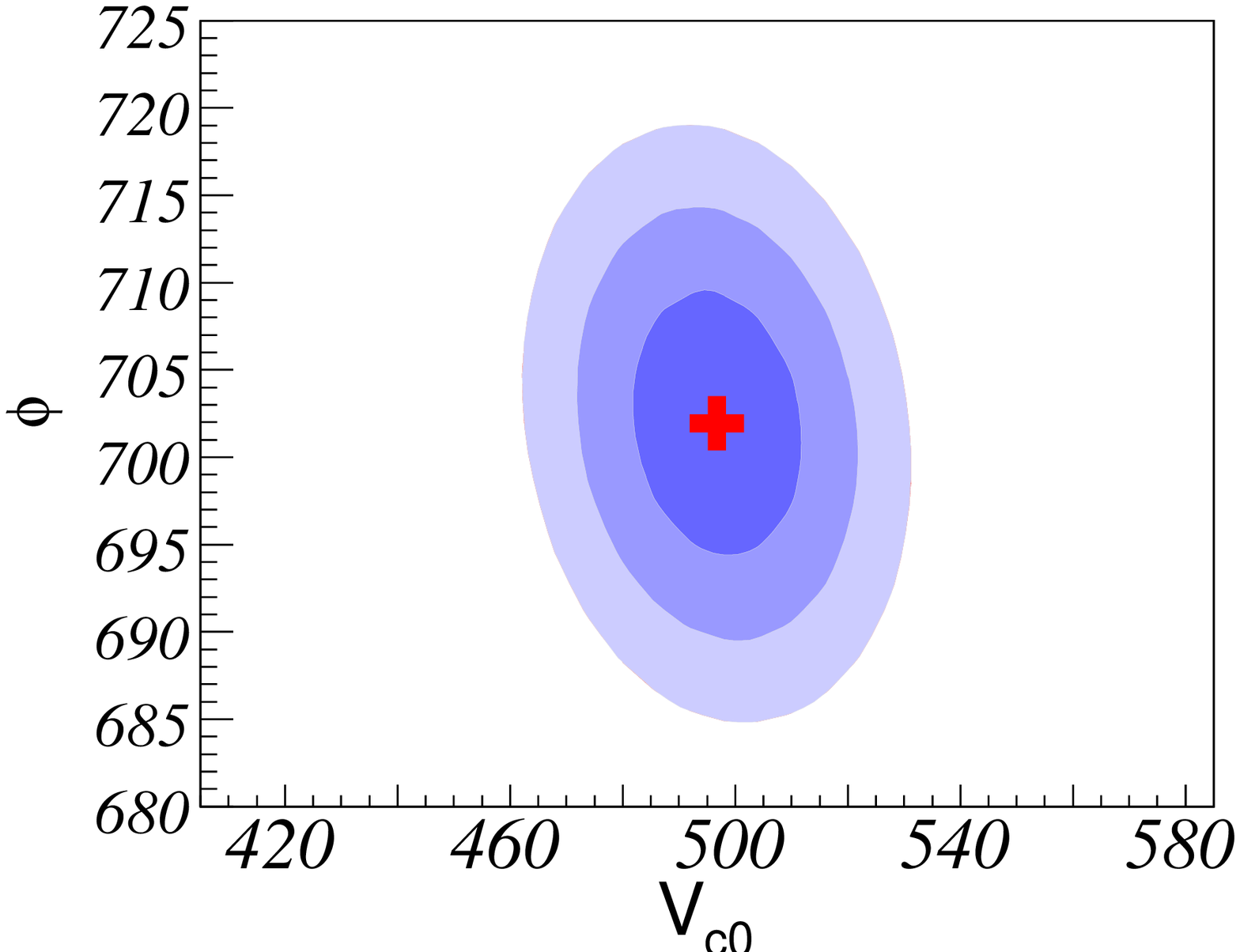}
\caption{
Two-dimensional correlation contours for
the combinations between the propagation parameters involving 
$Z_{h}$, $D_{0}/Z_{h}$, $\delta$, $V_{a}$, $V_{c0}$,$dV_{c}/dZ_h$, $N_p, \gamma_{1}^{p}, \rho^p_{br},\gamma_{2}^{p}$ and $\phi$.
The regions enclosing $68\%,95\%$ and $99\%$ C.L. are shown. The red plus in each plot indicates the best-fit value. 
}
 \label{fig:param_2d}
\end{figure}

\begin{figure}
\includegraphics[width=0.19\textwidth]{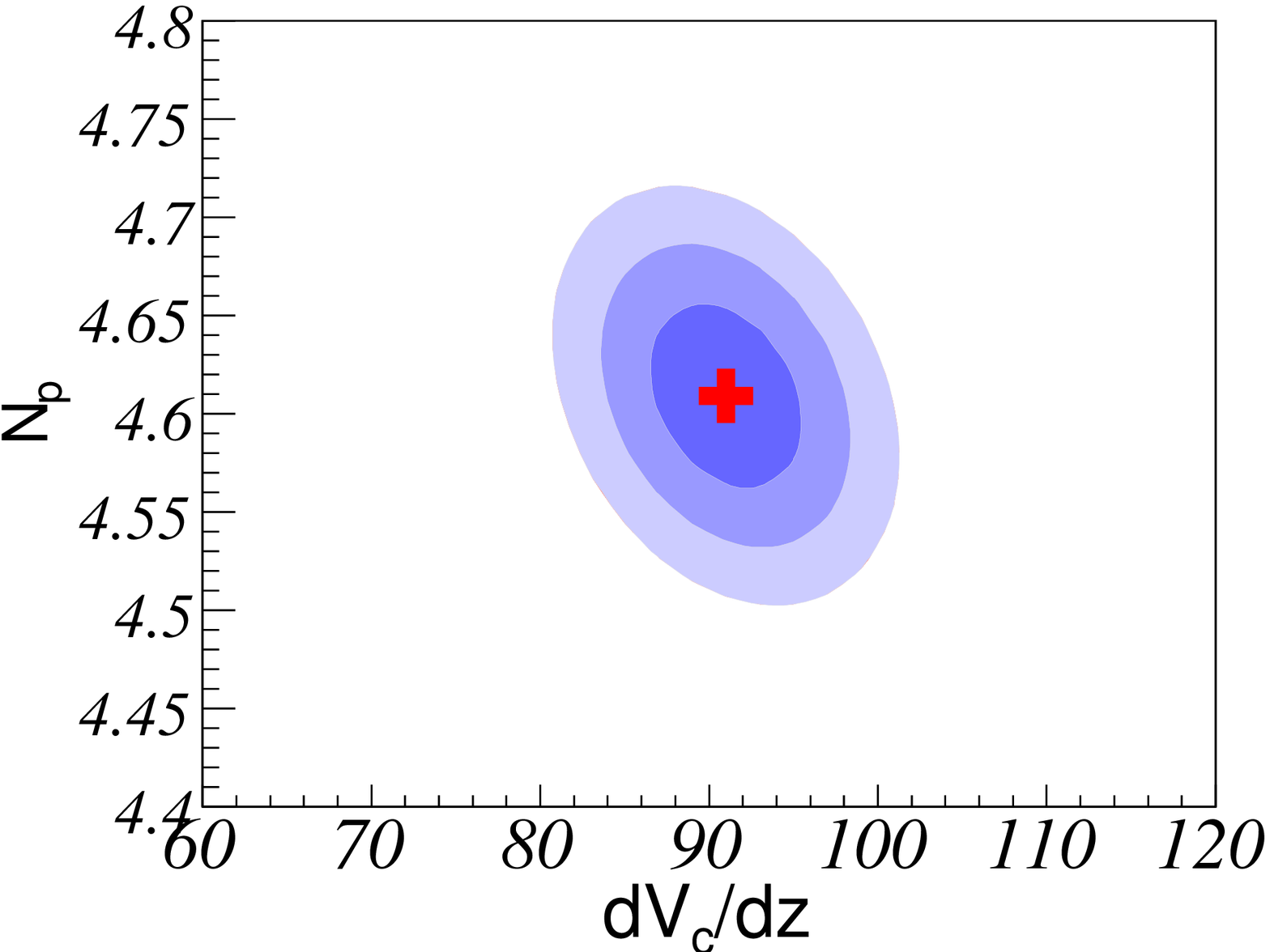}
\includegraphics[width=0.19\textwidth]{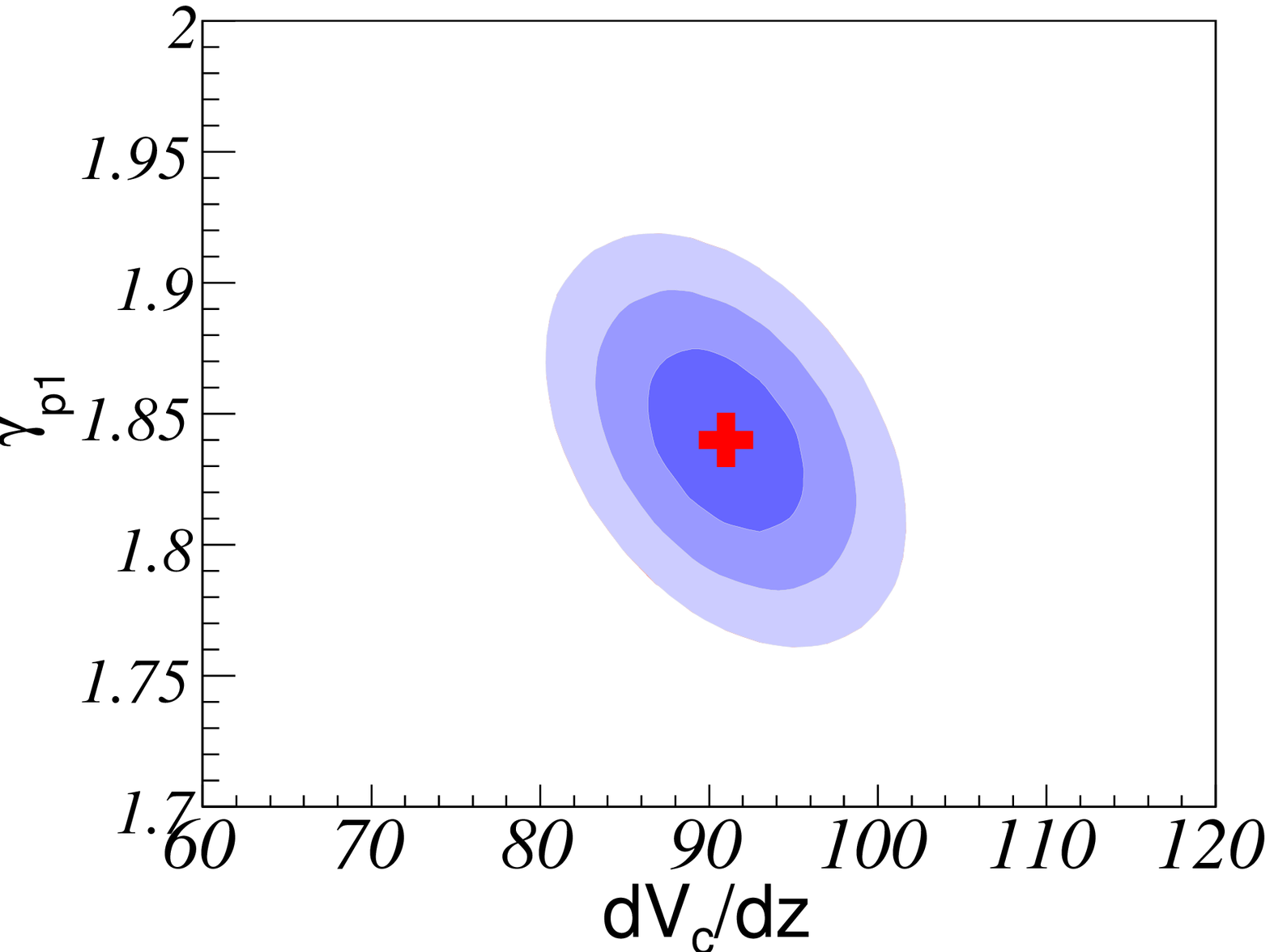}
\includegraphics[width=0.19\textwidth]{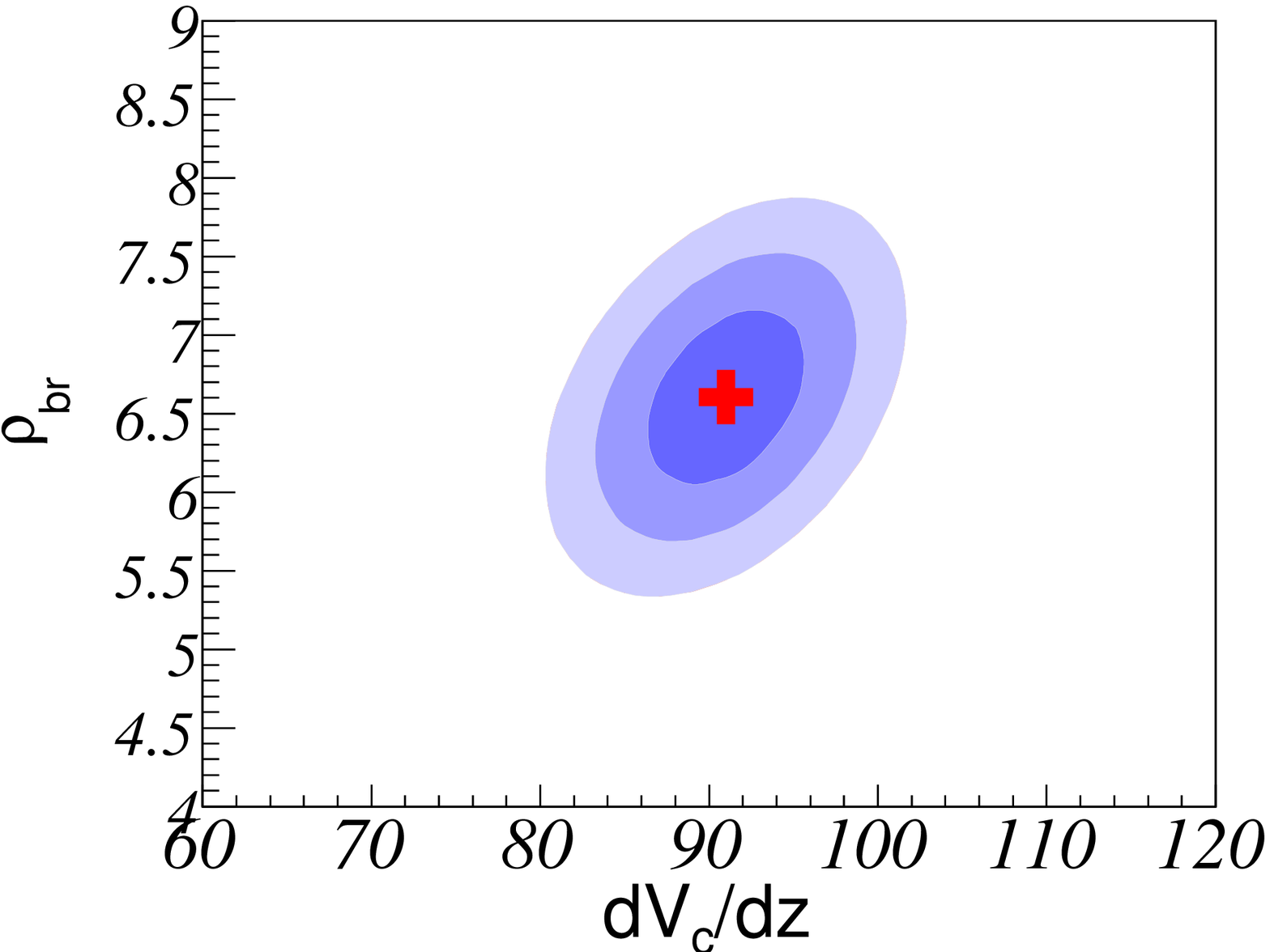}
\includegraphics[width=0.19\textwidth]{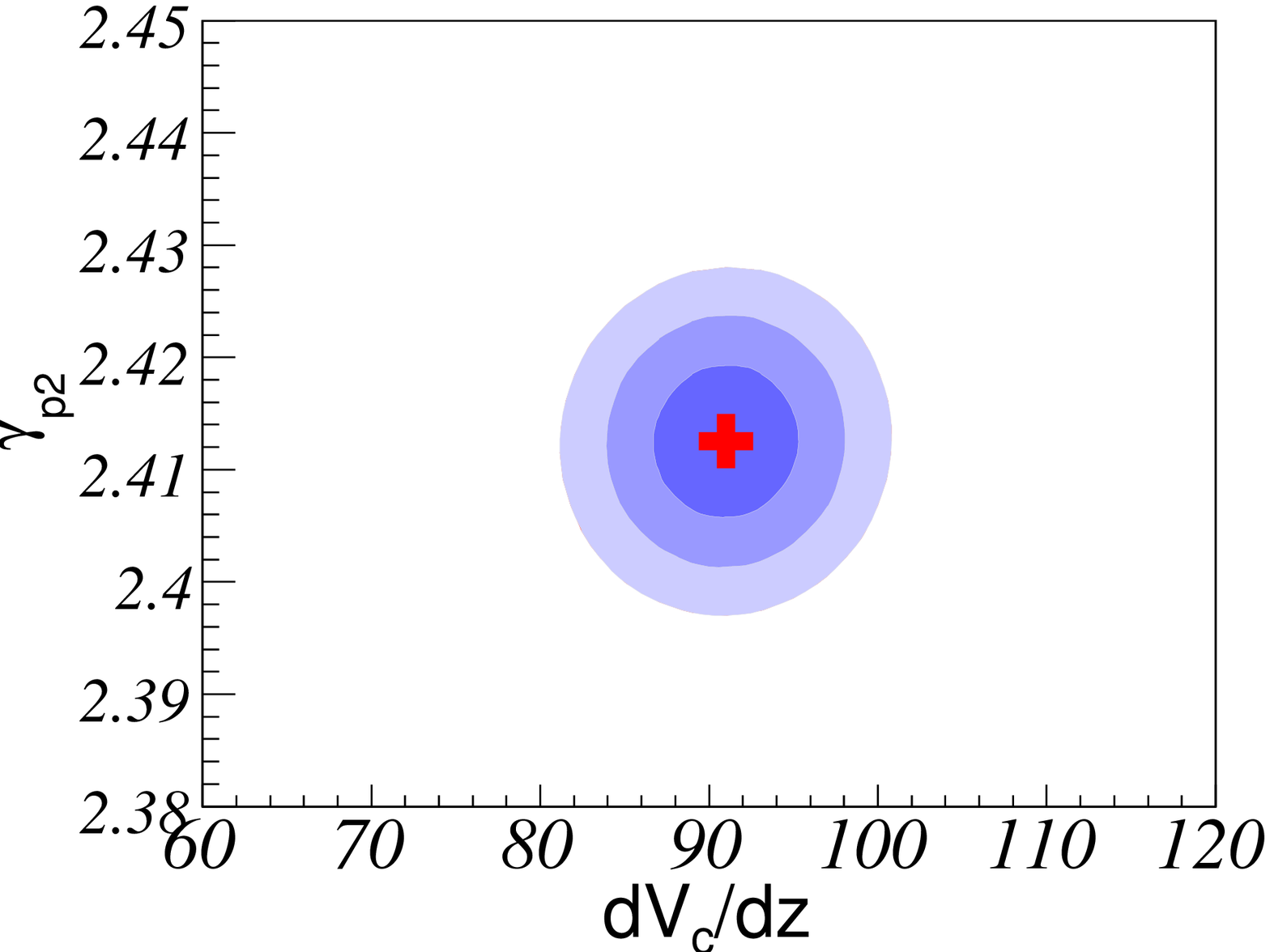}
\includegraphics[width=0.19\textwidth]{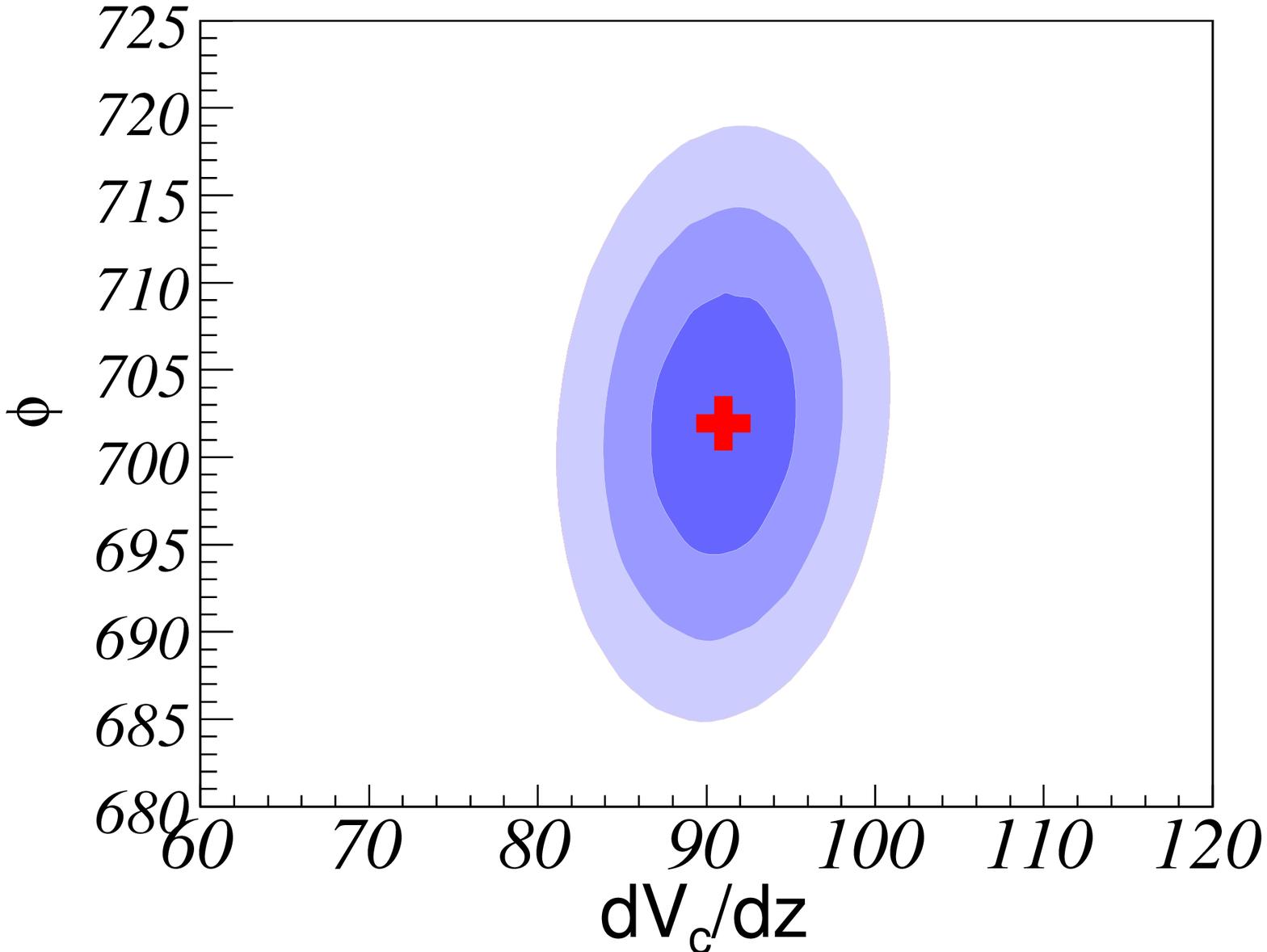}
\\
\includegraphics[width=0.19\textwidth]{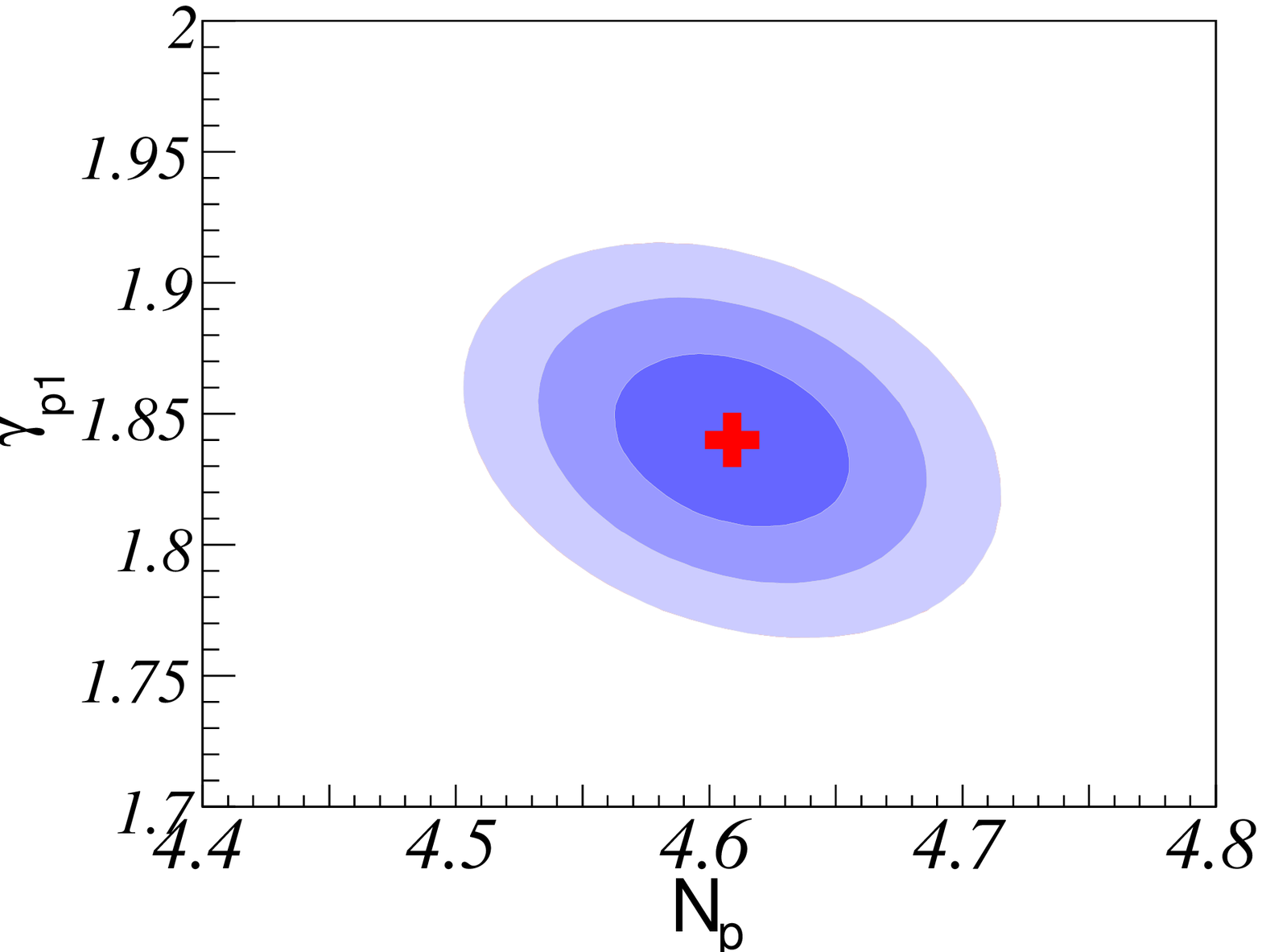}
\includegraphics[width=0.19\textwidth]{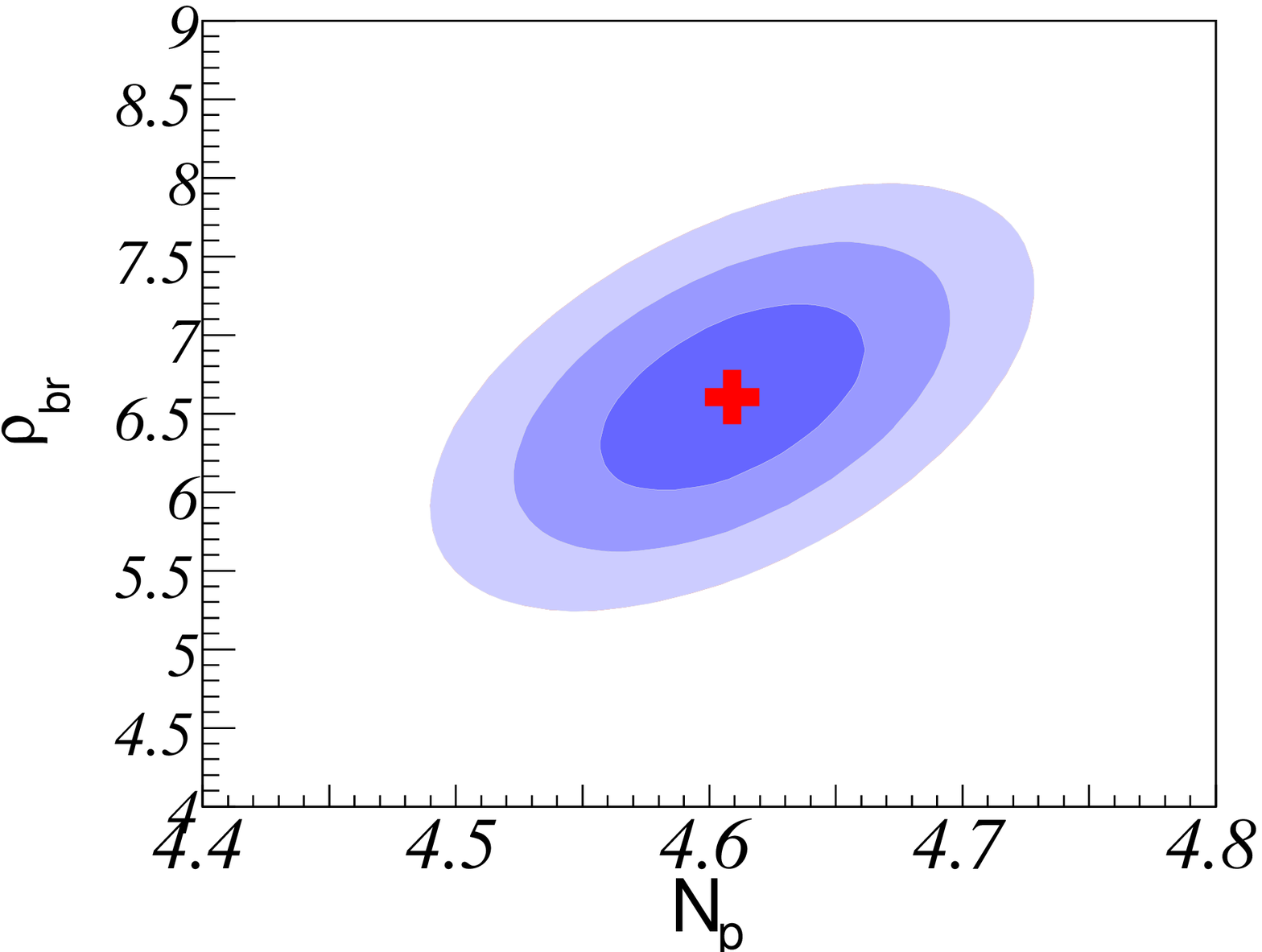}
\includegraphics[width=0.19\textwidth]{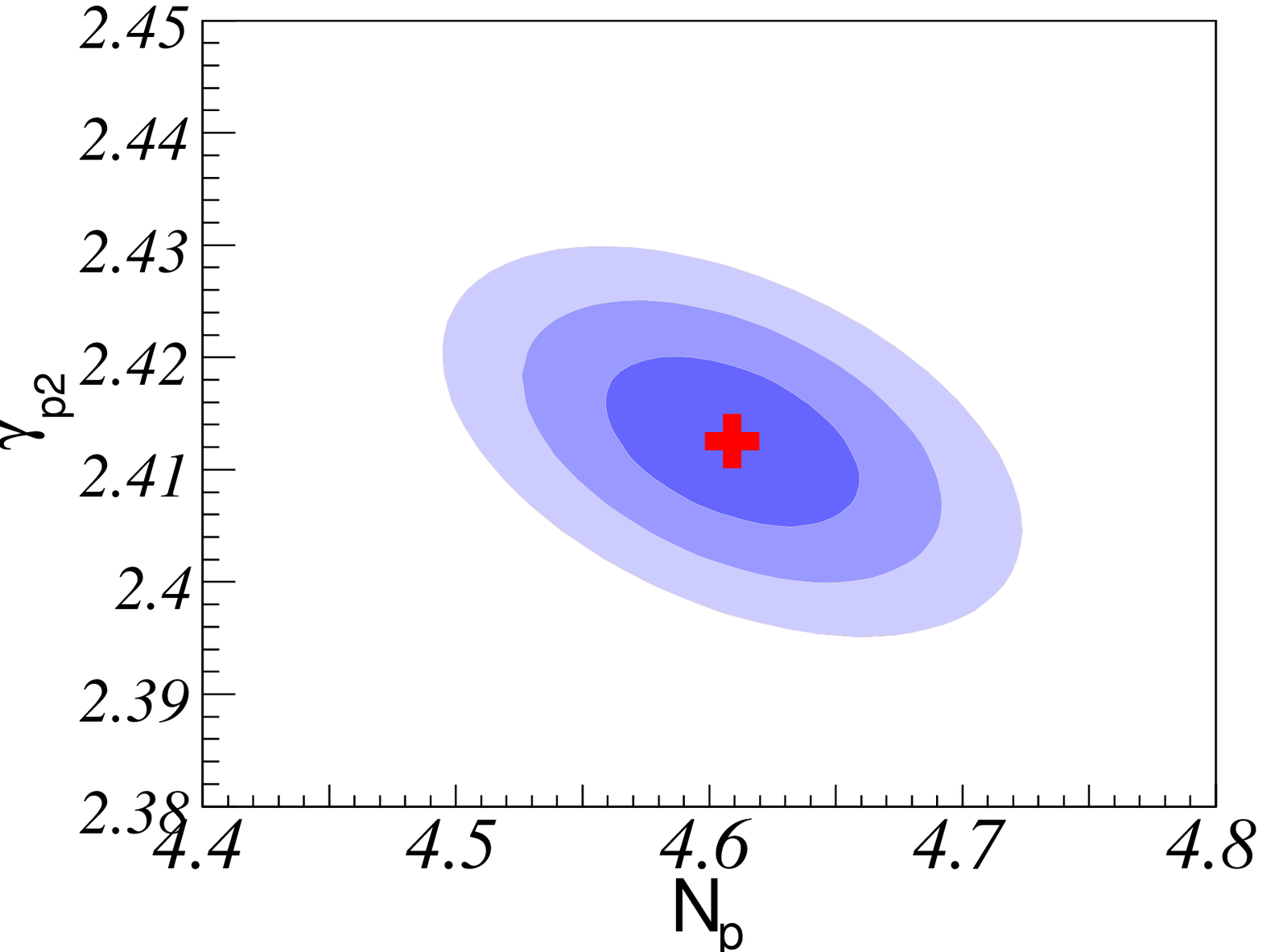}
\includegraphics[width=0.19\textwidth]{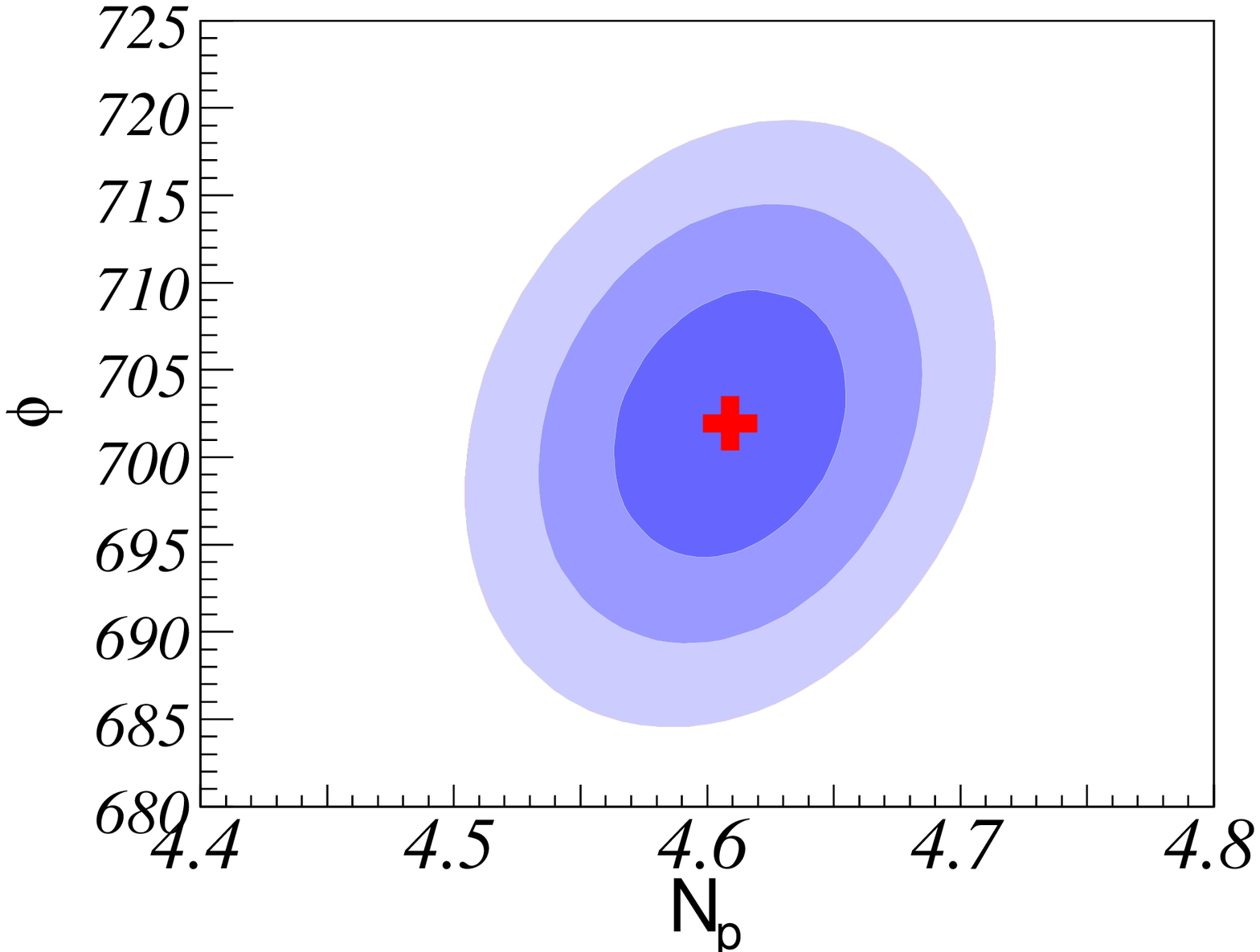}
\includegraphics[width=0.19\textwidth]{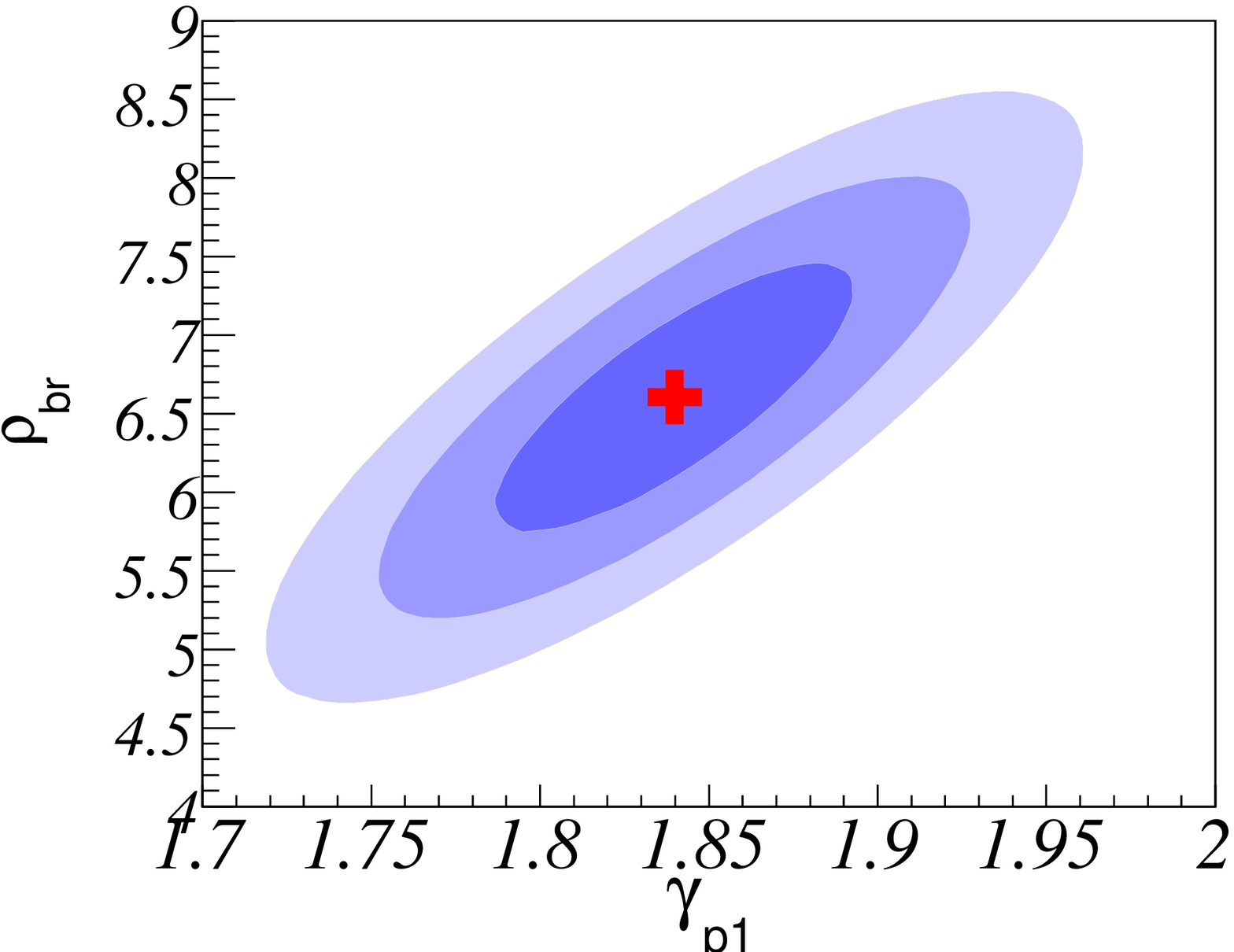}
\\
\includegraphics[width=0.19\textwidth]{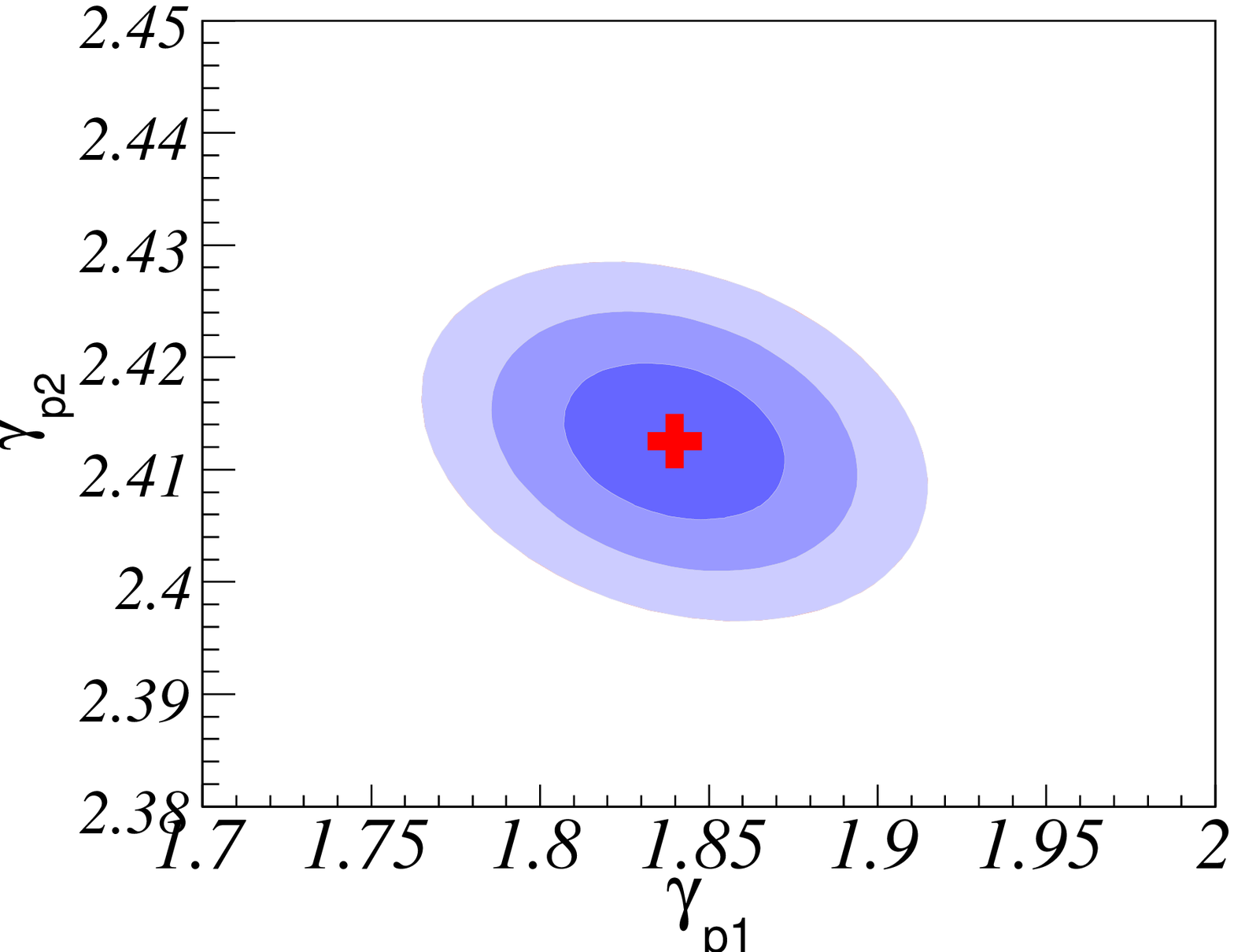}
\includegraphics[width=0.19\textwidth]{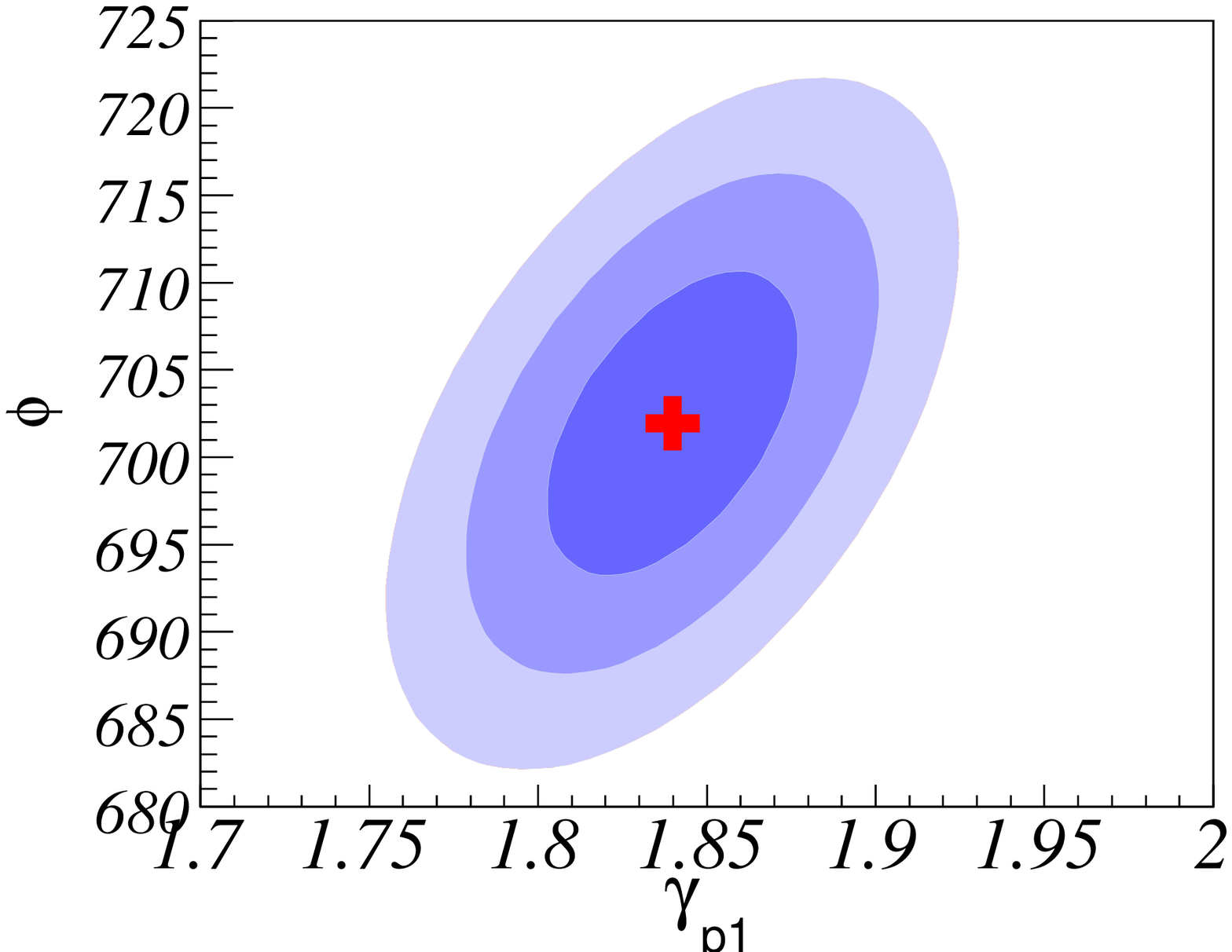}
\includegraphics[width=0.19\textwidth]{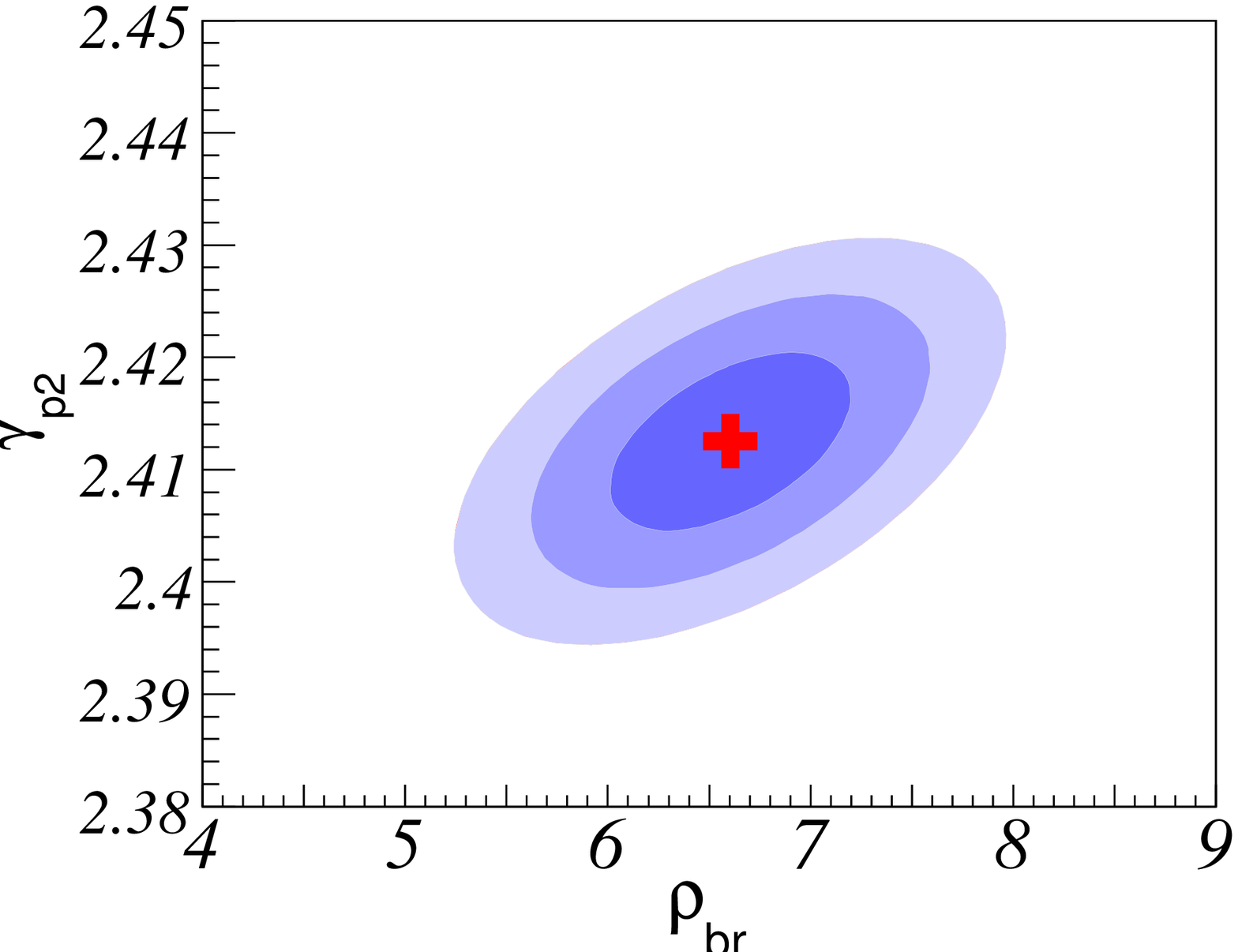}
\includegraphics[width=0.19\textwidth]{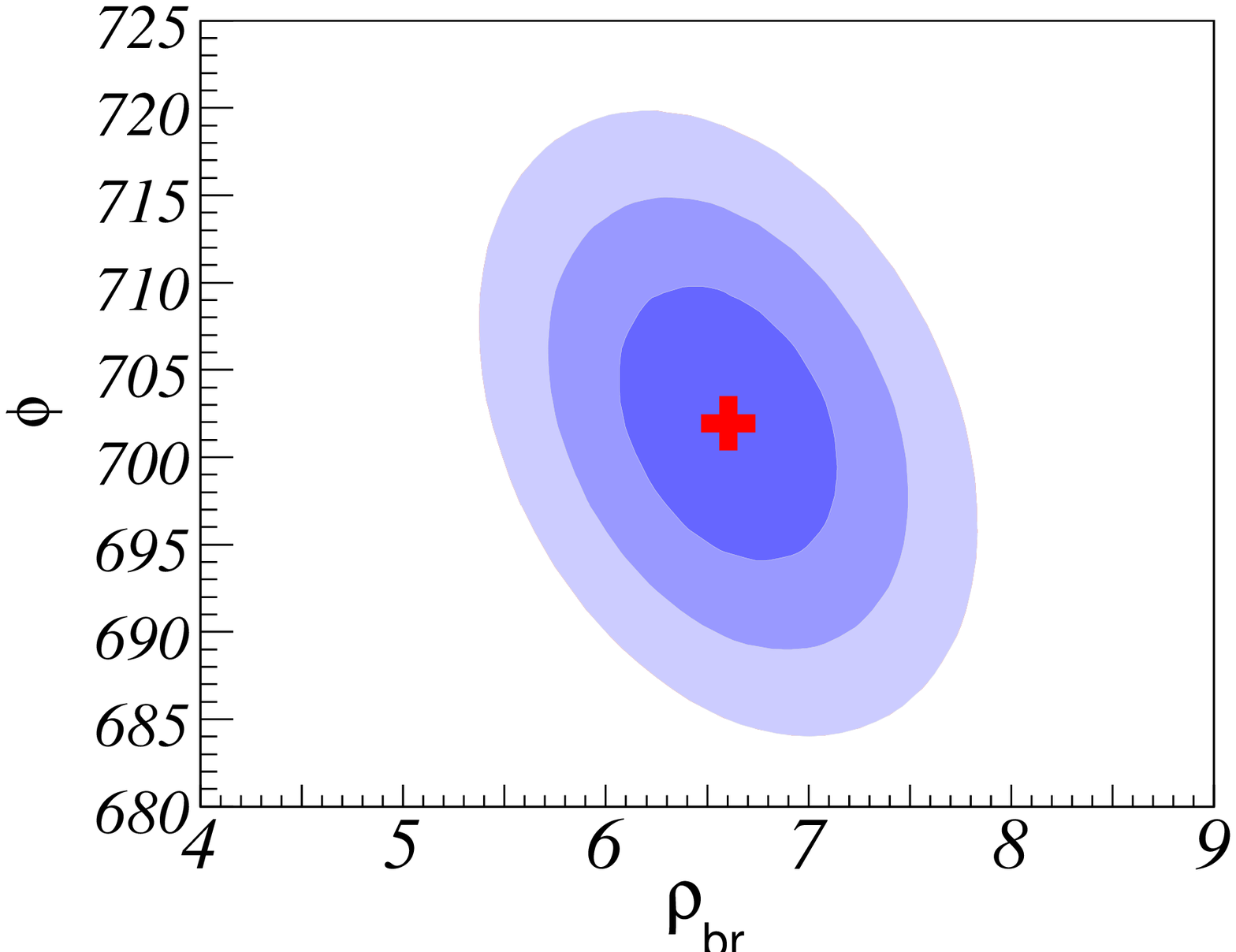}
\includegraphics[width=0.19\textwidth]{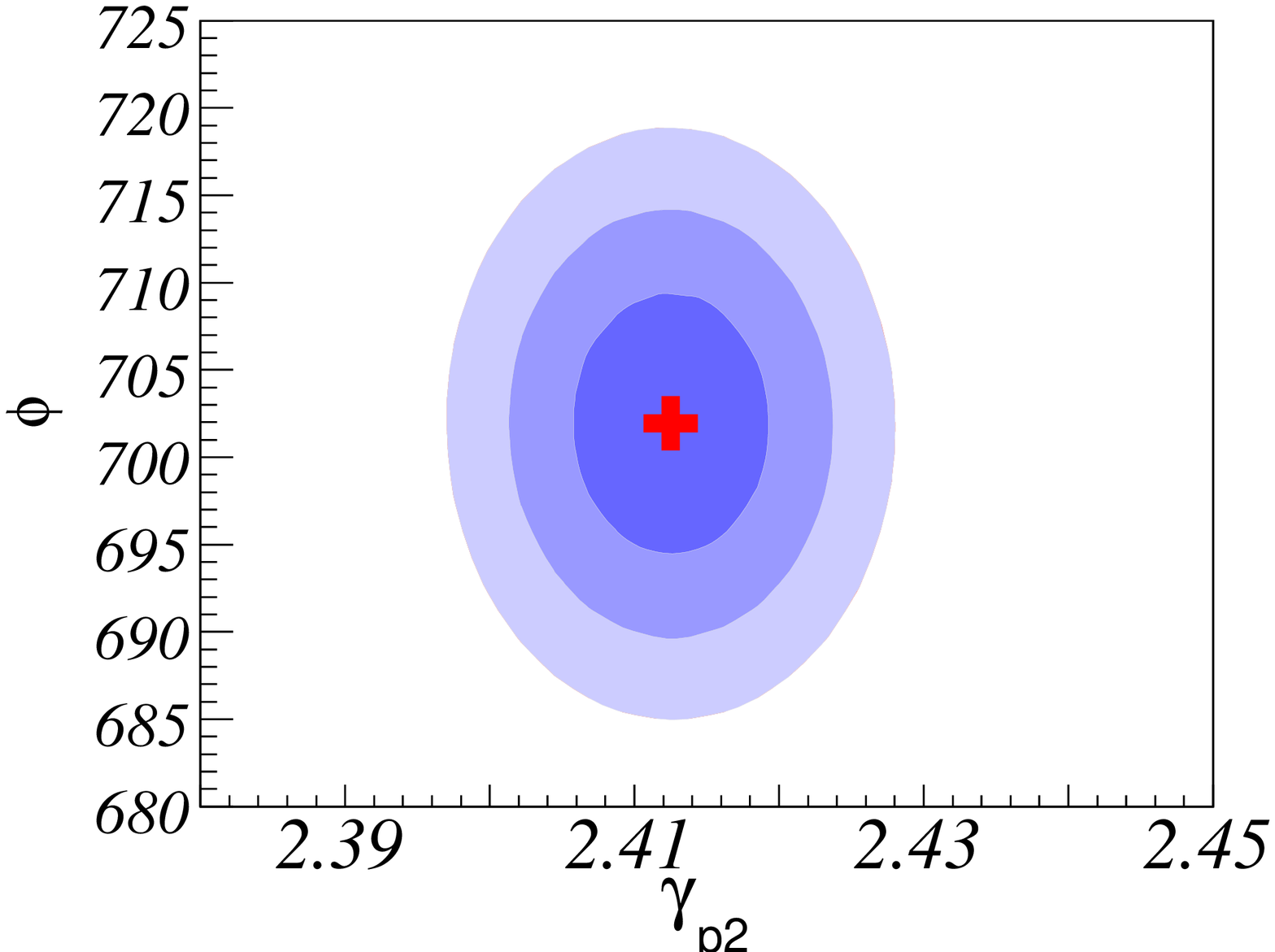}
\caption{
The same as Figure \ref{fig:param_2d}, the rest of all contours.
}
 \label{fig:param_2d2}
\end{figure}

\section{Results}\label{sec:results}
\subsection{The propagation parameter model constrained by AMS-02 data}\label{sec:results_1}
In Table \ref{tab:parameters}, the parameters are listed in all the models. These parameters are the best-fit values in the $\chi^2$ fitting by the MINUIT package.  
The best-fit values for $\alpha$ are much less than 1 in all the models. In fact, $\alpha$ approaches to vanish in the DCR and DC models, which indicates that the injection spectra of CR nucleons become very sharp near the referred rigidity. 
With the comparison of $\chi^2$ between the models, it is shown that the DCR model is favored by AMS-02 experiment and can well predict the fluxes of CR protons, antiprotons, positrons and the value of B/C. For the other models the details of the exclusion by AMS-02 data will be described in the following paragraphs.

\begin{table}[htb]
\begin{center}
\begin{tabular}{l|rrrr|r|rr}
\hline\hline
Models&$\chi^{2}_{\scriptsize\mbox{P,N=72}}$&$\chi^{2}_{\scriptsize\mbox{B/C,N=67}}$&$\chi^{2}_{\scriptsize\bar{\mbox{P}},N=10}$&$\chi^{2}_{\scriptsize\mbox{$e^+$},N=16}$&$\chi^{2}_{\scriptsize\mbox{ACE},N=4}$&$\chi^{2}$&$\chi^{2}$/\scriptsize\mbox{N}\\
\hline
DCR&22.2&50.16&12.94&13.01&\mbox{ }&98.31&0.596\\
DR&47.02&288.31&8.74&117.27&\mbox{ }&461.33&2.8\\
DC&424.6&414&594.5&1037.33&\mbox{ }&2470.4&14.97\\
DCR${}_V$&14.12&77.77&17.51&14.32&\mbox{ }&123.72&0.75\\
DCR${}_0$&49.17&91.42&382.62&185.58&\mbox{ }&708.79&4.3\\
DCR${}_1$&28.10&50.49&12.05&21.13&38.76&150.54&0.89\\
DCR${}_2$&123.57&109.26&356.98&1923.1&5.95&2518.9&14.9\\
 \hline\hline
\end{tabular} 
\end{center}
\caption{The best-fit  $\chi^2$ relevant to the models DCR, DR and DC. the subscript N of $\chi^2$ denotes the number of the experiment data points, for instance, the number of AMS-02 proton data points is N=72. The total $\chi^2$ and its value over the total data-points of the chosen experiment for each model are presented in the two tail columns.
}
\label{tab:chisquares}
\end{table} 
In Table \ref{tab:chisquares}, the best-fit $\chi^2$ relevant to the three types of models DCR, DR and DC are shown. With the estimation of the total $\chi^2$ over the point numbers of the experimental data($\chi^2$/N), DC, DR and DCR${}_{0}$ models are excluded obviously by AMS-02 data. The strong constraints on these models are derived partly from the latest released B/C data, which gets higher precision and has a more complex feature of spectra than the other experimental data. From the values in Table \ref{tab:chisquares} and \ref{tab:parameters}, it is also seen that in the DR model the predicted flux of CR positrons and B/C value have the remarkable deviation from AMS-02 data, though the solar modulations are separately considered for CR positrons and nucleons. 
In the DCR${}_{0}$ model, the hadronic interaction model is the Conventional model and that tension also does not been eliminated. 
In the DCR model, with an alternative interaction model, i.e. QGSJET-II-4, the predicted fluxes of CR antiprotons and positrons are well fitted to the AMS-02 data. It implies that the hadronic interaction model is a key to relax the predicted tension between CR positrons and antiprotons. As seen in Figure \ref{fig:zfactorAll}, for CR antiprotons, Z-factors relevant to QGSJET-II-4 model are greater than the Conventional model in the low energies. But for CR positrons, Z-factors relevant to QGSJET-II-4 model are slightly less than the Conventional model. 

In DCR${}_V$ model, Alfv$\grave{\mbox{e}}$n speed $V_{A}$ has the two different values relevant to CR positrons and CR nucleons, which are found in Table \ref{tab:parameters}. Comparing $\chi^2$ between the DCR and DCR${}_V$ models in Table \ref{tab:chisquares}, it is found that the differences of re-acceleration effect does not apparently improve the fluxes of CR positrons and CR nucleons to fit to AMS-02 data. At the meantime, as a bad result, the Galactic wind velocity $\mathbf{V}_{c}$ and the diffusion coefficient D${}_0$ are converged into the large values.
\begin{figure}
\begin{center}
\includegraphics[width=0.49\textwidth]{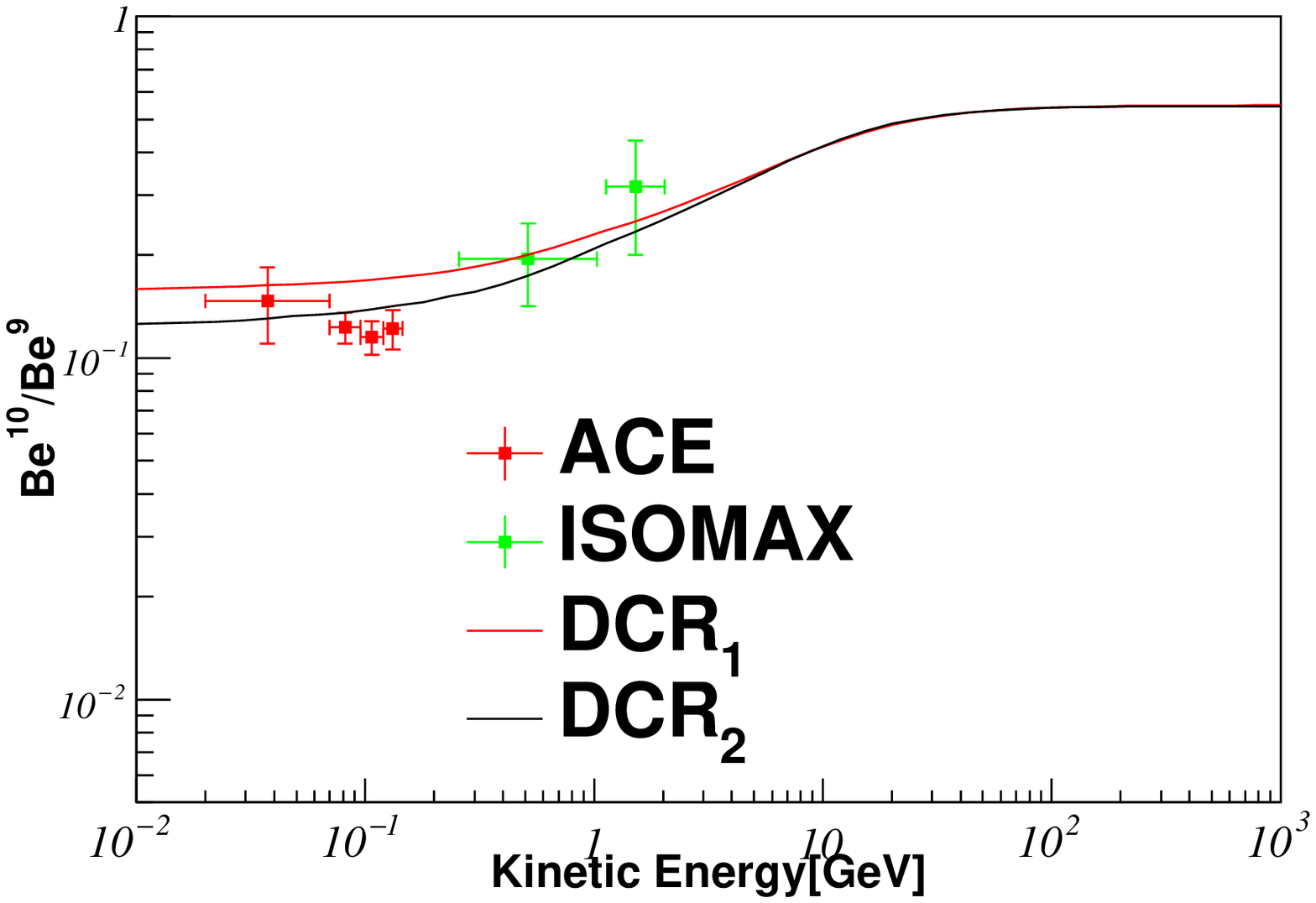}\includegraphics[width=0.49\textwidth]{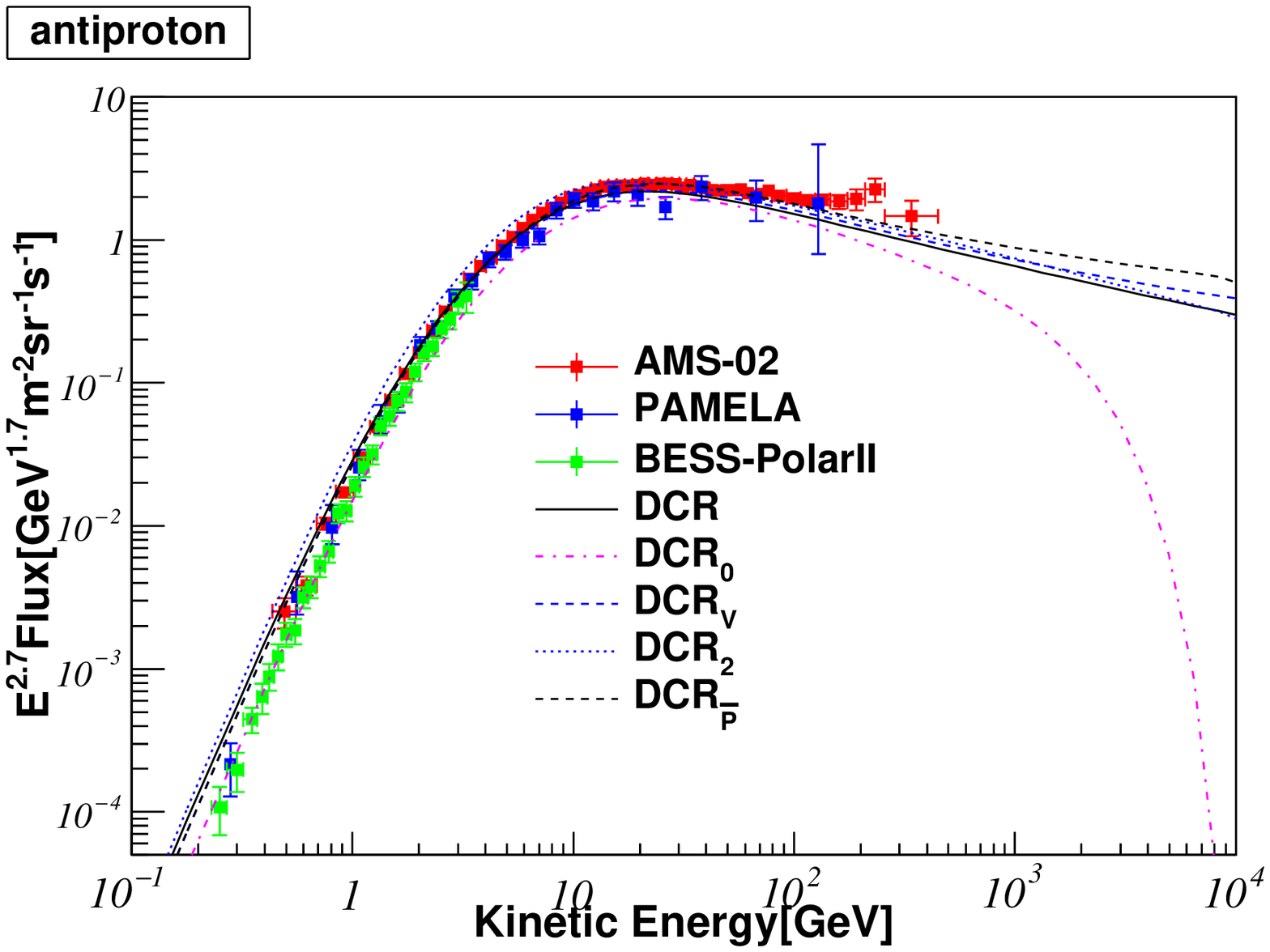}
\end{center}
\caption{(left) The predicted $\mbox{Be}^{10}/\mbox{Be}^{9}$ flux ratio from the re-fitting parameters by combining the ACE data (denoted as DCR${}_1$) and the chosed parameters relevant to $\chi^2/\mbox{N} <2$ only for the experimental data: CR protons, B/C and Be${}_{10}$/Be${}_9$ (denoted as DCR${}_2$) based on the DCR model. ACE \citet{yanasak2001} and ISOMAX \citet{Hams:2004rz} data are also shown.
(right) the flux of CR antiprotons predicted from the models: DCR, DCR${}_{0}$, DCR${}_{\bar{p}}$ (all the antiproton data from AMS-02 are included) and the measurements from AMS-02 \citet{Aguilar:2016kjl,Aguilar:2014mma,Aguilar:2014fea}, PAMELA \citet{Adriani:2010rc} and BESS-PolarII \citet{Abe:2011nx} experiments. 
}
\label{fig:BeComparison}
\end{figure}

In order to explore the correlations between the propagation parameters in DCR model, MCMC sampling is done using \software{CosmoMC} package, which call the functions of GALPROP in the sampling steps. Exception the smoothness parameter $\alpha$, there are 11 propagation parameters to be used. In the sampling result, the numbers of the sampling steps in the 28 chains amount to more than 428000 after burn-in. The details of the sampling methods are found in the previous paper \citet{Jin:2014ica}. For the contour drawing, the covariance matrix of 11 parameters is calculated with the sampling data. Based on the inverse of the covariance matrix, with the best-fit values of the propagation parameters in DCR model, the correlation contours between any two parameters are drawn in Figure \ref{fig:param_2d} and \ref{fig:param_2d2}. 

As seen in these contours, the Alfv$\grave{\mbox{e}}$n speed $V_{A}$, the Galactic wind velocity $\mathbf{V}_{c}$ and the diffusion coefficient D${}_0/Z_h$ are completely the positive correlations between the two parameters exception for the combination between the Alfv$\grave{\mbox{e}}$n speed $V_{A}$ and the diffusion coefficient D${}_0/Z_h$. Thus, it is clear that they are converged into the large values in the context.  

In order to check the value of $\mbox{Be}^{10}/\mbox{Be}^{9}$ compatible with ACE data, the propagation parameters are re-fitted by including the ACE data based on the DCR model.
The best-fit flux ratio from the refitted parameters and the predicted values from the chosen parameters of the DCR model are drawn in Figure \ref{fig:BeComparison}.
For ISOMAX data \citet{Hams:2004rz}, in DCR${}_1$ and DCR${}_2$ models the predicted values are both consistent.   
For ACE data \citet{yanasak2001}, as seen in Figure \ref{fig:BeComparison}, $\mbox{Be}^{10}/\mbox{Be}^{9}$ is compatible with ACE data in the DCR${}_2$ model, but deviate remarkably from ACE data in DCR${}_1$ model.  
In \ref{tab:chisquares} for Be$^{10}$/Be$^{9}$ from ACE data $\chi^2$/N is much greater than 2, but for the others from AMS-02 $\chi^2$/N are less than 2. 
Thus, as a result, there is a tension between Be$^{10}$/Be$^{9}$ and CR antiprotons and positrons.
Since CR protons and B/C data from AMS-02 do not favor the DR and DC model, Be$^{10}$/Be$^{9}$ data does not help to constrain the propagation parameters. 
In the propagation parameters constrained by CR protons, antiprotons, positrons, B/C and Be$^{10}$/Be$^{9}$ data, the large convective velocity, which appears in DCR${}_V$ model, is limited and do not help to predict the value of Be$^{10}$/Be$^{9}$ compatible with ACE data. 
That situation is derived from the high precision data from AMS-02, which is dominant in the chi-square fitting and depresses the constraint from ACE data with low accuracy. 

It is known that the spectra of CR protons from the measurement of AMS-02 experiment have two breaks with the kinetic energy increasing from 0.5 GeV to 2 TeV \citet{Aguilar:2015ooa}. 
The second break of the CR proton spectra means that the absolute index of CR proton spectra begins to decrease above 330 GeV, which is often called as CR hardening. 
From a comparison of $\chi^2$ for the CR protons in Table \ref{tab:chisquares}, it is indicated that the flux of CR protons with the multiple power indices is well predicted considering the re-acceleration and convection processes. Though the DCR${}_V$ model does not help to improve the prediction of the fluxes of CR positrons and antiprotons, the large $\mathbf{V}_{c}$ and D${}_0$ promote the $\chi^2$ relevant to CR protons decreasing from 22.2 to 14.12.
The experimental data of CR protons with high precision from AMS-02 has 72 points. The $\chi^2$ 14.12 means the flux of CR protons is better predicted in the DCR${}_V$ model, which justifies the convection effect for CR propagation.   
  
\begin{figure}
\includegraphics[width=0.49\textwidth]{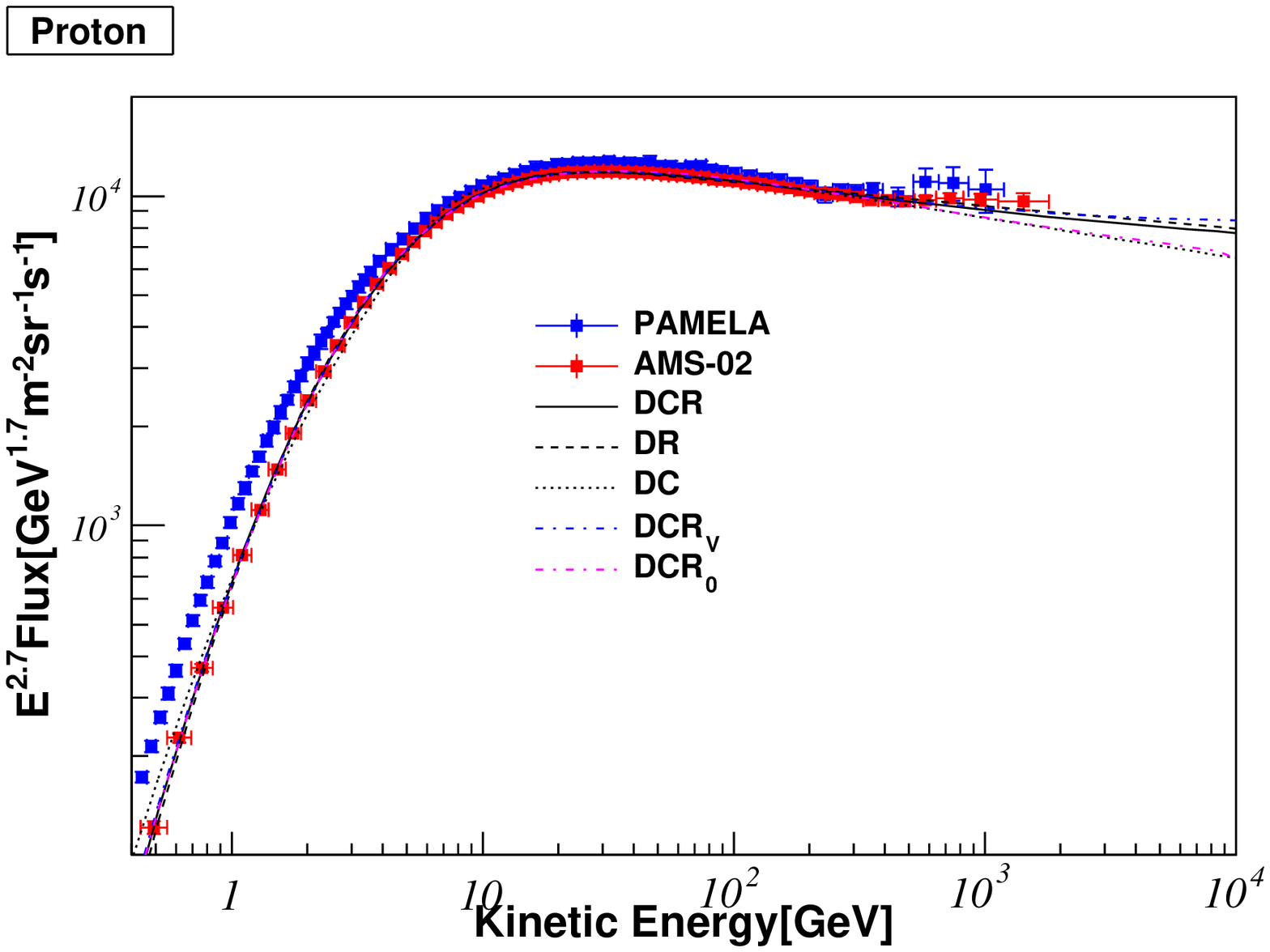}\includegraphics[width=0.49\textwidth]{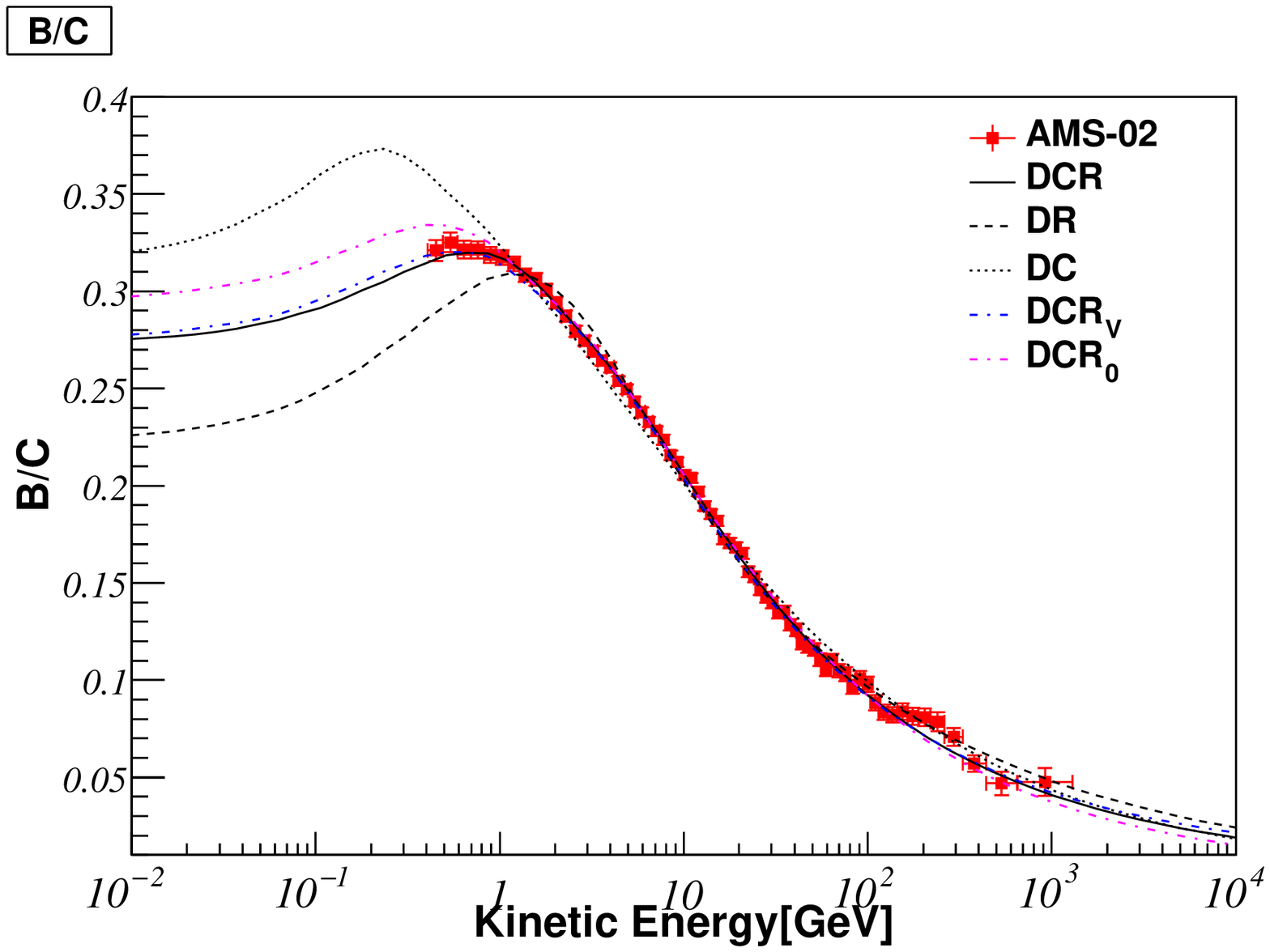}
\includegraphics[width=0.49\textwidth]{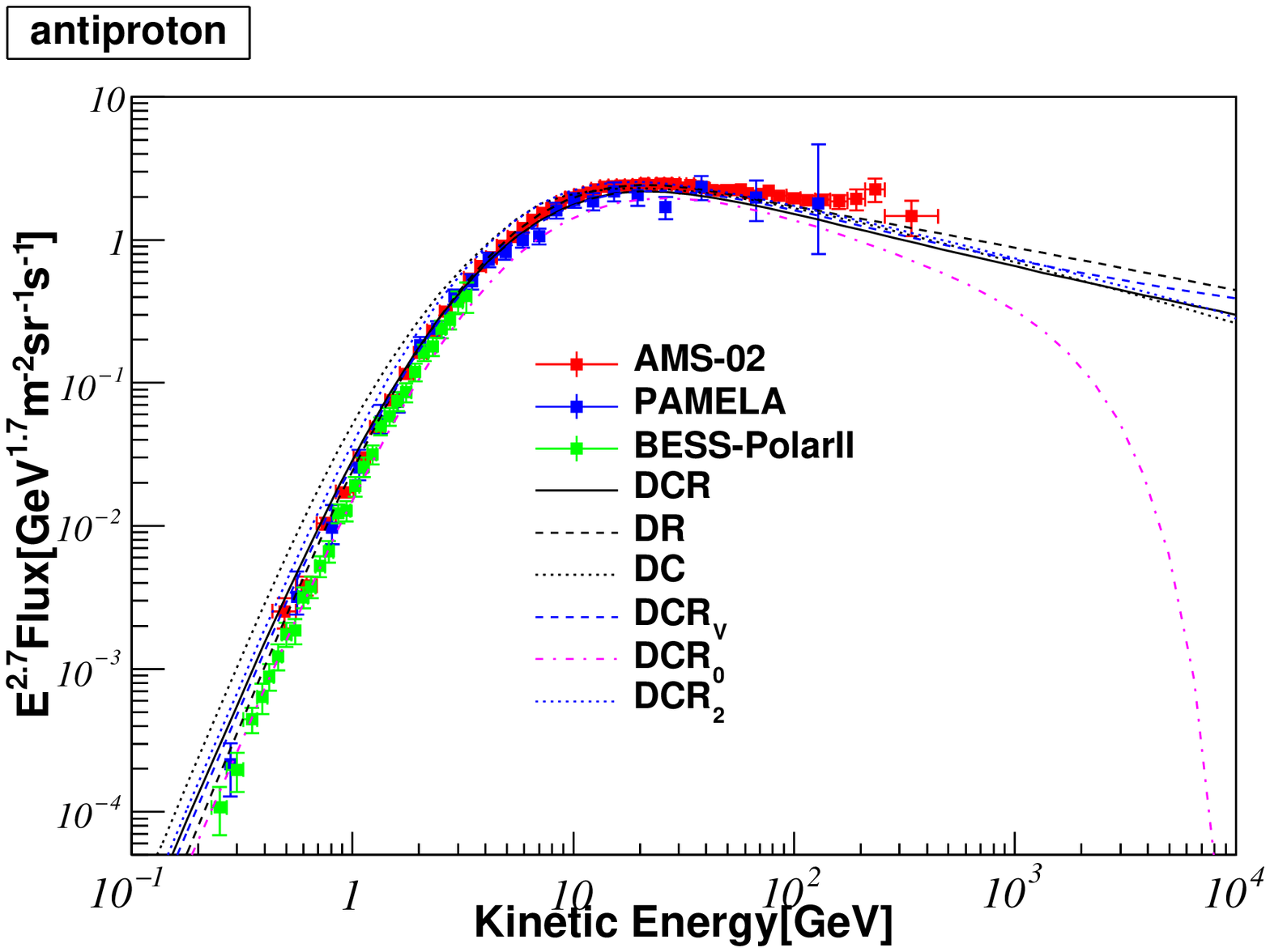}\includegraphics[width=0.49\textwidth]{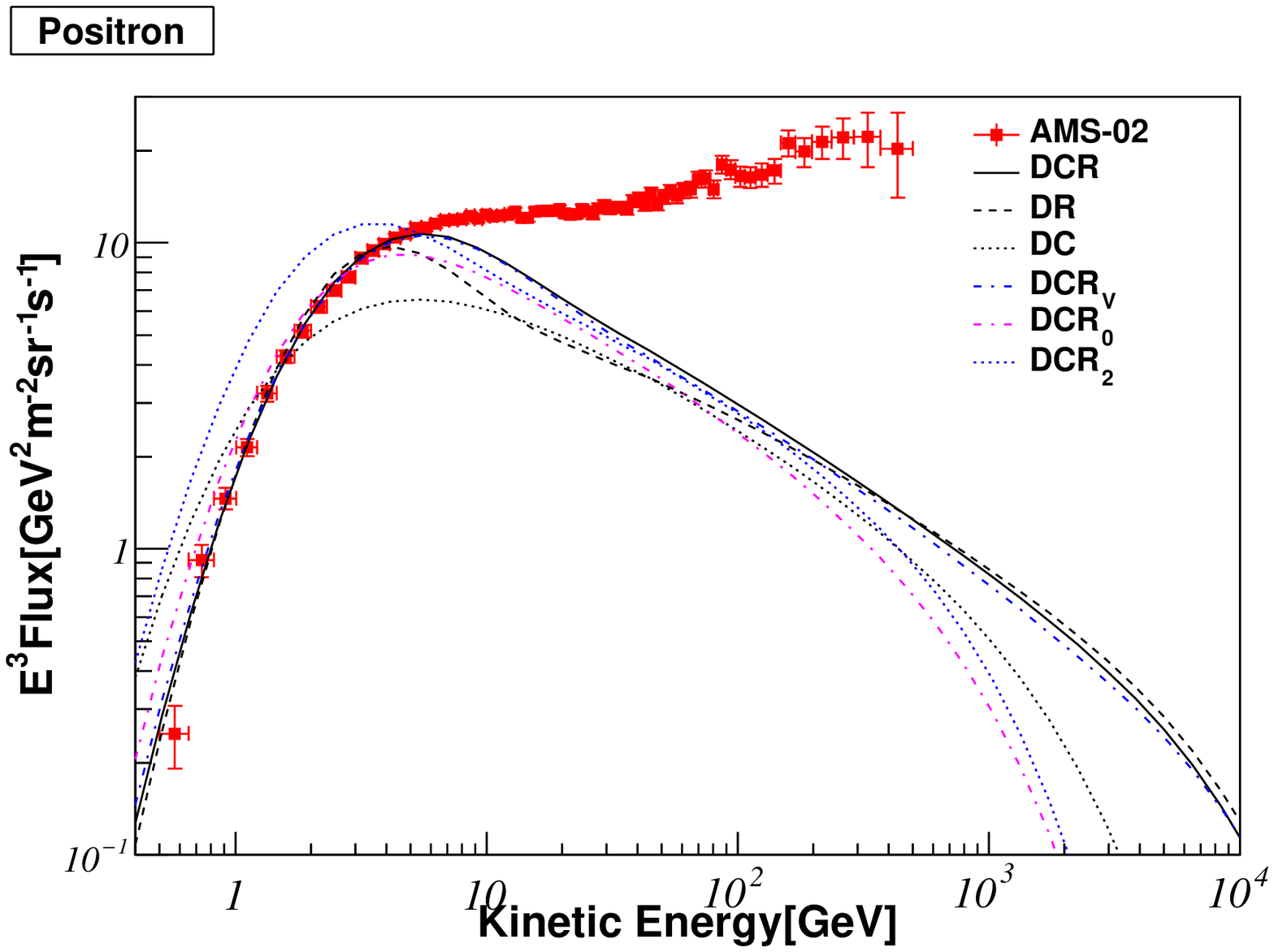}
\caption{In the first row, the flux of CR protons (left) predicted from the models DCR, DR and DC in comparison with the measurements from AMS-02 \citet{Aguilar:2015ooa} and PAMELA \citet{Adriani:2011cu} experiments;  the ratio of Boron to Carbon flux (right) predicted from the models: DCR, DR and DC, and the measurements from  AMS-02 \citet{Aguilar:2016vqr} experiment.
In the second row, the flux of CR antiprotons (left)  and  CR positrons (right) predicted from the models: DCR(DCR${}_{0}$), DR and DC, and the measurements from AMS-02 \citet{Aguilar:2016kjl,Aguilar:2014mma,Aguilar:2014fea}, PAMELA \citet{Adriani:2010rc} and BESS-PolarII \citet{Abe:2011nx} experiments. DCR${}_{0}$ model is relevant to the conventional hadronic interaction models and does not use the expression in Equation \eqref{zFactorsFitExpression}. DCR${}_{1}$ and DCR${}_{2}$ models are used to check the compatibility with ACE data. }
\label{fig:P_BC_Pbar_Positron}
\end{figure}

On the left of the first row of Figure \ref{fig:P_BC_Pbar_Positron}, the fluxes of CR protons are drawn. It is apparently seen that the fluxes of CR protons above 330 GeV have different trends changing with different models. As a result of the comparison among the models, DCR model can well predict the hardening flux of CR protons and the flux at the low energies is also well consistent with the AMS-02 data, which distinguishes from the DR model.
On the right of the first row in Figure \ref{fig:P_BC_Pbar_Positron}, the ratio of Boron to Carbon flux in the DC model is inconsistent with the AMS-02 data below 2 GeV, which indicates that the DC model cannot give a significant prediction.

In Figure \ref{fig:P_BC_Pbar_Positron}, the fluxes of CR antiprotons are plotted at the left of the second row. From the low energy to high one, except for the DC and DCR${}_{0}$ models, the DCR and DR models lead to a consistent prediction for the flux of CR antiprotons, though the tail data of AMS-02 show a bulge, which cannot completely be fitted. At the low energies, the best-fit flux of CR antiprotons prevents the excess interpretation in many papers.  In the DCR${}_{0}$ model, which is relevant to a conventional hadronic interaction model, the excess flux of CR antiprotons is remarkable in the low energies. In some papers, the access flux was interpreted as the contribution from dark matter annihilation. Nevertheless, at the high energies, the bulge relative to the predicted flux of CR antiprotons still exists and needs a further interpretation. We will make a detailed analysis in next paragraphs.

On the right of the second row in Figure \ref{fig:P_BC_Pbar_Positron}, the predicted fluxes of CR positrons are drawn.  In the models DC, DR and DCR${}_{0}$ there are the fully inconsistent prediction with AMS-02 data. In contrast, in the DCR model the flux of CR positrons may be well fitted to the AMS-02 data at the low energies. And the maximal energy range is up to about 10 GeV. Above 10 GeV, the CR positron access becomes significant and begins to increase with the energy raising. It implies that new sources, which are different from the secondary particle production, need to be considered.

\subsection{dark matter interpretations for CR antiproton and positron excesses based on AMS-02 data}
\begin{figure}
\includegraphics[width=0.49\textwidth]{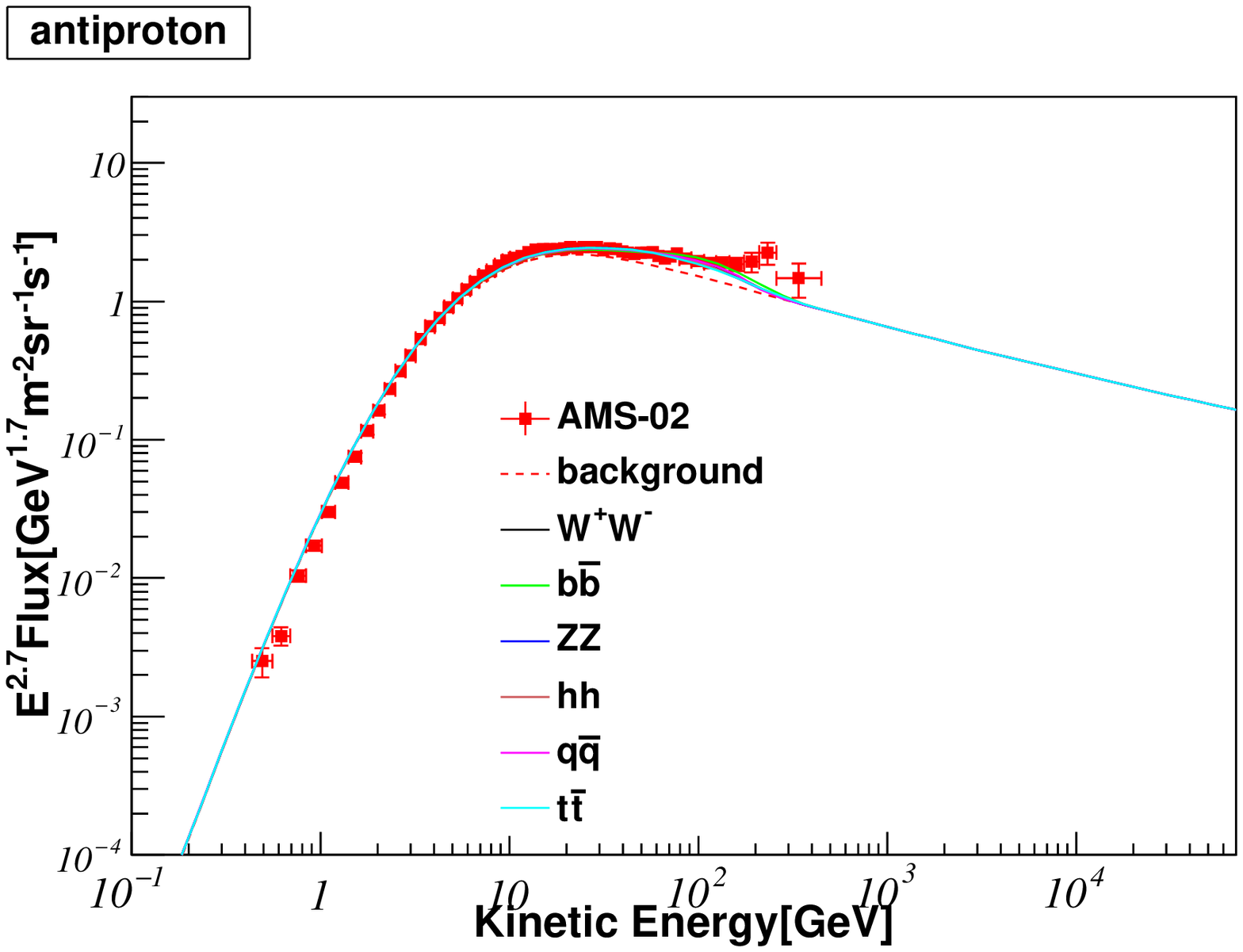}\includegraphics[width=0.49\textwidth]{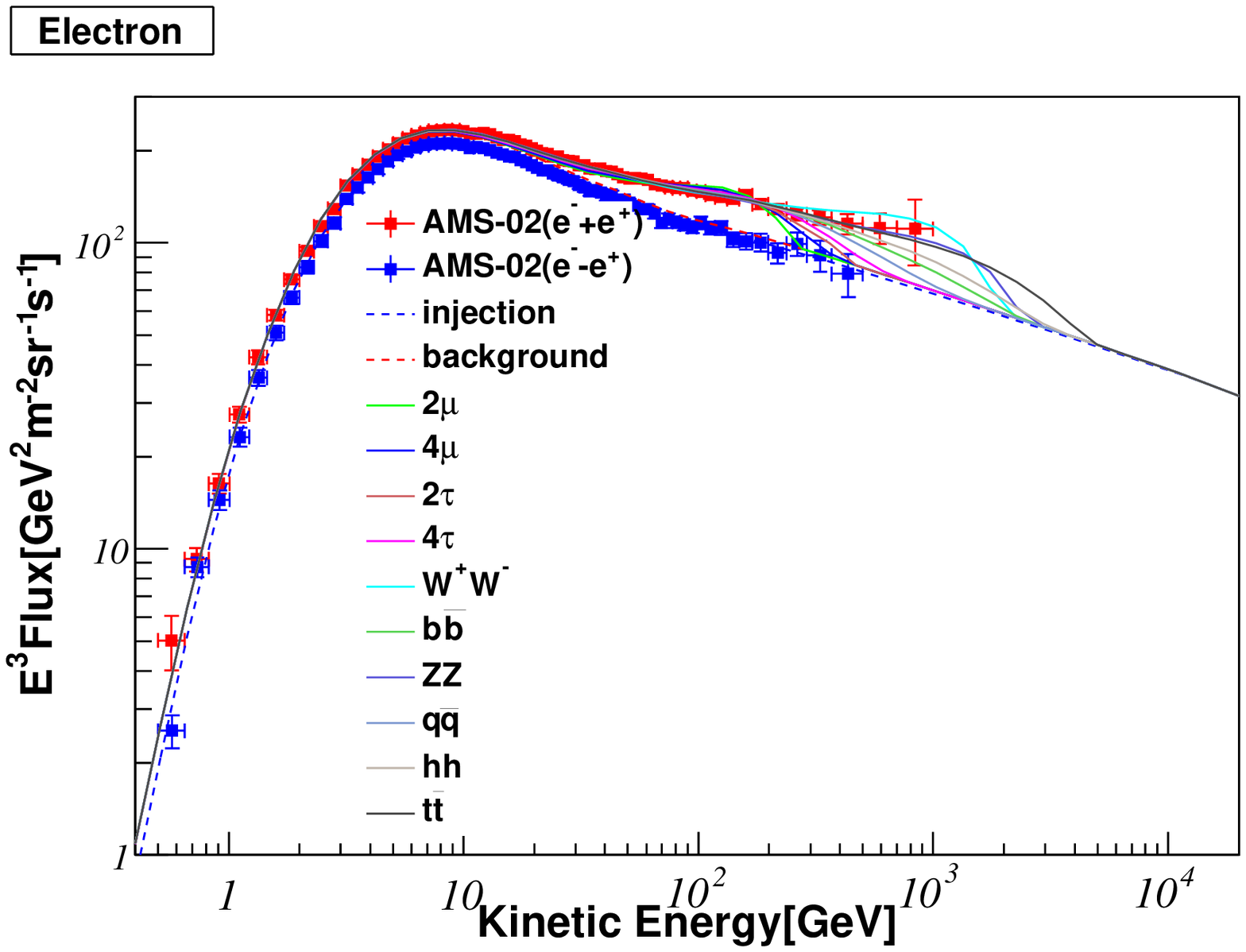}
\includegraphics[width=0.49\textwidth]{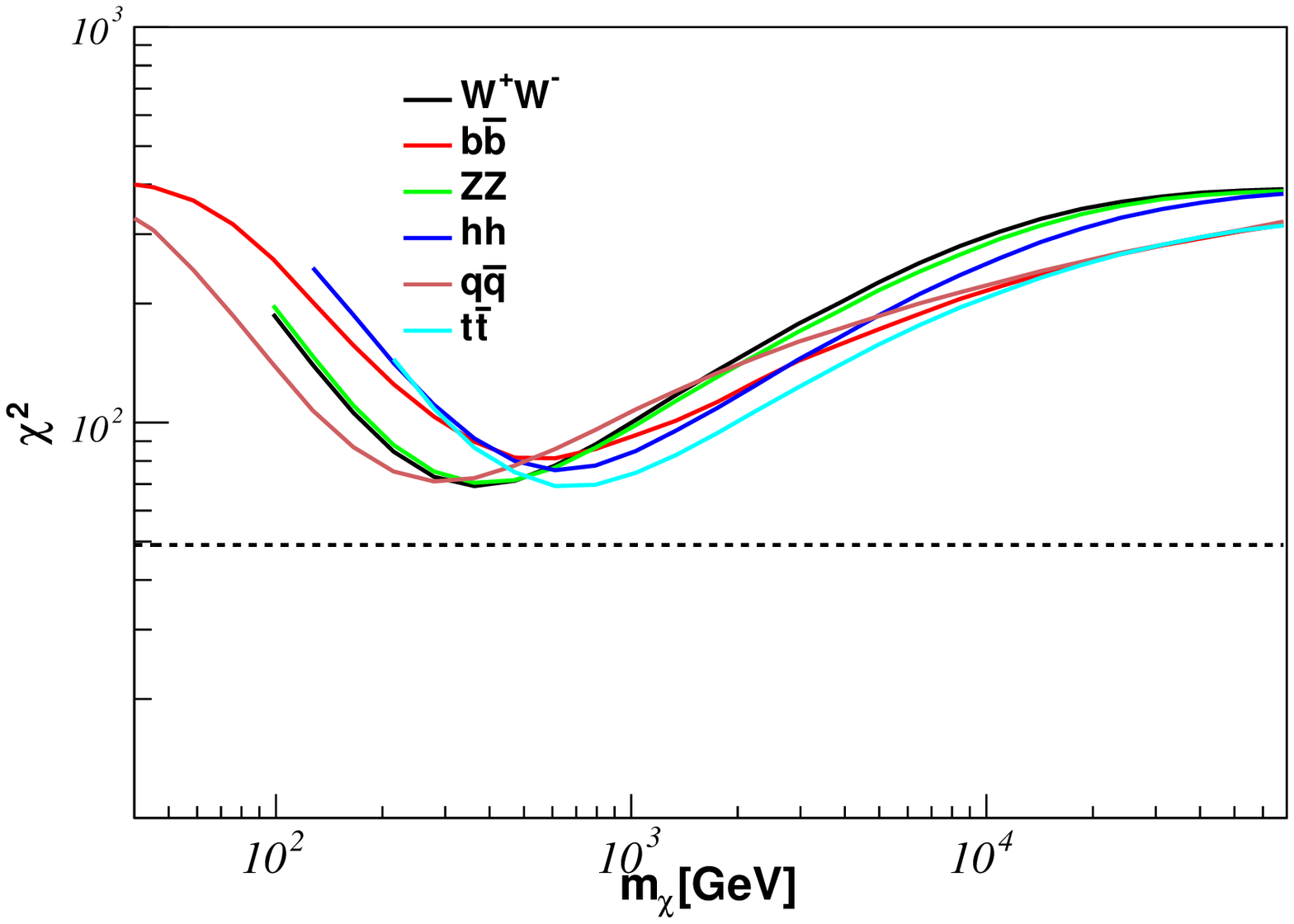}\includegraphics[width=0.49\textwidth]{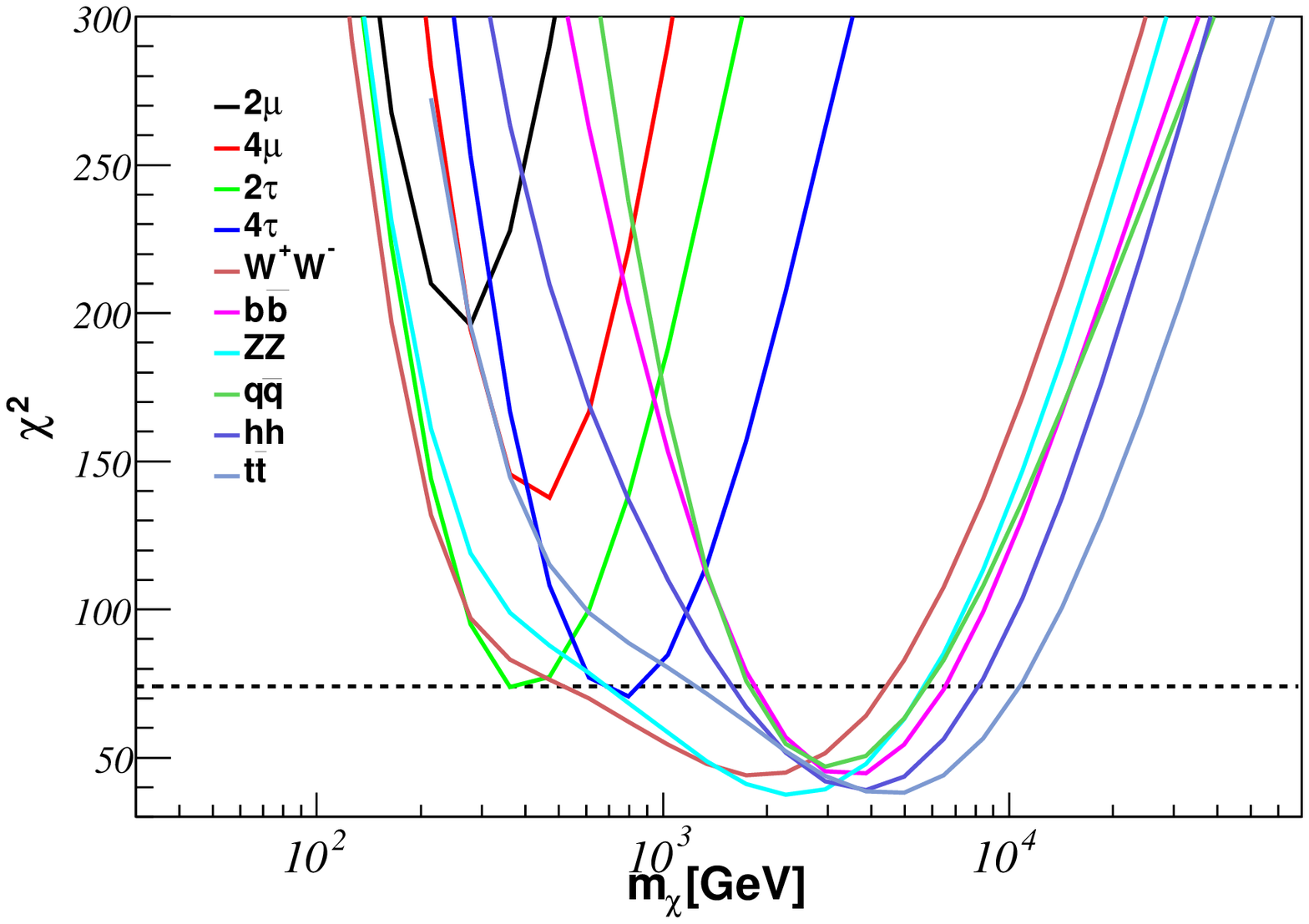}
\includegraphics[width=0.49\textwidth]{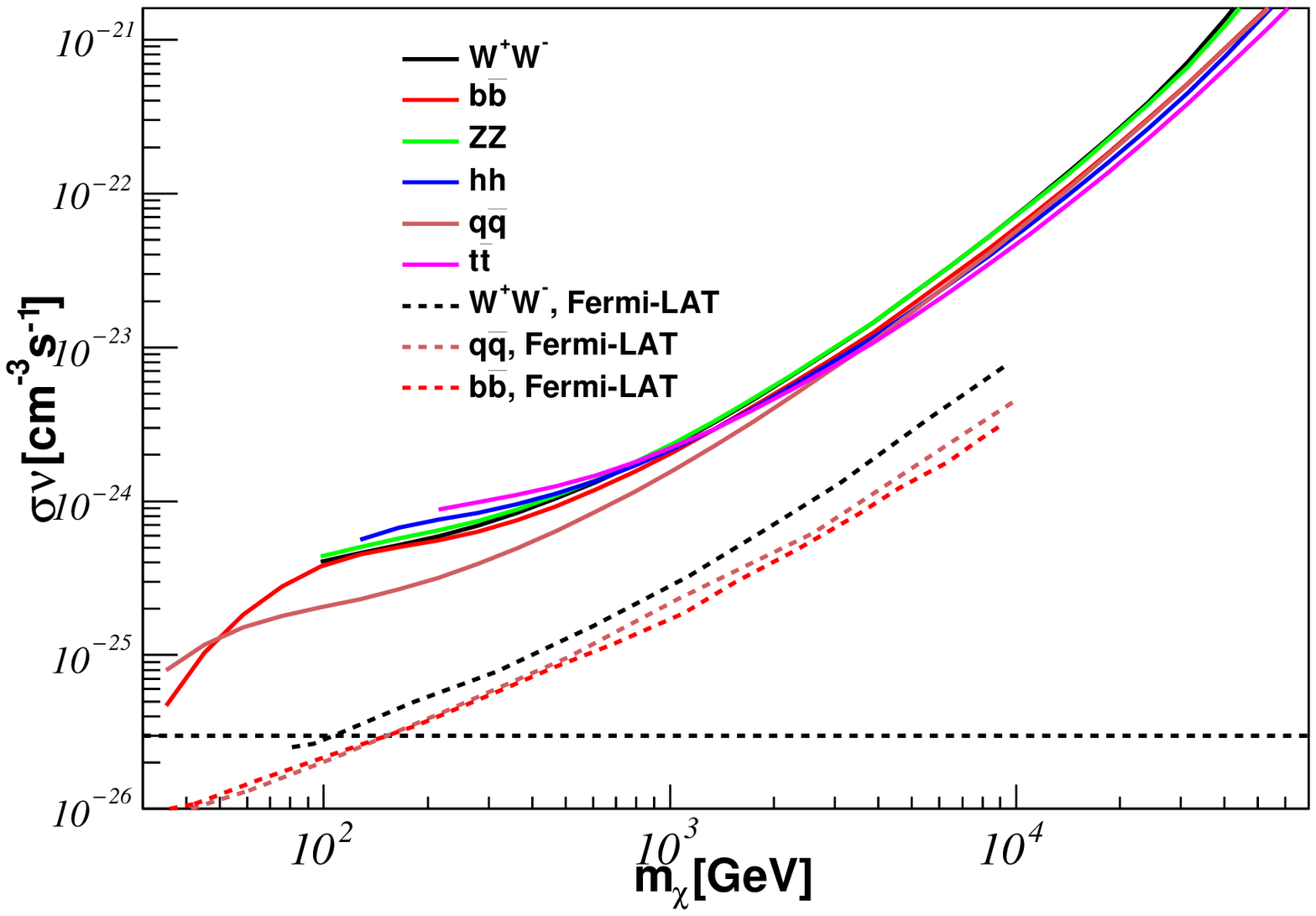}\includegraphics[width=0.49\textwidth]{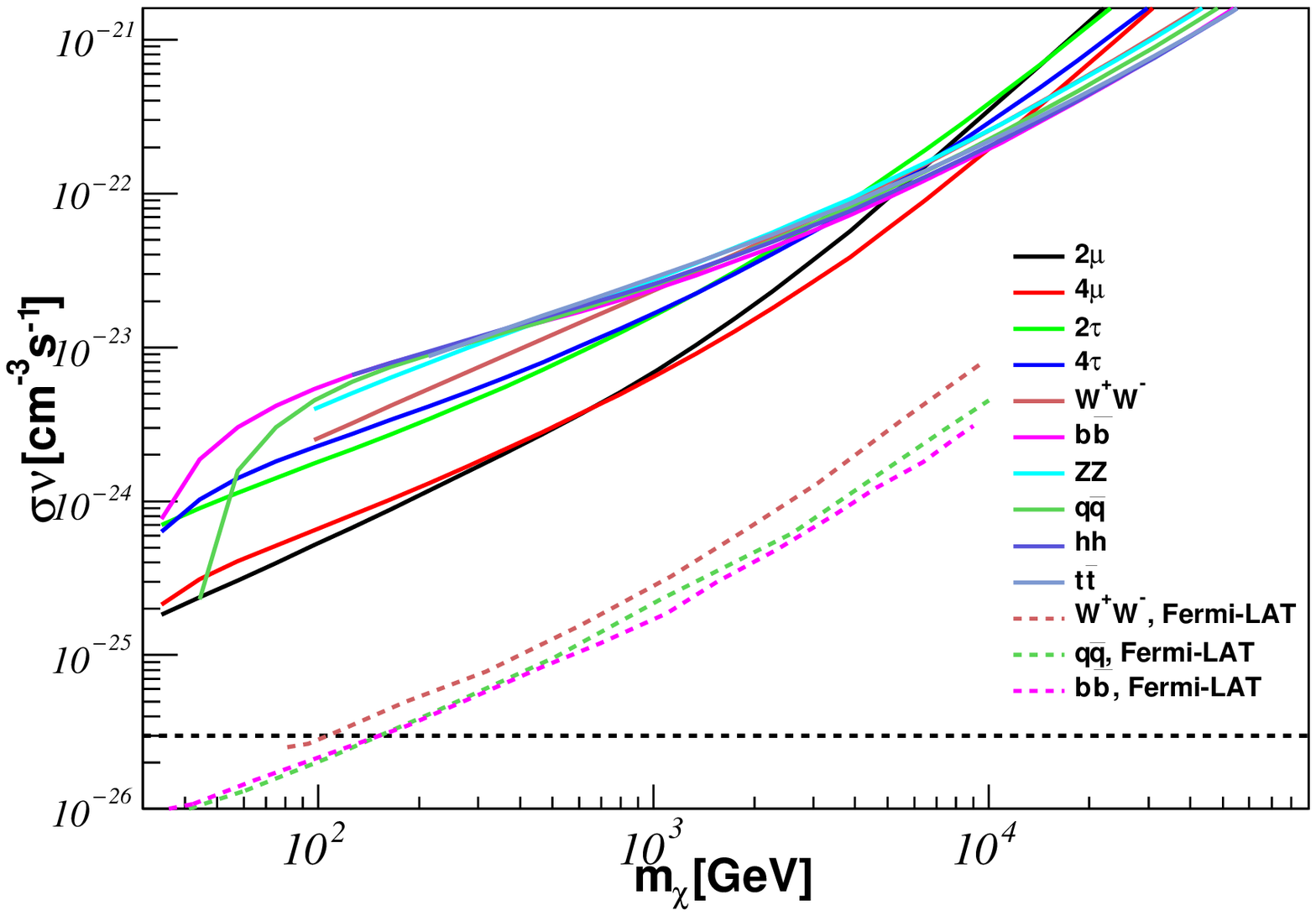}
\caption
{ All of rows: the left for CR antiprotons and the right for the total of CR electrons and positrons.
At the first row, the fluxes of CRs relevant to the astrophysical background, dark matter and AMS-02 experiments \citet{Aguilar:2016kjl,Aguilar:2014mma,Aguilar:2014fea}. ($e^--e^+$) denotes the primary electrons. ($e^-+e^+$) denotes CR electrons and positrons.
At the second row, $\chi^2$ change with the mass of dark matter increasing. The dash-line denotes number of the experimental data points.
At the last row, the mass-fixed best-fit annihilation cross-sections of dark matter via the channels of $b\bar{b}$, etc. The dash-line denotes the thermal cross-section $3\times10^{-26}\mbox{cm}{}^3\mbox{s}{}^{-1}$. The best-fit values relevant to the $q\bar q$, $b\bar b$ and $W^{+}W^{-}$ channels from the Fermi-LAT 6-year gamma-ray data of dwarf spheroidal satellite galaxies of the Milky Way are also shown \citet{Ackermann:2015zua}. 
}
\label{fig:DMCSFlux}
\end{figure}

In Figure \ref{fig:DMCSFlux}, all of the figures are drawn with the data calculated in the DCR model. 
On the left of the last row, the mass-fixed best-fit annihilation cross-sections of dark matter contributing to CR antiprotons are drawn. It is seen that the annihilation cross-sections of dark matter are different completely from the previous paper \citet{Jin:2015sqa}. That situation is relevant to the flux of CR antiprotons fitting to AMS-02 data at the low energies and weaker than AMS-02 data at the high energies. 
It is known that the dark matter interpretation of the positron excess take the large annihilation cross-sections, which does not match the weak excess of the CR antiprotons. As the astrophysical background flux of CR antiprotons in the DCR model is less than the other models, the large annihilation cross-sections of dark matter are predicted. That tension between the excess interpretations of CR antiprotons and positrons is relieved in a certain extent, which may give some hints about symmetry between the quark and lepton particles annihilated from dark matter.
 
In Figure \ref{fig:DMCSFlux}, it is seen that in the whole mass of the dark matter, the annihilation cross-sections are greater than the predicted values from Fermi-LAT gamma-ray data of dwarf spheroidal satellite galaxies of the Milky Way. That is also different completely from the other papers, which present the annihilation cross-sections from CR antiproton excess consistent with gamma-ray data of dwarf spheroidal satellite galaxies. 

In the second row of Figure \ref{fig:DMCSFlux}, the best-fit masses of dark matter are found near 600 GeV. In the different channels of the dark matter annihilation the predicted best-fit masses only are from 300 GeV to 700 GeV. The dark matter interpretation of CR antiproton excess has the weak effect to discriminate the dark matter annihilation channels.

Based on the change of $\chi^2$ with the dark mater mass shown on the left figure of the second row, the best-fit mass of dark matter are found by referring to the minimal $\chi^2$. And the fluxes of CR antiprotons from the total contribution of astrophysical background and dark matter annihilation are drawn on the left of the first row. As the annihilation spectra of dark matter contributing to CR antiprotons are not sharp as well as the CR electrons, the bulge in the tail data of AMS-02 does not strongly constrain the interpretations of dark matter.  As a result, for all the channels the predicted values do not match the fluxes of the last second point from AMS-02 data. 

In general, the analysis of positron excess is based on the total fluxes of CR electrons and positrons, which is different from the CR positron flux or fraction data of AMS-02 used in the previous papers. 
The relevant compatibility of the large annihilation cross section with the thermal relic density is discussed in terms of self-interacting dark matter  \citet{Liu:2011mn,Dev:2013hka,Liu:2013vha,Chen:2015uha,Li:2014vza,Li:2014wia,Liang:2016yjf} and in connection with collider physics  \citet{Huang:2015svl,Huang:2016pzz}.
Being similar to the CR antiprotons, the astrophysical backgrounds of the total electrons and positrons are also calculated in the DCR model. The injection spectra of the primary electrons are best-fit to the $e^{-}-e^{+}$ data of AMS-02 and the relevant power indices are found in Table \ref{tab:para_electronMinusPositon}. Apparently, based on the redefined background of CR electrons, the total fluxes of CR electrons and positrons also have the excess to CR background. On the right figure of first row, the flux of background is well consistent with the AMS-02 data at the low energies, but at the high energies, the experimental data of AMS-02 are manifestly in access of the background. 

In the last row of Figure \ref{fig:DMCSFlux}, the mass-fixed best-fit annihilation cross-sections of dark matter contributing to CR electrons and positrons via many channels: $2\mu$, $W^{+}W^{-}$, etc.,  are plotted. 
As seen in the figures, the mass-fixed best-fit values relevant to the lepton channels are different from the other channels with mass of dark matter increasing, and the trend of the differences between them is clear. 
The cross masses of dark matter between the channels are near 10 TeV, where the same annihilation cross-sections appear. 
In the second row of Figure \ref{fig:DMCSFlux}, the best-fit masses of dark matter are found easily from the change of $\chi^2$ with the mass of dark matter increasing. 
Except for $2\mu$ and $4\mu$ final states, the $\chi^2$ are less than the twice of the experimental data points, and the predicted masses of dark matter are from 400 GeV to 4 TeV. 
The different channels of dark matter annihilation may be discriminated by the best-fit masses of dark matter. That is obviously different from the excess interpretation CR antiprotons. 
The annihilation of heaviest dark matter is via the $t\bar{t}$ final state and the lightest one is relevant to $2\tau$. The total fluxes of CR electrons and positrons from the astrophysical background and the dark matter annihilation are drawn in the first row of Figure \ref{fig:DMCSFlux}. As seen in the figure, in the high energies, the predicted fluxes of CR electrons and positrons are consistent with AMS-02 data and the best fluxes fitting to AMS-02 data are relevant to the $ZZ$ final state. 

In the last row of Figure \ref{fig:DMCSFlux}, the differences of annihilation cross-sections of dark matter for all the annihilation channels are remarkable between the constraints from Fermi-LAT and AMS-02 data. The issues of the large annihilation cross-sections still exist. In details, though the limit lines of annihilation cross-sections relevant to $2\mu$ and $4\mu$ channels are lower than the other channels, the annihilation cross-sections still are one order of magnitude greater than the ones from Fermi-LAT gamma ray data of dwarf spheroidal satellite galaxies. 
The further exploration may be found in the papers \citet{Huang:2015svl,Huang:2016pzz}.

In this paper, as the excess is relevant to the experimental measurement of CR electrons and positrons with ignoring the charge polarity, the interpretation of the excess is easy to be promoted at higher energies for the upcoming data of DAMPE satellite experiment.    

\section{Conclusions}\label{sec:conclusion}

In the re-acceleration diffusion model, the propagation parameters were well constrained only with CR protons and B/C data of AMS-02 in the previous paper \citet{Jin:2014ica}. However, in those parameter models, the predicted fluxes of CR positrons and antiprotons at the low energies deviate significantly from the experimental data. For the latest AMS-02 data, the deviation is more remarkable. In the past analyses of many papers, the situation could not be improved completely. The analysis in ref. \citet{Trotta:2010mx} indicated that the under-predicted antiprotons may result from a general feature of the re-acceleration model. In our present considerations, the CR antiproton production has been recalculated using the QGSJET-II-4 model. As a result, in the re-acceleration diffusion model, the under-predicted flux of CR antiprotons can be enhanced to be consistent with AMS-02 data. Even so, the predicted flux of CR positrons and B/C deviate apparently from AMS-02 data in this model.

In the conventional model, the potential of the solar modulation is taken to be the same for the CR positrons and nucleons. In re-acceleration diffusion model, it has been shown that the prediction of CR positron flux cannot be improved even if the solar modulations are  different between the CR positrons and nucleons.
For the CR positrons, in re-acceleration diffusion model the over-predicted flux could not be depressed completely by QGSJET-II-4 model, which is explained from the Z-factor comparison in Figure \ref{fig:zfactorAll}. In this paper, using the diffusion model combining the convection and re-acceleration terms, the flux of CR positrons is predicted consistently. 

In the further exploration on the tension between fluxes of CR positrons and nucleons predicted by propagation models, it is found that the diffusion model including the convection and re-acceleration effects need be used for the propagation of CRs, and the secondary antiproton product is calculated using QGSJET-II-4 model. Based on two conditions, such a tension is relaxed completely and the predicted fluxes of CR positrons and antiprotons are both consistent with the latest AMS-02 data. In addition, the flux of CR protons with the hardening feature above 330 GeV is well predicted consistently. 

In the DCR model, the predicted value of $\mbox{Be}^{10}/\mbox{Be}^{9}$ is compatible with ACE data only in the propagation parameters constraint only by CR protons and B/C data from AMS-02 without CR antiproton and positron data. 
If the five groups of the experimental data: CR protons, antiprotons, positrons, B/C and Be10/Be9 are included to fit the best-fit parameters, the chi-square over DOF for Be10/Be9 is much greater than 2, but the others are less than 2.  
Thus, as a result, there is a tension between Be$^{10}$/Be$^{9}$ and CR antiprotons and positrons.
Since CR protons and B/C data from AMS-02 do not favor the DR and DC models, Be$^{10}$/Be$^{9}$ data from ACE does not help to constrain the propagation parameters. 

Based on the predicted background of CR positrons and antiprotons, the contribution to CRs from dark matter is analyzed. For CR antiprotons, the slight bulge in the tail data of AMS-02 does not strongly constrain the interpretations of dark matter contributing to CRs. The annihilation cross-sections of dark matter contributing to CR antiprotons are calculated by the constraints of the latest AMS-02 data. The result indicates that the limit lines of the annihilation cross-sections of dark matter are higher apparently than the ones from Fermi-LAT gamma ray data of dwarf spheroidal satellite galaxies. The best-fit masses of dark matter are found near 600 GeV. In the different channels of dark matter annihilation the predicted best-fit masses only are from 300 GeV to 700 GeV. 
The dark matter interpretation of CR antiproton excess has the weak effect to discriminate the dark matter annihilation channels. 
 
Our analysis on the positron excess is based on the total fluxes of CR electrons and positrons, which is different from the CR positron flux or fraction data of AMS-02 adopted in many papers. 
The injection spectra of the primary electrons are best-fit to the $e^{-}-e^{+}$ data of AMS-02. 
As a result, based on the redefined astrophysical background of CR electrons, the total fluxes of CR electrons and positrons are manifestly in excess. 
For the excess of the total fluxes of CR electrons and positrons, the differences of annihilation cross-sections of dark matter for all the annihilation channels are remarkable between the constraints from Fermi-LAT and AMS-02 data. The issues of the large annihilation cross-sections still exist.
Except for $2\mu$ and $4\mu$ final states, in the other channels, the best-fit masses of dark matter are from 400 GeV to 4 TeV.
The annihilation of heaviest dark matter is through the $t\bar{t}$ final state and the lightest one is via the $2\tau$ channel. The best flux fitting to AMS-02 data is relevant to the $ZZ$ final state.

\section*{Acknowledgments}

Y. L. Wu is grateful to S. Ting for insightful discussions. We thank P. Zuccon, A. Kounine, A. Oliva, and S. Haino for helpful discussions on the details of the AMS-02 detector. This work is supported in part by the National Basic Research Program of China (973 Program) under Grant No. 2010CB833000; and the National Nature Science Foundation of China (NSFC) under Grants No. 11335012, No. 11475237 and No.~11121064, and also by the Strategic Priority Research Program of the Chinese Academy of Sciences, Grant No. XDB23030100 as well as  the CAS Center for Excellence in Particle Physics (CCEPP).The numerical calculations were done using the HPC Cluster of SKLTP/ITP-CAS. We thank the referees of this paper for many suggestions, which are very useful to improve many parts in the paper.
\bibliography{../../Database/library}
\bibliographystyle{aasjournal}

\end{document}